\documentclass[11pt,a4paper]{article}
\pdfoutput=1
\usepackage{jcappub}
\usepackage{ifthen}
\usepackage{epsfig}
\usepackage{booktabs} 
\usepackage{natbib}
\usepackage{bbold}

\newcommand{\beqra}{\begin{flalign}}
\newcommand{\eeqra}{\end{flalign}}
\newcommand{\beq}{\begin{equation}}
\newcommand{\eeq}{\end{equation}}
\usepackage{ccicons} %new
\usepackage{listings}%new
\usepackage{color}   %new
\usepackage{braket}  %new
\usepackage{hyperref}%new
\usepackage{dsfont}  %new
\usepackage{leftidx} %new
\usepackage{bm}
\allowdisplaybreaks

\title{Form factors for dark matter capture by the Sun in effective theories}
\author[a]{Riccardo Catena}
\author[a]{and Bodo Schwabe}
\affiliation[a]{Institut f\"ur Theoretische Physik, Friedrich-Hund-Platz 1, 37077 G\"ottingen, Germany}
\emailAdd{riccardo.catena@theorie.physik.uni-goettingen.de}
\emailAdd{bodo.schwabe@theorie.physik.uni-goettingen.de}

\abstract{In the effective theory of isoscalar and isovector dark matter-nucleon interactions mediated by a heavy spin-1 or spin-0 particle, 8 isotope-dependent nuclear response functions can be generated in the dark matter scattering by nuclei. We compute the 8 nuclear response functions for the 16 most abundant elements in the Sun, i.e. H, $^{3}$He, $^{4}$He, $^{12}$C, $^{14}$N, $^{16}$O, $^{20}$Ne, $^{23}$Na, $^{24}$Mg, $^{27}$Al, $^{28}$Si, $^{32}$S, $^{40}$Ar, $^{40}$Ca, $^{56}$Fe, and $^{58}$Ni, through numerical shell model calculations. We use our response functions to compute the rate of dark matter capture by the Sun for all isoscalar and isovector dark matter-nucleon effective interactions, including several operators previously considered for dark matter direct detection only. We study in detail the dependence of the capture rate on specific dark matter-nucleon interaction operators, and on the different elements in the Sun. We find that a so far neglected momentum dependent dark matter coupling to the nuclear vector charge gives a larger contribution to the capture rate than the constant spin-dependent interaction commonly included in dark matter searches at neutrino telescopes. Our investigation lays the foundations for model independent analyses of dark matter induced neutrino signals from the Sun. The nuclear response functions obtained in this study are listed in analytic form in an appendix, ready to be used in other projects.}

\keywords{dark matter theory, dark matter experiments} 
\begin{document}
\maketitle

\section{Introduction}
The quest for dark matter is at a turning point. Data from direct, indirect and collider searches for dark matter with unprecedented exposure, resolution and extension in energy will finally be available during the next 5-10 years~\cite{Cakir:2014nba,Catena:2013pka,Ibarra:2012cc,Bergstrom:2012vd,Bringmann:2012ez,Baudis:2012ig,Bertone:2010at}. Efficient strategies to globally interpret these data in terms of dark matter particle mass and interaction properties are of prime importance in astroparticle physics~\cite{Strege:2014ija,Buchmueller:2013rsa,Bechtle:2012zk}. 

Effective theory methods have proven to be a very powerful tool in the analysis of collider data~\cite{Zhou:2013raa,Rajaraman:2011wf,Goodman:2010ku,Goodman:2010yf}, dark matter direct~\cite{Gluscevic:2014vga,Catena:2014uqa,Catena:2014epa,Catena:2014hla,Panci:2014gga,DelNobile:2013sia,Fitzpatrick:2012ib,Fornengo:2011sz,DelNobile:2011uf} and indirect~\cite{Chen:2013gya,Rajaraman:2012fu,Goodman:2010qn} detection experiments, and in combined studies of these different strategies~\cite{Fedderke:2014wda,Alves:2014yha,Alves:2015pea}. The main advantage of the effective theory approach to dark matter is that it allows for a model independent interpretation of the different observations when all relevant interaction operators are simultaneously explored in multidimensional statistical analyses, as for instance in~\cite{Catena:2014uqa,Catena:2014epa,Catena:2014hla}. In contrast, comparing a simplistic model for dark matter to observations, important physical properties can be missed, and spurious correlations among physical observables can be enforced by the inappropriately small number of model parameters.

In the context of effective theories for dark matter, the dark matter-nucleus interaction plays a special role, in that its exploration is complicated by non trivial properties related to the internal structure of the nuclei in analysis. The effective theory of dark matter-nucleon interactions~\cite{Fitzpatrick:2012ix,Fan:2010gt} predicts that 8 independent nuclear response functions - or form factors~- can be generated in the dark matter scattering by nuclei. The interpretation of any dark matter experiment probing the dark matter-nucleus interaction is unavoidably affected by the uncertainties within which the 8 nuclear response functions are known. Experiments of this type are dark matter direct detection experiments, and neutrino telescopes searching for solar neutrinos from dark matter annihilations. In this work we concentrate on the latter ones.

Dark matter can be captured by the Sun while scattering in the solar medium. Dark matter particles accumulated in the Sun might annihilate producing a flux of potentially observable energetic neutrinos~\cite{Edsjo:1997hp}. The solar neutrino flux from dark matter annihilations is strictly related to the rate of dark matter capture by the Sun (e.g. proportional to the latter, assuming equilibrium between capture and annihilation~\cite{Edsjo:1997hp}). It is therefore a function of the cross-section for dark matter-nucleus scattering, which in turn depends on the nuclear response functions computed in this work. For constant spin-independent dark matter-nucleon interactions, nuclear response functions for the most abundant element in the Sun are approximately known~\cite{Gondolo:2004sc}. For constant spin-dependent interactions, dark matter is assumed to scatter off Hydrogen only, and nuclei with a more complex structure are neglected. Finally, for momentum and velocity dependent dark matter-nucleon interactions, only simplified calculations have so far been performed in the literature. In Ref.~\cite{Liang:2013dsa}, for instance, the rate of dark matter capture by the Sun is calculated for 6 momentum/velocity dependent operators, considering dark matter scattering from Hydrogen only. Recently, momentum and velocity dependent dark matter-nucleon interactions have also been explored in the context of helioseismology~\cite{Vincent:2013lua,Lopes:2014aoa,Vincent:2014jia}.

Nuclear response functions for model independent analyses of dark matter direct detection experiments have been calculated in~\cite{Vietze:2014vsa,Fitzpatrick:2012ix} under the assumption of one-body dark matter-nucleon interactions. Two-body interactions have also been included in~\cite{Klos:2013rwa,Menendez:2012tm} in an investigation of spin-dependent dark matter-nucleus currents. In addition, two-body contributions to spin-independent dark matter-nucleon interactions have been claimed to be important in dark matter direct detection in~\cite{Prezeau:2003sv,Cirigliano:2013zta,Cirigliano:2012pq}. The nuclear response functions for isotopes of Xe, I, Ge, Na, and F found in~\cite{Fitzpatrick:2012ix} have been applied to complementary analyses of current direct detection experiments~\cite{Catena:2014uqa,Gresham:2014vja,DelNobile:2013sia}, and in studies of the prospects for dark matter direct detection~\cite{Catena:2014epa,Catena:2014hla}. 

In this paper we calculate the 8 nuclear response functions generated in the dark matter scattering by nuclei for the 16 most abundant elements in the Sun. We then use the novel response functions to calculate the rate of dark matter capture by the Sun within the general effective theory of isoscalar and isovector dark matter-nucleon interactions mediated by a heavy spin-1 or spin-0 particle. In the analysis, we comprehensively describe how the capture rate depends on specific dark matter-nucleon interaction operators, and on the elements in the Sun. This study constitutes the first step towards robust model independent analyses of dark matter induced neutrino signals from the Sun.

The paper is organized as follows. In Sec.~\ref{sec:astro} we provide the equations for computing the rate of dark matter capture by the Sun given an arbitrary dark matter-nucleon interaction. In Sec.~\ref{sec:eft} we review the effective theory of dark matter-nucleon interactions, while in Sec.~\ref{sec:obdme} we calculate the 8 nuclear response functions predicted by the theory for the most abundant elements in the Sun. We calculate the dark matter capture rate for all isoscalar and isovector dark matter-nucleon interactions in Sec.~\ref{sec:rate}, and we conclude in Sec.~\ref{sec:conc}. The dark matter response functions and the single-particle matrix elements needed in the analysis are listed in the Appendixes~\ref{sec:appDM} and \ref{sec:appME}, respectively. Finally, in Appendix~\ref{sec:appNuc} we provide the nuclear response functions of this work in analytic form.

\section{Dark matter capture by the Sun}
\label{sec:astro}
Dark matter particles of the galactic halo with interactions at the electroweak scale can be gravitationally captured by the Sun. For a dark matter particle of mass $m_\chi$ at a distance $R$ from the center of the Sun, the rate of scattering from a velocity $w$ to a velocity less than the local escape velocity $v(R)$ is given by~\cite{Gould:1987ir}
\begin{equation}
\Omega_{v}^{-}(w)= \sum_i n_i w\,\Theta\left( \frac{\mu_i}{\mu^2_{+,i}} - \frac{u^2}{w^2} \right)\int_{E_k u^2/w^2}^{E_k \mu_i/\mu_{+,i}^2} {\rm d}E\,\frac{{\rm d}\sigma_{i}}{{\rm d}E}\left(w^2,q^2\right)\,.
\label{eq:omega}
\end{equation}
In Eq.~(\ref{eq:omega}), $E_k=m_\chi w^2/2$, $d\sigma_i/dE$ is the differential cross-section for dark matter scattering by nuclei of mass $m_i$ and density $n_i(R)$ in the Sun, $q$ is the momentum transfer and $E=~q^2/(2m_i)$ the nuclear recoil energy. The sum in the scattering probability extends over the most abundant elements in the Sun, and the dimensionless parameters $\mu_i$ and $\mu_{\pm,i}$  are defined as follows 
\begin{equation}
\mu_i\equiv \frac{m_\chi}{m_i}\, \qquad\qquad \mu_{\pm,i}\equiv \frac{\mu_i\pm1}{2}\,.
\end{equation}
The velocity $u$ in Eq.~(\ref{eq:omega}) is the velocity of the dark matter particle at $R\rightarrow \infty$, where the Sun's gravitational potential is negligible. The relation between $u$ and  $w$ is $w=\sqrt{u^2+v(R)^2}$, and therefore $\Omega_{v}^{-}(w)$ depends on $R$.

In Eq.~(\ref{eq:omega}), we consider the general case in which the differential scattering-cross section depends both on the momentum transfer $q$, and on the dark matter-nucleus relative velocity $w$. We therefore relax the assumption of constant total cross-section, commonly made in this context. This generalization is important in the study of arbitrary dark matter-nucleus interactions, as we will see in the next sections.

Consider now a population of halo dark matter particles with speed distribution at infinity given by $f(u)$. A fraction of them will be captured by the Sun, with a differential capture rate per unit volume given by~\cite{Gould:1987ir}
\begin{equation}
\frac{{\rm d} C}{{\rm d}V} = \int_{0}^{\infty} {\rm d}u\, \frac{f(u)}{u}\, w\Omega_{v}^{-}(w) \,.
\end{equation}
The total capture rate takes the following form
\begin{equation}
C = 4\pi \int_{0}^{R_{\odot}} {\rm d} R\, R^2\,\frac{{\rm d} C}{{\rm d}V} \left(R\right)\,,
\label{eq:rate}
\end{equation}
where we integrate over a sphere of radius $R_{\rm \odot}$, corresponding to the volume of the Sun. The aim of this work is to evaluate Eq.~(\ref{eq:rate}) within the most general effective theory for dark matter-nucleon one-body interactions mediated by heavy spin-1 or spin-0 particles, using for each element in the Sun the appropriate nuclear response functions.

From Eq.~(\ref{eq:rate}), one can calculate the differential neutrino flux from dark matter annihilations in the Sun. It is given by~\cite{Edsjo:1997hp}
\begin{equation}
\frac{{\rm d \Phi_\nu}}{{\rm d E_\nu}} = \frac{\Gamma_A}{4\pi D^2} \sum_{f} B_{\chi}^{f} \, \frac{{\rm d N_\nu^f}}{{\rm d E_\nu}} \,
\label{eq:nuflux}
\end{equation}
where $E_{\nu}$ is the neutrino energy, $\Gamma_A$ the total dark matter annihilation rate, $B_{\chi}^{f}$ the branching ratio for dark matter pair annihilation into the final state $f$, and $D$ the distance from the observer to the center of the Sun. ${\rm d} N_\nu^f/{\rm d} E_\nu$ is the energy spectrum of neutrinos produced by dark matter annihilation into the final state $f$. In general, $\Gamma_A = (C/2) \tanh^2(t/\tau)$, where $t$ is the time variable, and $\tau$ the characteristic time scale for the equilibration of dark matter capture and annihilation.

In our calculations we consider the most abundant elements in the Sun, and use the densities $n_i(R)$ and the velocity $v(R)$ as implemented in the {\sffamily darksusy} code~\cite{Gondolo:2004sc}. Accordingly, we include in the analysis the following 16 elements: H, $^{3}$He, $^{4}$He, $^{12}$C, $^{14}$N, $^{16}$O, $^{20}$Ne, $^{23}$Na, $^{24}$Mg, $^{27}$Al, $^{28}$Si, $^{32}$S, $^{40}$Ar, $^{40}$Ca, $^{56}$Fe, and $^{58}$Ni. Finally, we assume the so-called standard halo model~\cite{Freese:2012xd}, with a Maxwell-Boltzmann speed distribution for $f(u)$, and a local standard of rest velocity of 220 km~s$^{-1}$. We leave an analysis of astrophysical uncertainties~\cite{Bozorgnia:2013pua,Catena:2011kv,Catena:2009mf} in the evaluation of Eq.~(\ref{eq:rate}) for future work.

\section{Dark matter-nucleus scattering}
\label{sec:eft}
In this section we review the theory of dark matter scattering by nucleons and nuclei~\cite{Fitzpatrick:2012ix}.
\subsection{Dark matter-nucleon effective interactions}
\label{sec:dmnu}
The amplitude for dark matter-nucleon elastic scattering, $\mathcal{M}$, is restricted by energy and momentum conservation, and respects Galilean invariance, i.e. the invariance under constant shifts of the tridimensional particle velocities. These restrictions determine how $\mathcal{M}$ depends on the momenta of the incoming and outgoing particles. Let us denote by ${\bf p}$ (${\bf p'}$) and  ${\bf k}$ (${\bf k'}$) the initial (final) dark matter and nucleon tridimensional momenta, respectively. Momentum conservation implies that only three out of these four momenta are independent in the scattering process. A possible choice of independent momenta is ${\bf p}$, ${\bf k}$, and ${\bf q}\equiv {\bf k}-{\bf k'}$, where ${\bf q}$ is the momentum transferred from the nucleon to the dark matter particle.    
Whereas ${\bf q}$ is Galilean invariant, ${\bf p}$ and ${\bf k}$ are not. Galilean invariance therefore implies that $\mathcal{M}$ must depend on the difference ${\bf v}\equiv {\bf p}/m_\chi -  {\bf k}/m_N $, rather than on ${\bf p}$ and ${\bf k}$ separately. ${\bf v}$ is the initial relative velocity between a dark matter particle of mass $m_\chi$ and a nucleon of mass $m_N$. In addition to particle momenta, $\mathcal{M}$ can depend on the dark matter particle and nucleon spins, $j_\chi$ and $j_N$, respectively. 

Any non-relativistic quantum mechanical Hamiltonian leading to a scattering amplitude obeying such restrictions can be expressed as a combination of the following five Hermitian operators
\begin{equation}
\mathbb{1}_{\chi N} \qquad\quad i{\bf{\hat{q}}} \qquad\quad {\bf{\hat{v}}}^{\perp} \qquad\quad {\bf{\hat{S}}}_{\chi} \qquad\quad {\bf{\hat{S}}}_{N}  \,.
\label{eq:5Op}
\end{equation}
The five operators in Eq.~(\ref{eq:5Op}) act on the two-particle Hilbert space spanned by tensor products of dark matter and nucleon
states, respectively $|{\bf p},j_\chi\rangle$ and $|{\bf k},j_N\rangle$.
The operator $\mathbb{1}_{\chi N}$ is the identity in this space, whereas ${\bf\hat{S}}_{\chi}$ and ${\bf\hat{S}}_{N}$ denote the dark matter particle and nucleon spin operators. Finally, $i{\bf{\hat{q}}}$ is the Hermitian transfer momentum operator, and ${\bf{\hat{v}^{\perp}}}$ the relative transverse velocity operator.
They are Galilean invariant, and characterized by the matrix elements 
\begin{eqnarray}
\label{eq:me1}
\langle {\bf p'},j_\chi;  {\bf k'},j_N| \,  i{\bf{\hat{q}}} \,|{\bf p},j_\chi;  {\bf k},j_N \rangle  &=& i {\bf q} \,e^{-i {\bf q} \cdot {\bf r}} \,(2\pi)^3\delta({\bf k'}+{\bf p'}-{\bf k}-{\bf p}) \\
\label{eq:me2}
\langle {\bf p'},j_\chi;  {\bf k'},j_N| \, {\bf{\hat{v}^{\perp}}} \, |{\bf{p}},j_\chi;  {\bf k},j_N \rangle  &=& \left({\bf v}+\frac{{\bf q}}{2\mu_N}\right) e^{-i {\bf q} \cdot {\bf r}} \,(2\pi)^3\delta({\bf k'}+{\bf p'}-{\bf k}-{\bf p})
\end{eqnarray} 
where $\mu_N$ is the reduced mass of the dark matter-nucleon system, and ${\bf r}$ is the position vector from the nucleon to the dark matter particle. 
Notice that energy conservation implies ${\bf v}\cdot{\bf q}=-q^2/(2\mu_N)$, and hence $ {\bf{v}^{\perp}}\cdot {\bf q}=0$, with ${\bf{v}^{\perp}} \equiv {\bf v}+{\bf q}/(2\mu_N)$. This justifies the use of the notation $\bf{\hat{v}^{\perp}}$. In Eqs.~(\ref{eq:me1}) and (\ref{eq:me2}) we adopt a non-relativistic normalization for single-particle states. 

\begin{table}[t]
    \centering
    \begin{tabular}{ll}
    \toprule
        $\hat{\mathcal{O}}_1 = \mathbb{1}_{\chi N}$ & $\hat{\mathcal{O}}_9 = i{\bf{\hat{S}}}_\chi\cdot\left(\hat{{\bf{S}}}_N\times\frac{{\bf{\hat{q}}}}{m_N}\right)$  \\
        $\hat{\mathcal{O}}_3 = i\hat{{\bf{S}}}_N\cdot\left(\frac{{\bf{\hat{q}}}}{m_N}\times{\bf{\hat{v}}}^{\perp}\right)$ \hspace{2 cm} &   $\hat{\mathcal{O}}_{10} = i\hat{{\bf{S}}}_N\cdot\frac{{\bf{\hat{q}}}}{m_N}$   \\
        $\hat{\mathcal{O}}_4 = \hat{{\bf{S}}}_{\chi}\cdot \hat{{\bf{S}}}_{N}$ &   $\hat{\mathcal{O}}_{11} = i{\bf{\hat{S}}}_\chi\cdot\frac{{\bf{\hat{q}}}}{m_N}$   \\                                                                             
        $\hat{\mathcal{O}}_5 = i{\bf{\hat{S}}}_\chi\cdot\left(\frac{{\bf{\hat{q}}}}{m_N}\times{\bf{\hat{v}}}^{\perp}\right)$ &  $\hat{\mathcal{O}}_{12} = \hat{{\bf{S}}}_{\chi}\cdot \left(\hat{{\bf{S}}}_{N} \times{\bf{\hat{v}}}^{\perp} \right)$ \\                                                                                                                 
        $\hat{\mathcal{O}}_6 = \left({\bf{\hat{S}}}_\chi\cdot\frac{{\bf{\hat{q}}}}{m_N}\right) \left(\hat{{\bf{S}}}_N\cdot\frac{\hat{{\bf{q}}}}{m_N}\right)$ &  $\hat{\mathcal{O}}_{13} =i \left(\hat{{\bf{S}}}_{\chi}\cdot {\bf{\hat{v}}}^{\perp}\right)\left(\hat{{\bf{S}}}_{N}\cdot \frac{{\bf{\hat{q}}}}{m_N}\right)$ \\   
        $\hat{\mathcal{O}}_7 = \hat{{\bf{S}}}_{N}\cdot {\bf{\hat{v}}}^{\perp}$ &  $\hat{\mathcal{O}}_{14} = i\left(\hat{{\bf{S}}}_{\chi}\cdot \frac{{\bf{\hat{q}}}}{m_N}\right)\left(\hat{{\bf{S}}}_{N}\cdot {\bf{\hat{v}}}^{\perp}\right)$  \\
        $\hat{\mathcal{O}}_8 = \hat{{\bf{S}}}_{\chi}\cdot {\bf{\hat{v}}}^{\perp}$  & $\hat{\mathcal{O}}_{15} = -\left(\hat{{\bf{S}}}_{\chi}\cdot \frac{{\bf{\hat{q}}}}{m_N}\right)\left[ \left(\hat{{\bf{S}}}_{N}\times {\bf{\hat{v}}}^{\perp} \right) \cdot \frac{{\bf{\hat{q}}}}{m_N}\right] $ \\                                                                               
    \bottomrule
    \end{tabular}
    \caption{Non-relativistic quantum mechanical operators constructed from Eq.~(\ref{eq:5Op}). Introducing the nucleon mass, $m_N$, in the equations all operators have the same mass dimension.} 
    \label{tab:operators}
\end{table}

Only 14 linearly independent quantum mechanical operators can be constructed from (\ref{eq:5Op}), if we demand that they are at most linear in $\hat{{\bf{S}}}_N$, $\hat{{\bf{S}}}_\chi$ and $\bf{\hat{v}^{\perp}}$. They are listed in Tab.~\ref{tab:operators}, and labelled as in \cite{Anand:2013yka}, where the operator $\hat{\mathcal{O}}_2={\bf{\hat{v}}}^{\perp}\cdot{\bf{\hat{v}}}^{\perp}$ was neglected since it cannot be a leading-order operator in effective theories. They are at most quadratic in the momentum transfer, with the exception of $\hat{\mathcal{O}}_{15}$, that is cubic in ${\bf{\hat{q}}}$. The restriction on the number of spin/transverse relative velocity operators is a constraint on the spin of the particle mediating the underlying relativistic interaction, that is assumed here to be less than or equal to 1. Tab.~\ref{tab:operators} also assumes that the mediating particle is heavy compared to the momentum transfer, i.e. long-range interactions are not included. 

The most general Hamiltonian density for dark matter-nucleon interactions mediated by a heavy spin-0 or spin-1 particle is hence a linear combination of 14 non-relativistic quantum mechanical operators, $\hat{\mathcal{O}}_k$, and is given by
\begin{eqnarray}
{\bf\hat{\mathcal{H}}}({\bf{r}}) &=& 2 \sum_{k=1}^{15} \left[ c_k^{p} \left( \frac{\mathbb{1}+ \tau_3}{2} \right) + c_k^{n} \left( \frac{\mathbb{1} - \tau_3}{2} \right) \right] \hat{\mathcal{O}}_{k}({\bf{r}}) \,.
\label{eq:H}
\end{eqnarray}
In Eq.~(\ref{eq:H}), $c^{p}_k$ and $c^{n}_k$ are the coupling constants for protons and neutrons as implemented in~\cite{Anand:2013yka}, and have dimension mass$^{-2}$. By construction, $c_2^p=c_2^n=0$. $\tau_3$ is the third Pauli matrix, and $\mathbb{1}$ denotes the $2\times2$ identity in isospin space. The matrices $(\mathbb{1} \pm \tau_{3})/2$ project a nucleon state into states of well defined isospin, i.e. protons and neutrons. As for the building blocks in Eq.~(\ref{eq:5Op}), the operators $\hat{\mathcal{O}}_k$ act on the two-particle Hilbert space spanned by tensor products of dark matter and nucleon states, $|{\bf p},j_\chi\rangle$ and $|{\bf k},j_N\rangle$, respectively. In the calculation of nuclear matrix elements for dark matter scattering by nuclei with well defined isospin quantum numbers, it is convenient to rewrite Eq.~(\ref{eq:H}) in terms of isoscalar and isovector coupling constants:
\begin{eqnarray}
{\bf\hat{\mathcal{H}}}({\bf{r}})&=& \sum_{\tau=0,1} \sum_{k=1}^{15} c_k^{\tau} \hat{\mathcal{O}}_{k}({\bf{r}}) \, t^{\tau} \,.
\label{eq:Hc0c1}
\end{eqnarray}
In Eq.~(\ref{eq:Hc0c1}) $t^0=\mathbb{1}$, $t^1=\tau_3$, and the isoscalar and isovector coupling constants, respectively, $c^0_k$ and $c^{1}_k$, are related to $c^{p}_k=(c^{0}_k+c^{1}_k)/2$ and $c^{n}_k=(c^{0}_k-c^{1}_k)/2$.

\subsection{Dark matter-nucleus effective interactions}
We construct the dark matter-nucleus interaction Hamiltonian density, $\hat{\mathcal{H}}_{\rm T}({\bf{r}})$, from Eq.~(\ref{eq:Hc0c1}) under the assumption of one-body dark matter-nucleon interactions. Within this assumption, 
$\hat{\mathcal{H}}_{\rm T}({\bf{r}})$ is the sum of $A$ terms of type (\ref{eq:Hc0c1}), one for each of the $A$ nucleons in the target nucleus
\begin{equation}
\hat{\mathcal{H}}_{\rm T}({\bf{r}}) = \sum_{i=1}^{A}  \sum_{\tau=0,1} \sum_{k=1}^{15} c_k^{\tau}\hat{\mathcal{O}}_{k}^{(i)}({\bf{r}}) \, t^{\tau}_{(i)} \,.
\label{eq:H_I}
\end{equation}
The Hermitian and Galilean invariant quantum mechanical operators $\hat{\mathcal{O}}_{k}^{(i)}({\bf{r}})$, $k=1,\dots,15$, are listed in Tab.~\ref{tab:operators}. We use the index $i$ to identify the specific nucleon to which dark matter couples in the scattering. Distinct nucleons are characterized by different isospin matrices $t^{\tau}_{(i)}$. 

Notice that within the effective theory approach reviewed here, the spin-dependent operators $\hat{\mathcal{O}}_{4}$ and $\hat{\mathcal{O}}_{6}$ are treated independently. In contrast, the non-relativistic limit of a contact axial-axial dark matter-nucleon interaction generates a linear combination of the two operators~\cite{Engel:1992}. Standard spin-dependent form factors used to interpret direct detection experiments account for this linear combination. The situation is different in the context of dark matter searches with neutrino telescopes. Here the operator $\hat{\mathcal{O}}_{4}$ is considered separately, in that for spin-dependent interactions dark matter is assumed to only scatter off Hydrogen with a constant cross-section, and obviously with no nuclear form factor.

In the single-particle state of the $i$th-nucleon, it is convenient to separate the motion of the nucleus center of mass from the intrinsic motion (relative to the nucleus center of mass) of the nucleon itself. This separation induces the following coordinate space representations for ${\bf{\hat{q}}}$ and ${\bf{\hat{v}}}^{\perp}$:
\begin{eqnarray}
\label{eq:qx}
{\bf{\hat{q}}} &=& - i \overleftarrow{\nabla}_{{\bf{x}}} \,\delta({\bf{x}}-{\bf{y}}+{\bf{r}}) - i\delta({\bf{x}}-{\bf{y}}+{\bf{r}}) \overrightarrow{\nabla}_{{\bf{x}}}  \\
\nonumber\\
\label{eq:vtvn}
{\bf{\hat{v}}}^{\perp} &=& {\bf{\hat{v}}}^{\perp}_{T} + {\bf{\hat{v}^{\perp}}}_{N}\,,
\end{eqnarray} 
with 
\begin{eqnarray}
\label{eq:xrep0}
{\bf{\hat{v}^{\perp}}}_{T} &=& \delta({\bf{x}}-{\bf{y}}+{\bf{r}}) \left( i \frac{ \overrightarrow{\nabla}_{{\bf{x}}}}{m_T}  - i \frac{\overrightarrow{\nabla}_{{\bf{y}}}}{m_\chi} \right)  + \frac{1}{2\mu_T}{\bf{\hat{q}}}
 \\ 
{\bf{\hat{v}^{\perp}}}_{N} &=& \frac{1}{2m_{N}}\left[i \overleftarrow{\nabla}_{{\bf{r}}} \,\delta({\bf{r}}-{\bf{r}}_i) - i\delta({\bf{r}}-{\bf{r}}_i) \overrightarrow{\nabla}_{{\bf{r}}}  \right] \,.
\label{eq:xrep}
\end{eqnarray} 
The operator $\nabla_{\bf{x}}$ acts on the nucleus center of mass wave function at ${\bf{x}}$, whereas $\nabla_{\bf{y}}$ acts on the dark matter particle wave function at ${\bf{y}}$.
In Eq.~(\ref{eq:xrep0}), $m_T$ is the target nucleus mass, and $\mu_T$ the dark matter-nucleus reduced mass.
Finally, the operator $\nabla_{\bf{r}}$ in Eq.~(\ref{eq:xrep}) acts on the constituent nucleon wave function at ${\bf{r}}$, where ${\bf{r}}$ denotes the radial coordinate of the dark matter particle in a frame with origin at the nucleus center of mass (notice that in Sec.~\ref{sec:dmnu}, {\bf{r}} was the position vector from the single nucleon to the dark matter particle). %$\langle {\bf{y}} |{\bf{p}}. \rangle$, and ${\bf{y}}_\chi$ denotes the dark matter particle position. 
Separating the center of mass motion from the intrinsic motion of the constituent nucleons, the only operator depending on the position of the nucleons relative to the nucleus centre of mass, ${\bf{r}}_i$, is ${\bf{\hat{v}^{\perp}}}_{N}({\bf{x}})$. Operators in Tab.~\ref{tab:operators} independent of ${\bf{\hat{v}^{\perp}}}_{N}({\bf{x}})$ act like the identity $\mathbb{1}_{i}$ on the $i$th-nucleon position ${\bf{r}}_i$. In the coordinate space representation $\mathbb{1}_{i}$ corresponds to $\delta({\bf{r}}-{\bf{r}}_i)$. 

Combining Eqs.~(\ref{eq:H_I}), (\ref{eq:vtvn}) and (\ref{eq:xrep}) with Tab.~\ref{tab:operators}, we can finally write the most general Hamiltonian density for dark matter-nucleus interactions mediated by heavy spin-0 or spin-1 particles as a combination of (one-body) charge and nuclear currents~\cite{Fitzpatrick:2012ix}:
\begin{eqnarray}
\hat{\mathcal{H}}_{\rm T}({\bf{r}}) &=& \sum_{\tau=0,1} \Bigg\{
\sum_{i=1}^A  \hat{l}_0^{\tau}~ \delta({\bf{r}}-{\bf{r}}_i)
+ \sum_{i=1}^A \hat{l}_{0A}^{\tau}~ \frac{1}{2m_N} \Bigg[i \overleftarrow{\nabla}_{\bf{r}} \cdot  \vec{\sigma}(i)\delta({\bf{r}}-{\bf{r}}_i) -i \delta({\bf{r}}-{\bf{r}}_i) \
\vec{\sigma}(i)  \cdot  \overrightarrow{\nabla}_{\bf{r}} \Bigg]  \nonumber \\
 &+& \sum_{i=1}^A  {\bf{\hat{l}}}_5^{\tau} \cdot \vec{\sigma}(i) \delta({\bf{r}}-{\bf{r}}_i) +   \sum_{i=1}^A {\bf{\hat{l}}}_M^{\tau} \cdot \frac{1}{2 m_N} \Bigg[i \overleftarrow{\nabla}_{\bf{r}}\delta({\bf{r}}-{\bf{r}}_i) -i \delta({\bf{r}}-{\bf{r}}_i)\overrightarrow{\nabla}_{\bf{r}} \Bigg]  \nonumber \\
&+& \sum_{i=1}^A {\bf{\hat{l}}}_E^{\tau} \cdot \frac{1}{2m_N} \Bigg[ \overleftarrow{\nabla}_{\bf{r}} \times \vec{\sigma}(i) \delta({\bf{r}}-{\bf{r}}_i) +\delta({\bf{r}}-{\bf{r}}_i)\
  \vec{\sigma}(i) \times \overrightarrow{\nabla}_{\bf{r}} \Bigg] \Bigg\} t^{\tau}_{(i)}
\label{eq:Hx}
\end{eqnarray}
where $\vec{\sigma}(i)$ denotes the set of three Pauli matrices representing the spin operator of the $i$th-nucleon in the target nucleus, and 
\begin{eqnarray}
\label{eq:ls}
\hat{l}_0^\tau &=& c_1^\tau + i  \left( {{\bf{\hat{q}}} \over m_N}  \times {\bf{\hat{v}}}_{T}^\perp \right) \cdot  {\bf{\hat{S}}}_\chi  ~c_5^\tau
+ {\bf{\hat{v}}}_{T}^\perp \cdot {\bf{\hat{S}}}_\chi  ~c_8^\tau + i {{\bf{\hat{q}}} \over m_N} \cdot {\bf{\hat{S}}}_\chi ~c_{11}^\tau \nonumber \\
\hat{l}_{0A}^{\tau} &=& -{1 \over 2}  \left[ c_7 ^\tau  +i {{\bf{\hat{q}}} \over m_N} \cdot {\bf{\hat{S}}}_\chi~ c_{14}^\tau \right] \nonumber \\
{\bf{\hat{l}}}_5^{\tau}&=& {1 \over 2} \left[ i {{\bf{\hat{q}}} \over m_N} \times {\bf{\hat{v}}}_{T}^\perp~ c_3^\tau + {\bf{\hat{S}}}_\chi ~c_4^\tau
+  {{\bf{\hat{q}}} \over m_N}~{{\bf{\hat{q}}} \over m_N} \cdot {\bf{\hat{S}}}_\chi ~c_6^\tau
+   {\bf{\hat{v}}}_{T}^\perp ~c_7^\tau + i {{\bf{\hat{q}}} \over m_N} \times {\bf{\hat{S}}}_\chi ~c_9^\tau + i {{\bf{\hat{q}}} \over m_N}~c_{10}^\tau \right. \nonumber \\
 && \left.  +  {\bf{\hat{v}}}_{T}^\perp \times {\bf{\hat{S}}}_\chi ~c_{12}^\tau
+i  {{\bf{\hat{q}}} \over m_N} {\bf{\hat{v}}}_{T}^\perp \cdot {\bf{\hat{S}}}_\chi ~c_{13}^\tau+i {\bf{\hat{v}}}_{T}^\perp {{\bf{\hat{q}}} \over m_N} \cdot {\bf{\hat{S}}}_\chi ~ c_{14}^\tau+{{\bf{\hat{q}}} \over\
 m_N} \times {\bf{\hat{v}}}_{T}^\perp~ {{\bf{\hat{q}}} \over m_N} \cdot {\bf{\hat{S}}}_\chi ~ c_{15}^\tau  \right]\nonumber \\
{\bf{\hat{l}}}_M^{\tau} &=&   i {{\bf{\hat{q}}} \over m_N}  \times {\bf{\hat{S}}}_\chi ~c_5^\tau - {\bf{\hat{S}}}_\chi ~c_8^\tau \nonumber \\
{\bf{\hat{l}}}_E^{\tau} &=& {1 \over 2} \left[  {{\bf{\hat{q}}} \over m_N} ~ c_3^\tau +i {\bf{\hat{S}}}_\chi~c_{12}^\tau - {{\bf{\hat{q}}} \over  m_N} \times{\bf{\hat{S}}}_\chi  ~c_{13}^\tau-i 
{{\bf{\hat{q}}} \over  m_N} {{\bf{\hat{q}}} \over m_N} \cdot {\bf{\hat{S}}}_\chi  ~c_{15}^\tau \right] \,.
\end{eqnarray}
Inspection of Eq.~(\ref{eq:Hx}) shows that dark matter couples to the constituent nucleons through  the nuclear vector and axial charges (first line in Eq.~(\ref{eq:Hx})), the nuclear spin and convection currents (second line in Eq.~(\ref{eq:Hx})), and the nuclear spin-velocity current (last line in Eq.~(\ref{eq:Hx})).  The 14 dark matter-nucleon interaction operators in Tab.~\ref{tab:operators} contribute to these couplings in different ways. For instance, the constant spin-independent operator $\mathcal{O}_1$ contributes to the vector charge coupling through the operator $\hat{l}_0^\tau$, whereas the constant spin-dependent operator $\mathcal{O}_4$ contributes to the nuclear spin current coupling through the operator ${\bf{\hat{l}}}_5^{\tau}$.

The interaction Hamiltonian relevant for nuclear matrix element calculations is finally obtained by integrating the Hamiltonian density (\ref{eq:H_I}) over space coordinates 
\begin{eqnarray}
H_{\rm T}= \int d^{3}{\bf{r}}\, \hat{\mathcal{H}}_{\rm T}({\bf{r}}) \,.
\label{eq:Hfinal}
\end{eqnarray}
This latter integration eliminates the delta functions $\delta({\bf{r}}-{\bf{r}}_i)$ in Eq.~(\ref{eq:Hx}).

\subsection{Dark matter-nucleus scattering cross-section}
From the dark matter-nucleus interaction Hamiltonian~(\ref{eq:Hfinal}), one can calculate the amplitude for transitions between initial, $|i\rangle$, and final, $|f\rangle$, scattering states. We denote initial nuclear states by $|{\bf{k}}_T,J,M_J,T,M_T\rangle$, where $J$ and $T$ are the nuclear spin and isospin, respectively, and $M_J$ and $M_T$ the associated magnetic quantum numbers. With this notation $|i\rangle~=~|{\bf{k}}_T,J,M_J,T,M_T\rangle \otimes |{\bf{p}},j_\chi,M_\chi\rangle$ ($M_\chi$ is the spin magnetic quantum number of the dark matter particle, omitted in previous equations for simplicity), and an analogous expression applies to $|f\rangle$. We can therefore write
\begin{equation}
\langle f |H_{\rm T} |i\rangle = (2\pi)^3 \delta({\bf k}'_T+{\bf p}'-{\bf k}_T-{\bf p}) \,i \mathcal{M}_{NR}
\label{eq:M0}
\end{equation}
with 
\begin{eqnarray}
i \mathcal{M}_{NR}%_{\chi T} 
&=&
  \langle J,M_J, T, M_T | \sum_{\tau=0,1} \left[
  \langle \hat{l}_0^\tau\rangle~ \sum_{i=1}^A  ~e^{-i {\bf{q}} \cdot {\bf{r}}_i}   \right. \nonumber \\
&+& \langle \hat{l}_{0A}^{\tau}\rangle~ \sum_{i=1}^A ~ {1 \over 2m_N} \left(i \overleftarrow{\nabla}_{{\bf{r}}_i} \cdot  \vec{\sigma}(i)~ e^{-i {\bf{q}} \cdot {\bf{r}}_i}  
-ie^{-i {\bf{q}} \cdot {\bf{r}}_i}  \vec{\sigma}(i)  \cdot   \overrightarrow{\nabla}_{{\bf{r}}_i}   \right)  \nonumber \\
&+&  \langle{\bf{\hat{l}}}_5^\tau\rangle \cdot  \sum_{i=1}^A  ~\vec{\sigma}(i)~e^{-i {\bf{q}} \cdot {\bf{r}}_i} \nonumber \\
&+&  \langle{\bf{\hat{l}}}_M^\tau\rangle \cdot  \sum_{i=1}^A  ~ {1 \over 2m_N} \left(i \overleftarrow{\nabla}_{{\bf{r}}_i} e^{-i {\bf{q}} \cdot {\bf{r}}_i}  -i e^{-i {\bf{q}} \cdot {\bf{r}}_i} 
\overrightarrow{\nabla}_{{\bf{r}}_i} \right)  \nonumber \\
&+&  \langle{\bf{\hat{l}}}_E^\tau\rangle \left. \cdot  \sum_{i=1}^A ~ {1 \over 2m_N} \left( \overleftarrow{\nabla}_{{\bf{r}}_i} \times \vec{\sigma}(i) e^{-i {\bf{q}} \cdot {\bf{r}}_i}  +e^{-i {\bf{q}} \cdot {\bf{r}}_i}  \vec{\sigma}(i) \times \overrightarrow{\nabla}_{{\bf{r}}_i} \right)   \right] t^\tau_{(i)} ~ | J, M_J,T,M_T \rangle \,. \nonumber\\
\label{eq:M}
\end{eqnarray}
In Eqs.~(\ref{eq:M0}) and (\ref{eq:M}) we use the result
\begin{eqnarray} 
\langle {\bf{k}}'_T;{\bf{p}}',j_\chi,M_\chi |\,{\bf{\hat{l}}}[{\bf{\hat{q}}},{\bf{\hat{v}}}^{\perp}_T,{\bf{\hat{S}}}_\chi]\,| {\bf{k}}_T;{\bf{p}},j_\chi,M_\chi \rangle &=&e^{-i {\bf{q}} \cdot {\bf{r}}} (2\pi)^3\delta({\bf k}'_T+{\bf p}'-{\bf k}_T-{\bf p})  \nonumber\\ &\times& \langle j_\chi,M_\chi |\,{\bf{\hat{l}}}[{\bf{q}},{\bf{v}}^{\perp}_T,{\bf{\hat{S}}}_\chi]\,| j_\chi,M_\chi \rangle \,,
\label{eq:lme}
\end{eqnarray}
and the notation $\langle {\bf{\hat{l}}} \rangle \equiv \langle j_\chi,M_\chi |\,{\bf{\hat{l}}}[{\bf{q}},{\bf{v}}^{\perp}_T,{\bf{\hat{S}}}_\chi]\,| j_\chi,M_\chi \rangle$, with ${\bf{\hat{l}}}$ equal to one of the operators $\hat{l}_0^\tau$, $\hat{l}_{0A}^{\tau}$, ${\bf{\hat{l}}}_5^\tau$, ${\bf{\hat{l}}}_M^\tau$, and ${\bf{\hat{l}}}_E^\tau$. Notice that on the right hand side of Eq.~(\ref{eq:lme}), ${\bf{q}}$ and ${\bf{v}}^{\perp}_T\equiv {\bf{v}} + {\bf{q}}/(2\mu_T)$ replace, respectively, ${\bf{\hat{q}}}$ and ${\bf{\hat{v}}}^{\perp}_T$, in agreement with Eqs.~(\ref{eq:me1}) and (\ref{eq:me2}). From now on $|{\bf{v}}|=w$ denotes the dark matter-nucleus relative velocity in the Sun. Importantly, each line in the transition amplitude (\ref{eq:M}) is equal to the product of a term containing information on the kinematics of the scattering and on the dark matter-nucleon coupling strength, i.e. $\langle {\bf{\hat{l}}}\rangle $, and a term given by a nuclear matrix element.

In order to evaluate the nuclear matrix elements in Eq.~(\ref{eq:M}), we perform a multipole expansion of the nuclear charges and currents using a spherical unit vector basis ${\bf {e}}_\lambda$ with z-axis along ${\bf{q}}$, and the identities
\begin{eqnarray}
e^{i {\bf{q}} \cdot {\bf{r}}_i} &=& \sum_{L=0}^\infty \sqrt{4 \pi(2L+1)}~ i^L j_L(q r_i) Y_{L0}(\Omega_{{\bf{r}}_i}) \nonumber \\
e^{i {\bf{q}} \cdot {\bf{r}}_i} {\bf{e}}_{0} &=& \sum_{L=0}^\infty \sqrt{4 \pi(2L+1)}~ i^{L-1} {\overrightarrow{\nabla}_{{\bf{r}}_i} \over q\
} j_L(qr_i) Y_{L0}(\Omega_{{\bf{r}}_i})\nonumber\\
e^{i {\bf{q}} \cdot {\bf{r}}_i} {\bf{e}}_{\lambda} &=& \sum_{L = 1}^\infty \sqrt{2 \pi(2L+1)}~ i^{L-2} \left[
\lambda j_L(qr_i)  {\bf Y}_{LL1}^\lambda(\Omega_{{\bf{r}}_i}) +  { \overrightarrow{\nabla}_{{\bf{r}}_i} \over q} \times j_L(qr_i) {\bf Y}_{LL1}^\lambda
(\Omega_{{\bf{r}}_i}) \right],  \lambda=\pm 1\nonumber\\
\label{eq:Y}
\end{eqnarray}
together with 
\begin{equation}
{\bf{A}}=\sum_{\lambda=0,\pm1} \left( {\bf{A}} \cdot {\bf{e}}_{\lambda} \right) {\bf{e}}^{\dagger}_{\lambda}\,,
\end{equation}
that holds for any vector ${\bf{A}}$, given a spherical unit vector basis ${\bf {e}}_\lambda$. The vector spherical harmonics in Eq.~(\ref{eq:Y}) are defined in terms of Clebsch-Gordan coefficients and scalar spherical harmonics:
\begin{equation}
{\bf Y}^M_{LL'1}(\Omega_{{\bf{r}}_i}) = \sum_{m\lambda} \langle L'm1\lambda|L'1LM \rangle
Y_{L'm}(\Omega_{{\bf{r}}_i}) \, {\bf e}_\lambda \,.
\end{equation}
They obey the identity ${\bf Y}^{\lambda\dagger}_{LL'1}=-(-1)^{\lambda}{\bf Y}^{-\lambda}_{LL'1}$.
The multipole expansion of the nuclear spin current, for instance, leads to
\begin{eqnarray}
\langle{\bf{\hat{l}}}_5^\tau\rangle \cdot  \sum_{i=1}^A  ~\vec{\sigma}(i)~e^{-i {\bf{q}} \cdot {\bf{r}}_i} &=& \sum_{L=0}^{\infty}
\sqrt{4\pi (2L +1)}(-i)^{L}  i \Sigma''_{L0;\tau} (q)  (\langle{\bf{\hat{l}}}_5^\tau\rangle \cdot {\bf{e}}_0) \nonumber \\
&-&   \sum_{L=1}^{\infty}
\sqrt{2\pi (2L +1)}(-i)^{L} \sum_{\lambda=\pm1} \Big(\lambda \Sigma_{L-\lambda;\tau}(q)+i\Sigma'_{L-\lambda;\tau}(q)\Big)  (\langle{\bf{\hat{l}}}_5^\tau\rangle \cdot {\bf{e}}_\lambda) \,, \nonumber\\
\label{eq:l5}
\end{eqnarray}
with 
\begin{eqnarray}
\Sigma'_{LM;\tau}(q) &=& -i \sum_{i=1}^{A} \left[ \frac{1}{q} \overrightarrow{\nabla}_{{\bf{r}}_i} \times {\bf{M}}_{LL}^{M}(q {\bf{r}}_i)  \right] \cdot \vec{\sigma}(i) t^{\tau}_{(i)}\nonumber\\
\Sigma''_{LM;\tau}(q) &=&\sum_{i=1}^{A} \left[ \frac{1}{q} \overrightarrow{\nabla}_{{\bf{r}}_i} M_{LM}(q {\bf{r}}_i)  \right] \cdot \vec{\sigma}(i) t^{\tau}_{(i)}
\label{eq:S1S2}
\end{eqnarray}
where ${\bf{M}}_{LL}^{M}(q {\bf{r}}_i)=j_{L}(q r_i){\bf Y}^M_{LL1}(\Omega_{{\bf{r}}_i})$ , and $M_{LM}(q {\bf{r}}_i)=j_{L}(q r_i)Y_{LM}(\Omega_{{\bf{r}}_i})$. Assuming that nuclear ground states are eigenstates of $P$ and $CP$, only multipoles that transform as even-even under $P$ and $CP$ contribute to the square modulus of the transition amplitude. With this assumption 
$\Sigma_{LM;\tau}(q)$ does not contribute at all, and is therefore not defined here. Expressions similar to Eq.~(\ref{eq:l5}) can be derived for the remaining charges and currents. 
Besides the two operators in Eq.~(\ref{eq:S1S2}), four additional nuclear response operators contribute to the transition probability, namely 
\begin{eqnarray}
M_{LM;\tau}(q) &=& \sum_{i=1}^{A} M_{LM}(q {\bf{r}}_i) t^{\tau}_{(i)}\nonumber\\
\Delta_{LM;\tau}(q) &=&\sum_{i=1}^{A}  {\bf{M}}_{LL}^{M}(q {\bf{r}}_i) \cdot \frac{1}{q}\overrightarrow{\nabla}_{{\bf{r}}_i} t^{\tau}_{(i)} \nonumber\\
\tilde{\Phi}^{\prime}_{LM;\tau}(q) &=& \sum_{i=1}^A \left[ \left( {1 \over q} \overrightarrow{\nabla}_{{\bf{r}}_i} \times {\bf{M}}_{LL}^M(q {\bf{r}}_i) \right) \cdot \left(\vec{\sigma}(i) \times {1 \over q} \overrightarrow{\nabla}_{{\bf{r}}_i} \right) + {1 \over 2} {\bf{M}}_{LL}^M(q {\bf{r}}_i) \cdot \vec{\sigma}(i) \right]~t^\tau_{(i)} \nonumber \\
\Phi^{\prime \prime}_{LM;\tau}(q ) &=& i  \sum_{i=1}^A\left( {1 \over q} \overrightarrow{\nabla}_{{\bf{r}}_i}  M_{LM}(q {\bf{r}}_i) \right) \cdot \left(\vec{\sigma}(i) \times \
{1 \over q} \overrightarrow{\nabla}_{{\bf{r}}_i}  \right)~t^\tau_{(i)} \,.
\label{eq:MDP}
\end{eqnarray}
Squaring the amplitude (\ref{eq:M}), summing (averaging) the result over final (initial) spin configurations, and demanding that only multipoles transforming as even-even under $P$ and $CP$ contribute, one finally obtains~\cite{Fitzpatrick:2012ix} 
\begin{eqnarray}
P_{\rm tot}({w}^2,{q}^2)&\equiv&{1 \over 2j_\chi + 1} {1 \over 2J + 1} \sum_{\rm spins} |\mathcal{M}_{NR}|^2 \nonumber \\  &=& {4 \pi \over 2J + 1} 
\sum_{ \tau=0,1} \sum_{\tau^\prime = 0,1} \Bigg\{ \Bigg[ R_{M}^{\tau \tau^\prime}\left({v}^{\perp 2}_{T}, {{q}^{2} \over m_N^2}\right)~W_{M}^{\tau \tau^\prime}(y)   \nonumber\\
&+& R_{\Sigma^{\prime \prime}}^{\tau \tau^\prime}\left({v}^{\perp 2}_{T}, {{q}^{2} \over m_N^2}\right)   ~W_{\Sigma^{\prime \prime}}^{\tau \tau^\prime}(y) 
+   R_{\Sigma^\prime}^{\tau \tau^\prime}\left({v}^{\perp 2}_{T}, {{q}^{2} \over m_N^2}\right) ~ W_{\Sigma^\prime}^{\tau \tau^\prime}(y) \Bigg]  \nonumber\\  
&+& {{q}^{2} \over m_N^2} ~\Bigg[R_{\Phi^{\prime \prime}}^{\tau \tau^\prime}\left({v}^{\perp 2}_{T}, {{q}^{2} \over m_N^2}\right) ~ W_{\Phi^{\prime \prime}}^{\tau \tau^\prime}(y)  +  R_{ \Phi^{\prime \prime}M}^{\tau \tau^\prime}\left({v}^{\perp 2}_{T}, {{q}^{2} \over m_N^2}\right)  ~W_{ \Phi^{\prime \prime}M}^{\tau \tau^\prime}(y) \nonumber\\
&+&   R_{\tilde{\Phi}^\prime}^{\tau \tau^\prime}\left({v}^{\perp 2}_{T}, {{q}^{2} \over m_N^2}\right) ~W_{\tilde{\Phi}^\prime}^{\tau \tau^\prime}(y) 
+   R_{\Delta}^{\tau \tau^\prime}\left({v}^{\perp 2}_{T}, {{q}^{2} \over m_N^2}\right) ~ W_{\Delta}^{\tau \tau^\prime}(y) \nonumber\\
 &+&  R_{\Delta \Sigma^\prime}^{\tau \tau^\prime}\left({v}^{\perp 2}_{T}, {{q}^{2} \over m_N^2}\right)  ~W_{\Delta \Sigma^\prime}^{\tau \tau^\prime}(y)   \Bigg]  \Bigg\}  \,,
\label{eq:Ptot}
\end{eqnarray}
where the dark matter response function $R_{M}^{\tau \tau^\prime}$, $R_{\Sigma^{\prime \prime}}^{\tau \tau^\prime}$, $R_{\Sigma^\prime}^{\tau \tau^\prime}$, $R_{\Phi^{\prime \prime}}^{\tau \tau^\prime}$, $R_{\Phi^{\prime\prime}M}^{\tau \tau^\prime}$, $R_{\tilde{\Phi}^\prime}^{\tau \tau^\prime}$, $R_{\Delta}^{\tau \tau^\prime}$ and $R_{\Delta \Sigma^\prime}^{\tau \tau^\prime}$ are quadratic combinations of the matrix elements $\langle {\bf{\hat{l}}} \rangle$ and are defined in Appendix~\ref{sec:appDM}. They depend on the momentum transfer, the dark matter-nucleus relative velocity, as well as on the dark matter-nucleon interaction strength. 

The nuclear response functions in Eq.~(\ref{eq:Ptot}) are defined as follows   
\begin{equation}
W_{AB}^{\tau \tau^\prime}(y)= \sum_{L\in S_{AB}}  \langle J,T,M_T ||~ A_{L;\tau} (q)~ || J,T,M_T \rangle \langle J,T,M_T ||~ B_{L;\tau^\prime} (q)~ || J,T,M_T \rangle \,
\label{eq:W}
\end{equation}
where $A$ and $B$ correspond to pairs of operators in Eqs.(\ref{eq:S1S2}) and (\ref{eq:MDP}). When $A=B$, only one letter is used. $S_{AB}=\{0,2,\dots\}$ for the pairs of operators $A=B=M_{LM;\tau}$,  $A=B=\Phi^{\prime\prime}_{LM;\tau}$ and $A=\Phi^{\prime\prime}_{LM;\tau}$, $B=M_{LM;\tau}$. $S_{AB}=\{1,3,\dots\}$ for the pairs of operators $A=B=\Sigma^{\prime}_{LM;\tau}$, $A=B=\Sigma^{\prime\prime}_{LM;\tau}$, $A=B=\Delta_{LM;\tau}$, and $A=\Delta_{LM;\tau}$, $B=\Sigma^{\prime}_{LM;\tau}$. Finally, $S_{AB}=\{2,4,\dots\}$ for the pair of operators $A=B=\tilde{\Phi}^{\prime}_{LM;\tau}$. 
The integer numbers in $S_{AB}$ select multipoles transforming as even-even under $P$ and $CP$. Notice that only two interference terms, i.e. $A\neq B$, can satisfy this requirement and thereby appear in Eq.~(\ref{eq:Ptot}).

The nuclear response functions in Eq.~(\ref{eq:W}) are expressed in terms of matrix elements reduced in the spin magnetic quantum number $M_J$. The reduction of a tensor operator $A_{LM;\tau}$ of rank $L$ is done by the Wigner-Eckart theorem
\begin{equation}
\langle J,M_J |\,{A}_{LM;\tau}\,|J,M_J\rangle =(-1)^{J-M_J}\left(
\begin{array}{ccc} J&L&J\\
-M_J&M&M_J 
\end{array} 
\right)
\langle  J  ||\,{A}_{L;\tau}\,|| J  \rangle \,,
\label{eq:red}
\end{equation}
and it involves Wigner $3j$-symbols which cancel in Eq.~(\ref{eq:W}) after summing over spin configurations because of their orthonormality. In the next section, we will evaluate our nuclear response functions using the {\sffamily Mathematica} package of Ref.~\cite{Anand:2013yka}, which assumes the harmonic oscillator basis with length parameter $b=\sqrt{41.467/(45 A^{-1/3}-25A^{-2/3})}$~fm for the single-particle states. In this case, the nuclear response functions in Eq.~(\ref{eq:W}) only depend on the dimensionless variable $y=(bq/2)^2$.

For the $i$th-element in the Sun, we can finally write the dark matter-nucleus differential cross-section as follows 
\begin{equation}
\frac{{\rm d} \sigma_i}{{\rm d} E}(w^2,q^2) = \frac{m_T}{2\pi w^2} \,P_{\rm tot}({w}^2,{q}^2) \,,
\label{eq:sigma}
\end{equation}
which constitutes the particle physics input in the calculation of the rate of dark matter capture by the Sun.

\section{Nuclear matrix element calculation}
\label{sec:obdme}
In this section we calculate the reduced nuclear matrix elements that appear in Eq.~(\ref{eq:W}) for the most abundant elements in the Sun. We list analytic expressions for the associated nuclear response functions in Appendix \ref{sec:appNuc}. These expressions can be used by the reader in analyses of dark matter induced neutrino signals from the Sun. We perform this calculation using the {\sffamily Mathematica} package introduced in \cite{Anand:2013yka}, which requires as an input the one-body density matrix elements (OBDME) for ground-state to ground-state transitions of the target nuclei in analysis. We compute these OBDME using the {\sffamily Nushell@MSU} program~\cite{NuShell}, which allows for fast nuclear structure calculations based on the nuclear shell model.

In order to relate the nuclear matrix elements in Eq.~(\ref{eq:W}) to the underlying OBDME, we expand the nuclear operators in Eqs.~(\ref{eq:S1S2}) and (\ref{eq:MDP}), here collectively denoted by $A_{LM;\tau}$, in a complete set of spherically symmetric single-particles states, $|\alpha\rangle$. Here we assume the nuclear harmonic oscillator model for the radial part of the wave functions associated with the states $|\alpha\rangle$. Within this assumption, single-particle states can be labelled by their principal, angular momentum and spin quantum numbers, respectively $n_\alpha$, $l_\alpha$ and $s_\alpha$, and by their total spin and isospin, respectively $j_{\alpha}$ and $t_{\alpha}$: $|\alpha\rangle=|n_\alpha(l_\alpha s_\alpha=1/2)j_\alpha m_{j_\alpha};t_{\alpha}=1/2,m_{t_\alpha}\rangle$. Here $m_{j_\alpha}$ and $m_{t_\alpha}$ denote the total spin and isospin magnetic quantum numbers, whereas $|\alpha|$ represents the set of all non magnetic quantum numbers, i.e. $|\alpha\rangle=||\alpha|,m_{j_\alpha};m_{t_{\alpha}}\rangle$. With this notation, the nuclear operators in Eqs.~(\ref{eq:S1S2}) and (\ref{eq:MDP}) can be expanded as follows
 
\begin{eqnarray}
  A_{LM;\tau} &=&\;  \sum_{\alpha\beta}\langle \alpha|~A_{LM;\tau}~|\beta\rangle \,a_{\alpha}^{\dagger}a_{\beta} \nonumber\\
&=&\; \sum_{|\alpha||\beta|}\langle\left|\alpha\right|\vdots\vdots A_{L;\tau}\vdots\vdots \left|\beta\right|\rangle\frac{[a_{|\alpha|}^{\dagger}\otimes\tilde{a}_{|\beta|}]_{LM;\tau}}{\sqrt{(2L+1)(2\tau+1)}} \,,
\label{eq:keyeq}
\end{eqnarray}  
where $\tilde{a}_{|\beta|,m_{j_\beta},m_{t_{\beta}}}\equiv (-1)^{j_\beta-m_{j_\beta}+1/2-m_{t_\beta}}\,a_{|\beta|,-m_{j_\beta},-m_{t_{\beta}}}$, and $\vdots\vdots$ denotes a matrix element reduced in spin and isospin according to Eq.~(\ref{eq:red}). The creation and annihilation operators $a^{\dagger}_{\alpha}$ and $\tilde{a}_\beta$ transform as tensors under spin and isospin transformations, and their tensor product admits  the following representation 
\begin{eqnarray}
[a_{|\alpha|}^{\dagger}\otimes\tilde{a}_{|\beta|}]_{LM;\tau} &=&  \sqrt{(2L+1)(2\tau+1)}\sum_{m_{j_\alpha}m_{t_{\alpha}}m_{j_\beta}m_{t_{\beta}}}(-1)^{\,j_{\alpha}-m_{j_\alpha}+t_{j_\alpha}-m_{t_{\alpha}}}\nonumber\\
&\times&\begin{pmatrix}
    j_{\alpha}&L&j_{\beta}\\
    -m_{\alpha}&M&m_{\beta}
  \end{pmatrix}\begin{pmatrix}
    t_{\alpha}&\tau&t_{\beta}\\
    -m_{t_{\alpha}}&0&m_{t_{\beta}}
  \end{pmatrix}a_{\alpha}^{\dagger}a_{\beta}
\,.
\end{eqnarray}
The reduced nuclear matrix elements in Eq.~(\ref{eq:W}) can be further reduced in nuclear isospin, and hence written as 
\begin{eqnarray}
\langle J, T, M_T  ||~ A_{LM;\tau} ~ || J, T, M_T \rangle &=& (-1)^{T-M_T}
\begin{pmatrix}
    T&\tau&T\\
    -M_T&0&M_T
  \end{pmatrix} \nonumber\\
&\times& \sum_{|\alpha||\beta|}\langle\left|\alpha\right|\vdots\vdots A_{L;\tau}\vdots\vdots \left|\beta\right|\rangle\frac{\langle J, T \vdots\vdots ~[a_{|\alpha|}^{\dagger}\otimes\tilde{a}_{|\beta|}]_{L;\tau}
~\vdots\vdots J, T\rangle}{\sqrt{(2L+1)(2\tau+1)}} \,.\nonumber\\
\end{eqnarray}
Using the definition of ground-state to ground-state OBDME, namely,
\begin{equation}
\psi^{L;\tau}_{|\alpha||\beta|} \equiv \frac{\langle J, T\vdots\vdots ~[a_{|\alpha|}^{\dagger}\otimes\tilde{a}_{|\beta|}]_{L;\tau}
~\vdots\vdots J, T\rangle}{\sqrt{(2L+1)(2\tau+1)}} \,,
\label{eq:OBDME}
\end{equation}
we can finally write the reduced nuclear matrix elements in Eq.~(\ref{eq:W}) as follows
\begin{eqnarray}
\langle J, T, M_T  ||~ A_{LM;\tau} ~ || J, T, M_T \rangle 
&=& (-1)^{T-M_T}
\begin{pmatrix}
    T&\tau&T\\
    -M_T&0&M_T
  \end{pmatrix} 
   \sum_{|\alpha||\beta|}\psi^{L;\tau}_{|\alpha||\beta|} \,\langle \left|\alpha\right|\vdots\vdots A_{L;\tau}\vdots\vdots \left|\beta\right|\rangle \,, \nonumber\\
   \label{eq:master}
\end{eqnarray}
which is the master equation for nuclear matrix element calculations based on the assumption of one-body dark matter-nucleon interactions. Since the nuclear operators $A_{LM;\tau}$ depend on isospin through the matrices $t^{\tau}_{(i)}$ only, the doubly reduced matrix elements in Eq.~(\ref{eq:master}) can be further simplified as follows
\begin{equation}
\langle \left|\alpha\right|\vdots\vdots A_{L;\tau}\vdots\vdots \left|\beta\right|\rangle = \sqrt{2(2\tau+1)} \,\langle n_{\alpha}(l_\alpha1/2)j_\alpha ||\,A_{L}\,|| n_{\beta}(l_\beta1/2)j_\beta  \rangle\,,
\label{eq:math}
\end{equation}
where $A_L$ is the part of the operator $A_{L;\tau}$ acting on nuclear spin and space coordinates.  
In Appendix~\ref{sec:appME}, we provide explicit expressions for the reduced matrix elements on the right hand side of Eq.~(\ref{eq:math}), which in the case of the harmonic oscillator single-particle basis are known analytically, and depend on the momentum transfer through the variable $y$ defined above. The {\sffamily Mathematica} package in Ref.~\cite{Anand:2013yka} provides an efficient implementations of these expressions.

We now move on to the OBDME calculation. In this computation, the multipole number $L$ is bounded from above, i.e. $L\le 2J$, whereas $\tau=0,1$. In contrast, the indexes $|\alpha|$ and $|\beta|$ in principle span a complete set of infinite single-particle quantum numbers. The nuclear shell model provides a robust framework to restrict the set of relevant $|\alpha|$ and $|\beta|$ in the OBDME definition (\ref{eq:OBDME}), and to consistently truncate the infinite sums in Eq.~(\ref{eq:master}). 

\begin{table}
  \centering
  \begin{tabular}[!h]{|c|c|c|c|c|c|r|c|}
\hline
    Element&$2J$&$2T$&P&core-orbits&valence-orbits&Hamiltonian&restrictions\\
\hline
\hline
%${}^{1}$H&1&1&+&none&s-p-sd-pf&wbt~\cite{Warburton:1992rh}&none\\
%\hline
${}^{3}$He&1&1&+&none&s-p-sd-pf&wbt~\cite{Warburton:1992rh}&none\\
\hline
${}^{4}$He&0&0&+&none&s-p-sd-pf&wbt~\cite{Warburton:1992rh}&none\\
\hline
${}^{12}$C&0&0&+&s&p&pewt~\cite{Warburton:1992rh}&none\\
\hline
${}^{14}$N&2&0&+&s&p&pewt~\cite{Warburton:1992rh}&none\\
\hline
${}^{16}$O&0&0&+&none&s-p-sd-pf&wbt~\cite{Warburton:1992rh}&$0d_{3/2}1s_{1/2}1p\,0f$\\
\hline
${}^{20}$Ne&0&0&+&s-p&sd&w~\cite{Wildenthal:1984mf}&none\\
\hline
${}^{23}$Na&3&1&+&s-p&sd&w~\cite{Wildenthal:1984mf}&none\\
\hline
${}^{24}$Mg&0&0&+&s-p&sd&w~\cite{Wildenthal:1984mf}&none\\
\hline
${}^{27}$Al&5&1&+&s-p&sd&w~\cite{Wildenthal:1984mf}&none\\
\hline
${}^{28}$Si&0&0&+&s-p&sd&w~\cite{Wildenthal:1984mf}&none\\
\hline
${}^{32}$S&0&0&+&s-p&sd&w~\cite{Wildenthal:1984mf}&none\\
\hline
${}^{40}$Ar&0&4&+&s-p&sd-pf&sdpfnow~\cite{Nummela:2001xh}&$1p_{1/2}1p_{3/2}0f_{5/2}$\\
\hline
${}^{40}$Ca&0&0&+&s-p&sd-pf&sdpfnow~\cite{Nummela:2001xh}&$1p_{1/2}1p_{3/2}0f_{5/2}$\\
\hline
${}^{56}$Fe&0&4&+&s-p-sd&pf&gx1~\cite{Honma:2004xk}&$1p_{1/2}0f_{5/2}$\\
\hline
${}^{58}$Ni&0&2&+&s-p-sd&pf&gx1~\cite{Honma:2004xk}&$1p_{1/2}0f_{5/2}$\\
\hline
  \end{tabular}
  \caption{Summary of element specific input parameters needed for the calculation of the OBDME via the {\sffamily Nushell@MSU} code. We use the notation of~\cite{Brown:2001zz} in defining the major shells. For each element in the Sun, we use a model space comprising the core-orbits and valence-orbits reported in this table. In the ``restrictions'' column, we list the energetic orbits not included in the calculation in order to make the computation numerically feasible. The interaction Hamiltonians in the next to last column are described in the review~\cite{Brown:2001zz}, and in the corresponding references.}
\label{tab:inputs}
\end{table}

In the nuclear shell model nucleons occupy single-particle states degenerate in the total spin magnetic quantum number called sub-shells, or orbits. Sub-shells are solutions of the Schr\"odinger equation for a given nuclear potential (e.g. harmonic oscillator potential, Woods-Saxon potential, etc\dots) and reflect a choice of single-particle basis. Orbits are labeled with conventions similar to those used for atomic orbitals, e.g. the orbit $0p_{1/2}$ has ``principal quantum number'' 0, orbital angular momentum 1 and total spin 1/2. Groups of energetically close sub-shells form major shells of progressively increasing energy. The set of fully occupied major shells forms the nuclear core. For instance, the orbits $0s_{1/2}$, $0p_{3/2}$ and $0p_{1/2}$ divide into the $s$ and $p$ major shells, and together form the core of, e.g, $^{20}$Ne, which contains 8 proton/neutron pairs. Analogously, the orbits $1s_{1/2}$, $0d_{3/2}$, and $0d_{5/2}$ form the $sd$ major shell, and the orbits $1p_{1/2}$, $1p_{3/2}$, $0f_{5/2}$, and $0f_{7/2}$ the $pf$ major shell. Nucleons that are not in the nuclear core are called valence nucleons. Not all orbits are accessible to valence nucleons since sizable energy gaps separate adjacent major shells. 
Restrictions on the number of nucleons allowed in the most energetic orbits are often imposed in order to reduce the computational effort. 
The set of orbits that are actually accessible in a calculation constitutes the so-called model space. 
Therefore, the original $A$-nucleon problem characterized by the bare nuclear interaction is simplified to a many-body problem restricted to the model space, and subject to an effective Hamiltonian. Effective Hamiltonians for nuclear shell model calculations can be computed microscopically from nuclear forces, or fitted empirically to observations.
Within this framework, the sums in Eq.~(\ref{eq:master}) only extend over orbits in the assumed model space, since the remaining states do not contribute by construction. We refer to~\cite{Brown:2001zz,Caurier:2004gf} for a more extended introduction to the nuclear shell model.
\begin{figure}[t]
\begin{center}
\begin{minipage}[t]{0.49\linewidth}
\centering
\includegraphics[width=\textwidth]{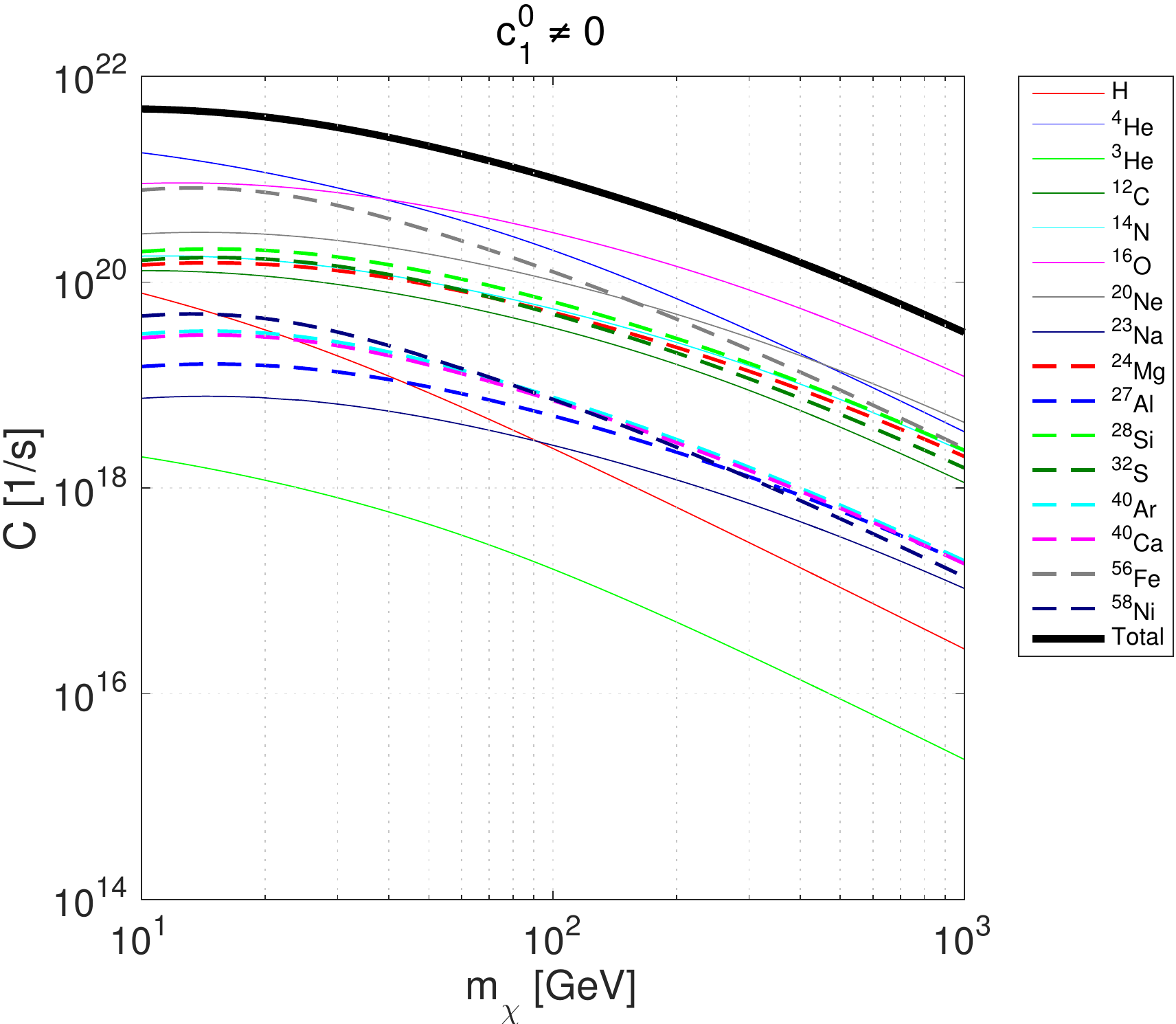}
\end{minipage}
\begin{minipage}[t]{0.49\linewidth}
\centering
\includegraphics[width=\textwidth]{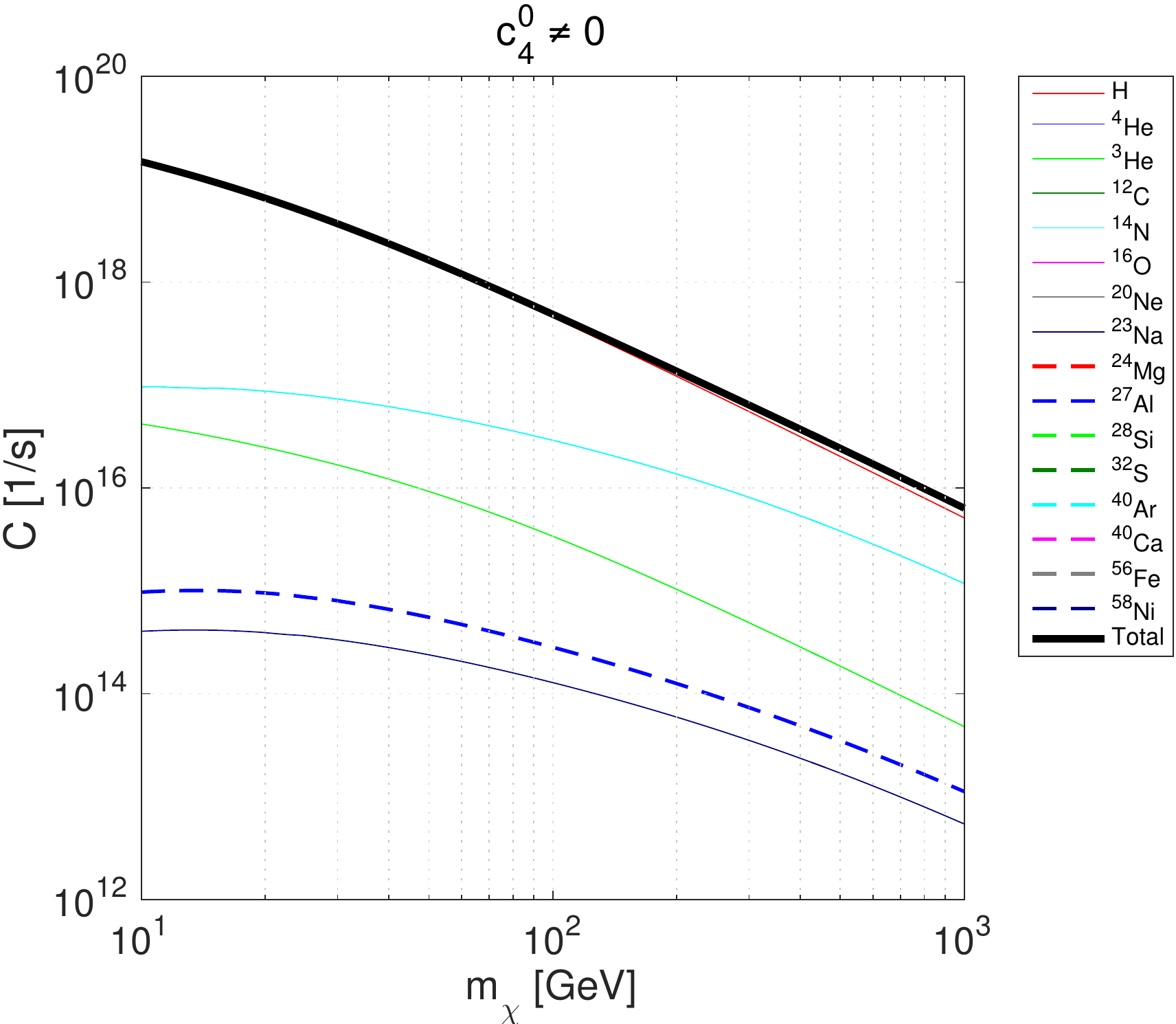}
\end{minipage}
\begin{minipage}[t]{0.49\linewidth}
\centering
\includegraphics[width=\textwidth]{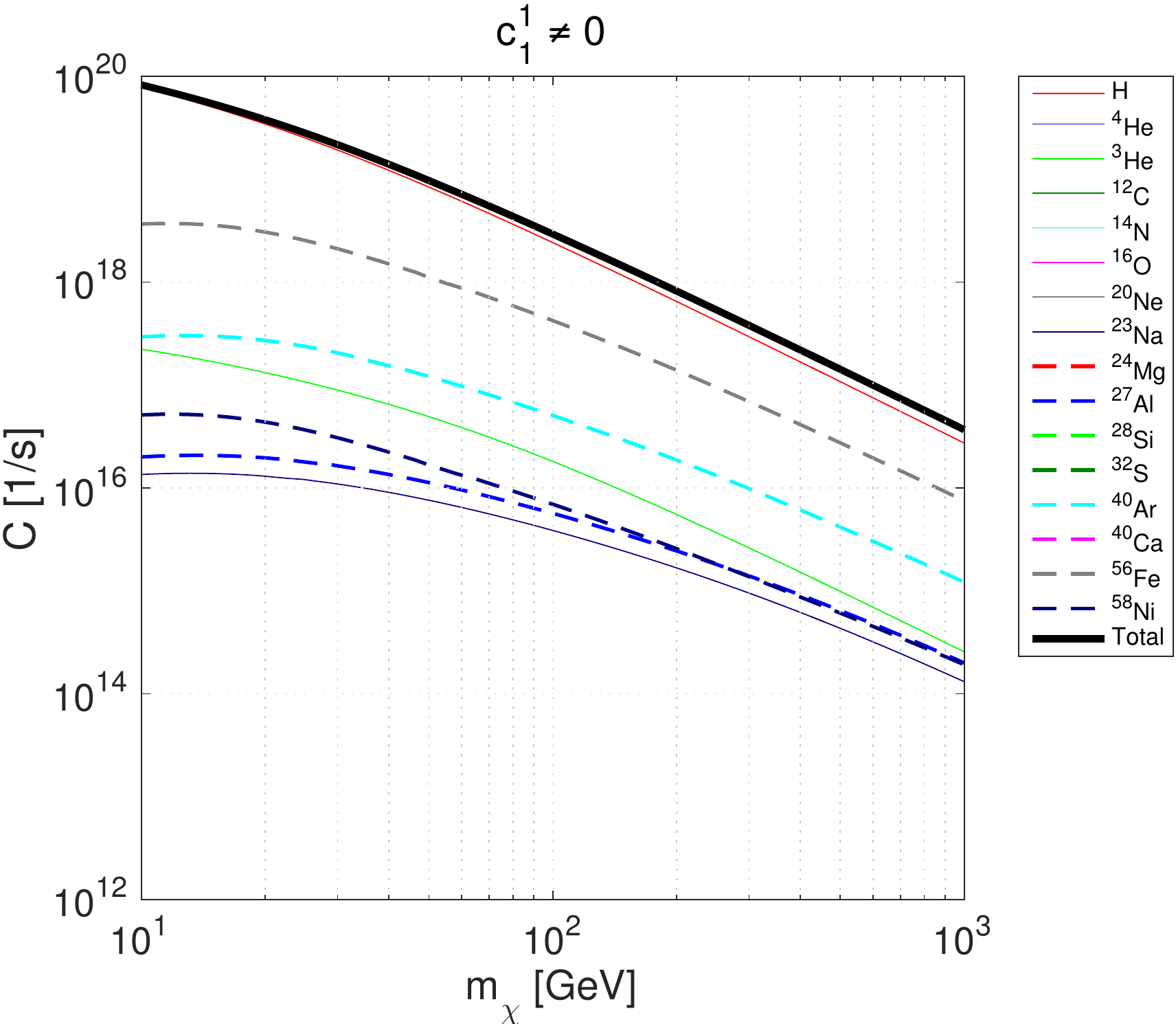}
\end{minipage}
\begin{minipage}[t]{0.49\linewidth}
\centering
\includegraphics[width=\textwidth]{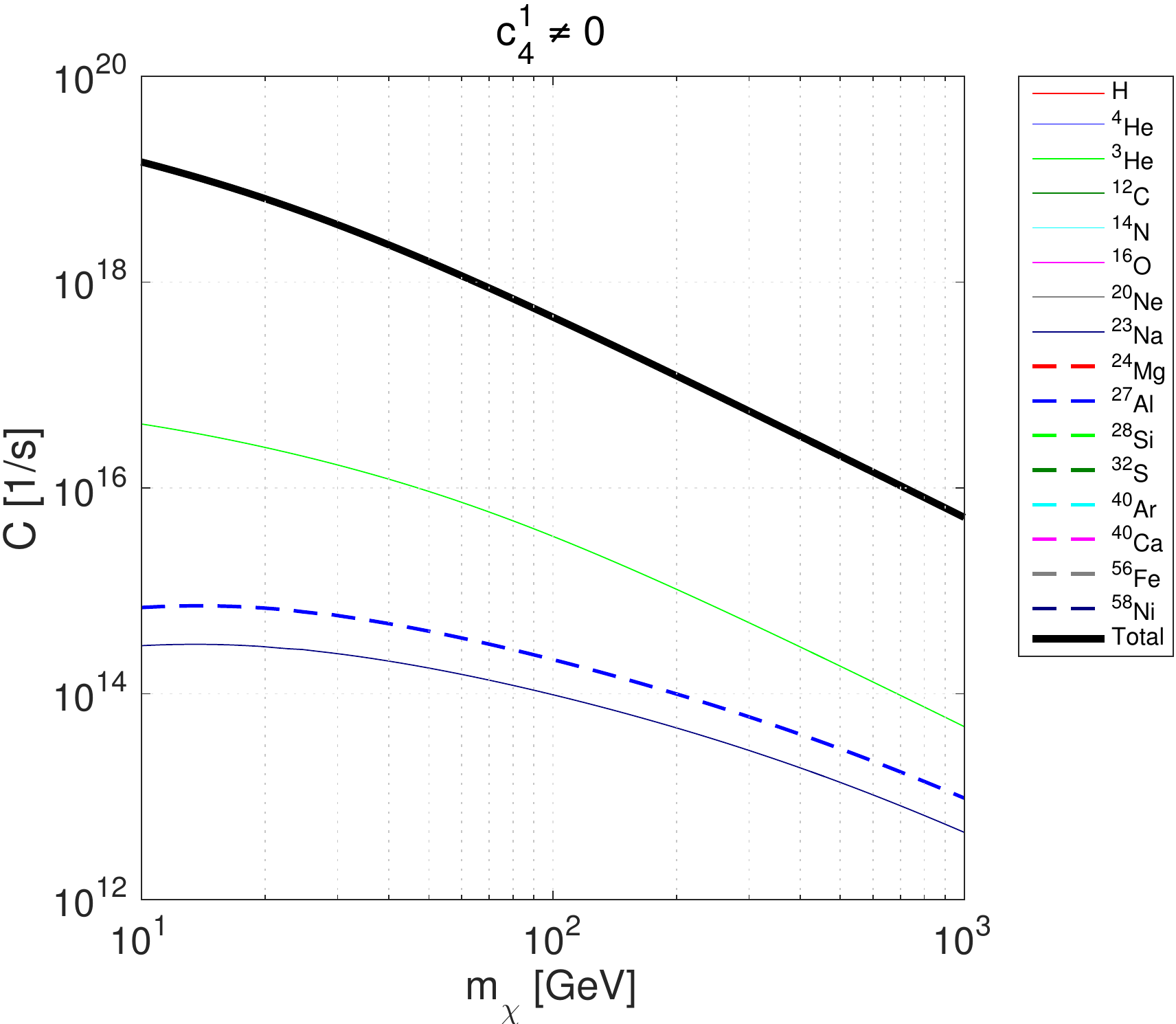}
\end{minipage}
\end{center}
\caption{Dark matter capture rate $C$ as a function of the dark matter particle mass $m_\chi$ for $c_1^0\neq0$ (top left panel), $c_4^0\neq0$ (top right panel), $c_1^1\neq0$ (bottom left panel), and $c_4^1\neq0$ (bottom right panel). We report the total capture rate (thick black line), and partial capture rates specific to the 16 most abundant elements in the Sun. Conventions for colors and lines are those in the legends. }
\label{fig:c1c4}
\end{figure}
\begin{figure}[t]
\begin{center}
\begin{minipage}[t]{0.49\linewidth}
\centering
\includegraphics[width=\textwidth]{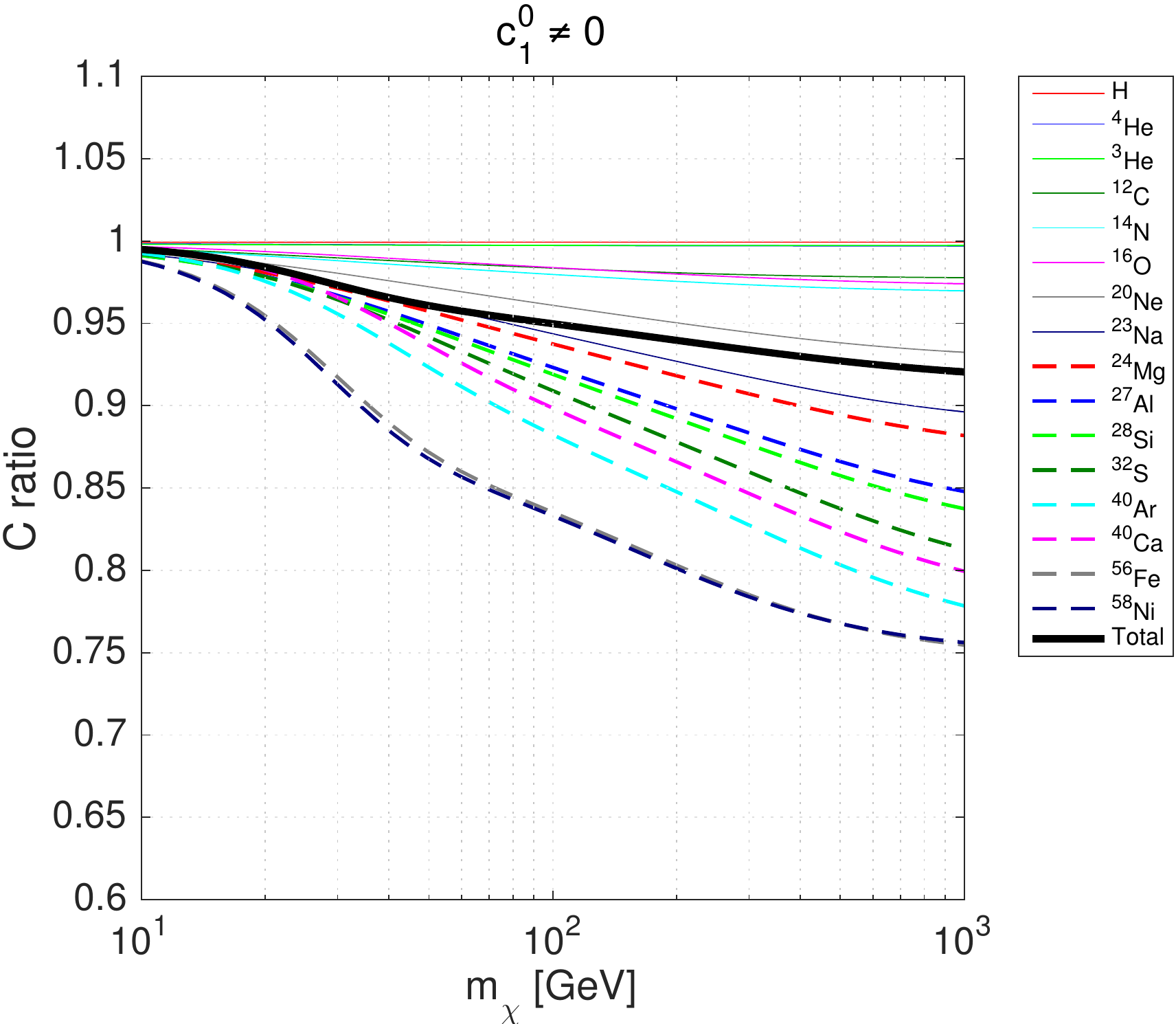}
\end{minipage}
\begin{minipage}[t]{0.48\linewidth}
\centering
\includegraphics[width=\textwidth]{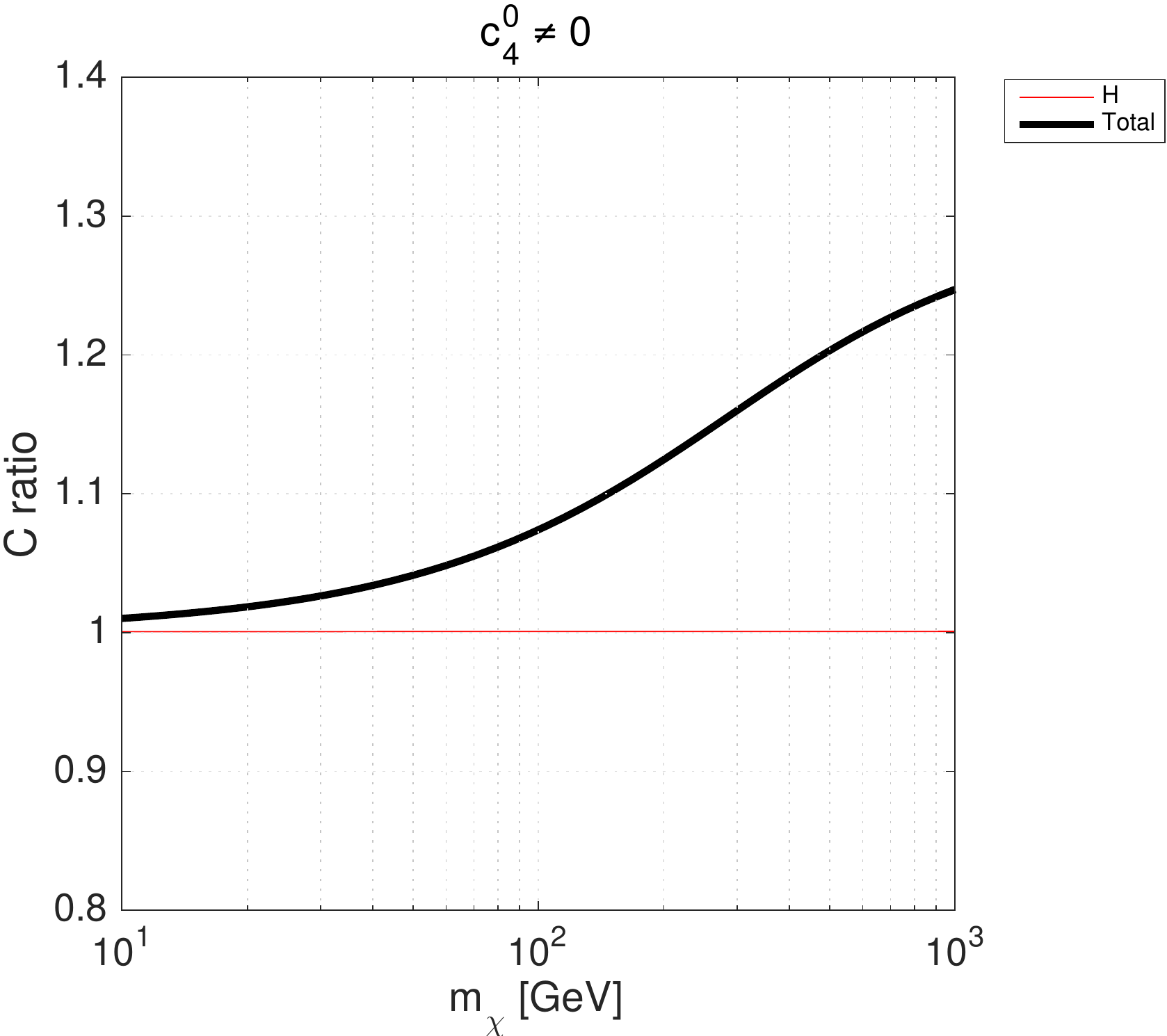}
\end{minipage}
\end{center}
\caption{{\it Left panel.} Ratio of the capture rate of this work for $c_1^0\neq0$ to the capture rate computed with {\sffamily darksusy} for spin-independent dark matter interactions. We report the ratio of total rates (thick black line), and the ratio of partial rates specific to the 16 most abundant elements in the Sun. The two total rates differ by at most 8\%. {\it Right panel.} Same as in the left panel, but for $c_4^0\neq0$. In this case the comparison can be performed for the total rates and for H only, since elements heavier than H are not included in {\sffamily darksusy} for dark matter spin-dependent interactions.}
\label{fig:comp}
\end{figure}

The OBDME for orbits corresponding to the nuclear core can be analytically calculated. Only multipoles of nuclear response operators with $L=\tau=0$ contribute, since in a nuclear core all orbits are fully occupied. One finds~\cite{Walecka1}
\begin{equation}
\psi^{L;\tau}_{|\alpha| |\beta|}=\sqrt{2(2J+1)(2T+1)(2j_\alpha+1)}\, \delta_{|\alpha| |\beta|}\delta_{\tau0}\,\delta_{L0}\,.
\label{eq:core}
\end{equation}
The calculation of the OBDME for the remaining orbits in the model space instead requires a numerical approach.
We address this problem using the {\sffamily Nushell@MSU} program~\cite{NuShell,Brown:2001zz}. 
This code mainly relies on three inputs: the target nucleus spin, isospin and parity; the Hamiltonian for valence nucleon interactions (several options are provided with the code); the model space, including restrictions on the number of nucleons in the most energetic orbits. The assumptions made in our calculations are listed in Tab.~\ref{tab:inputs}, and closely follow the guidelines provided in Ref.~\cite{Brown:2001zz}, and references therein. Assigned these inputs, the code first calculates the many-body ground-state wave function of the valence nucleon system diagonalizing the assumed interaction Hamiltonian. Then it evaluates the overlap of this wave function with the single-particle states $|\alpha \rangle$ according to Eq.~(\ref{eq:OBDME}). 
The OBDME that we obtain for $^{23}$Na, $^{28}$Si and $^{19}$F using the {\sffamily Nushell@MSU} w-interaction negligibly differ from those in the code~\cite{Anand:2013yka} (here we use $^{19}$F for comparison only, but it does not enter in our calculation). The remaining interactions in Tab.~\ref{tab:inputs} were studied in~\cite{Warburton:1992rh,Nummela:2001xh,Honma:2004xk}. For instance, in the full $pf$ model space the gx1 interaction was found to successfully describe binding energies, electro-magnetic transitions, and excitation spectra of Iron, and of various Nickel isotopes~\cite{Honma:2004xk}. The major limitation of our numerical OBDME calculation hence resides in the use of model space restrictions. We adopt such restrictions because of limits in the available computing power: {\sffamily Nushell@MSU} only runs on Windows machines, whereas our cluster for extensive calculations has a Unix architecture. For the gx1 interaction, the impact of restrictions on observable quantitates has been discussed in~\cite{Honma:2004xk}. For the isotopes $^{56}$Fe and $^{58}$Ni, for example, a restriction of the model space where 5 or more nucleons are allowed to be excited from the $f_{7/2}$ orbit to higher orbits implies an underestimation of the binding energy of the order of a few percent. 

Ultimately, the nuclear structure calculations performed here have to be considered explorative, due to the restrictions in Tab.~\ref{tab:inputs}, and to the fact that more sophisticated interaction Hamiltonians could in principle be considered. At the same time, we are not aware of nuclear structure calculations of comparable complexity in the context of dark matter capture by the Sun.

Before concluding this section, we comment on the OBDME calculation for Hydrogen. In the present analysis, Hydrogen constitutes a special case, in that it consists of a single valence nucleon system with no-core. Its OBDME can be trivially calculated as follows~\cite{Walecka1} 
\begin{equation}
\psi^{L;\tau}_{|\alpha| |\beta|}= \delta_{|\alpha| |\gamma|} \delta_{|\beta| |\gamma|}\,,
\end{equation}
where $|\gamma|$ corresponds to the $0s_{1/2}$ orbit. 

Using the OBDME resulting from the methods outlined above, we evaluate the reduced nuclear matrix elements in Eq.~(\ref{eq:W}), and hence the dark matter-nucleus scattering cross-section (\ref{eq:sigma}) for the most abundant elements in the Sun. This cross-section will allow us to calculate the rate of dark matter capture by the Sun (\ref{eq:rate}) for all interaction operators in Tab.~\ref{tab:operators}, as we will see next. The nuclear response functions that we obtain in this analysis, i.e. Eq.~(\ref{eq:W}), are listed in Appendix~\ref{sec:appNuc}, and can be used by the reader for other projects.

\begin{figure}[t]
\begin{center}
\begin{minipage}[t]{0.49\linewidth}
\centering
\includegraphics[width=\textwidth]{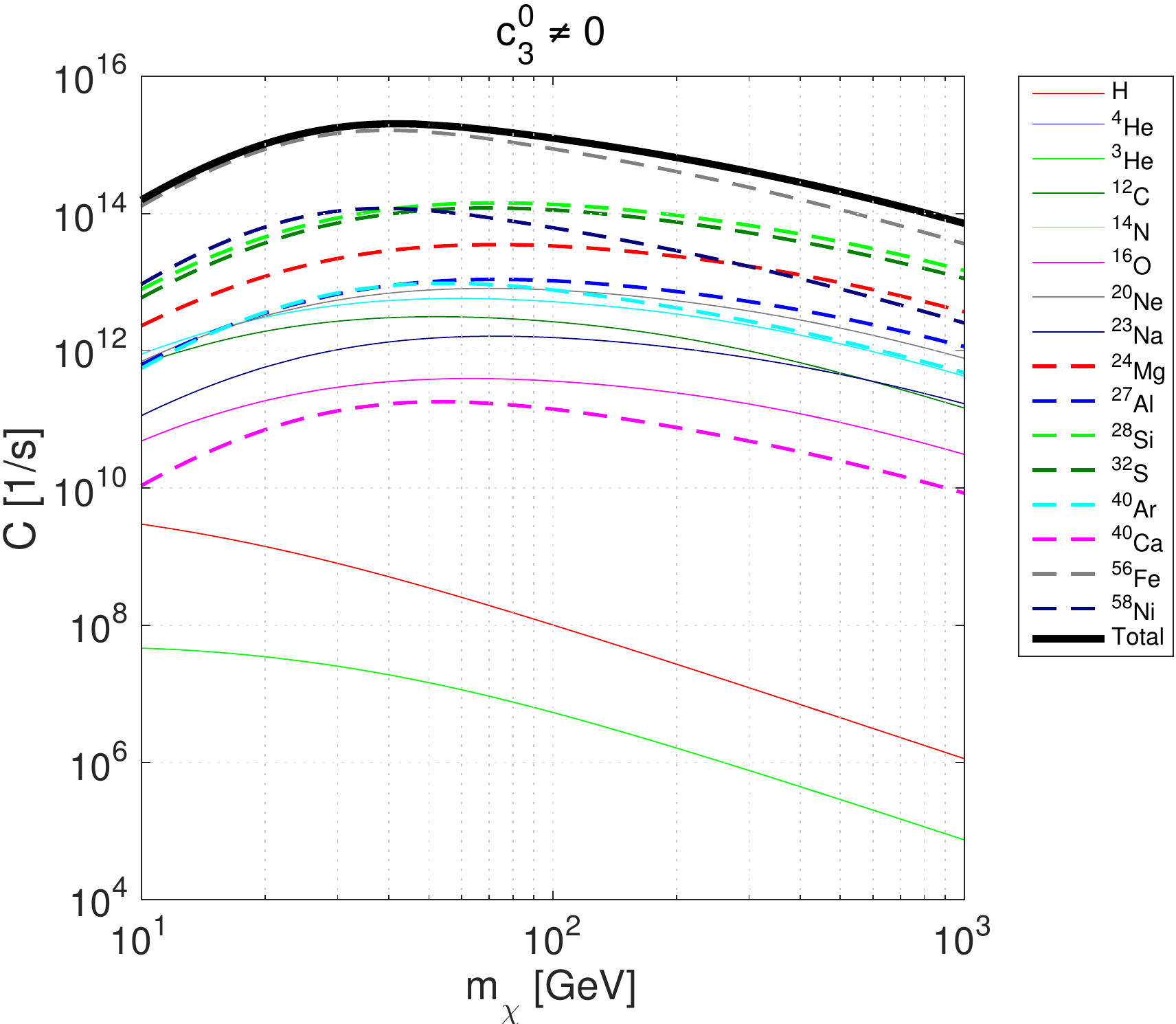}
\end{minipage}
\begin{minipage}[t]{0.49\linewidth}
\centering
\includegraphics[width=\textwidth]{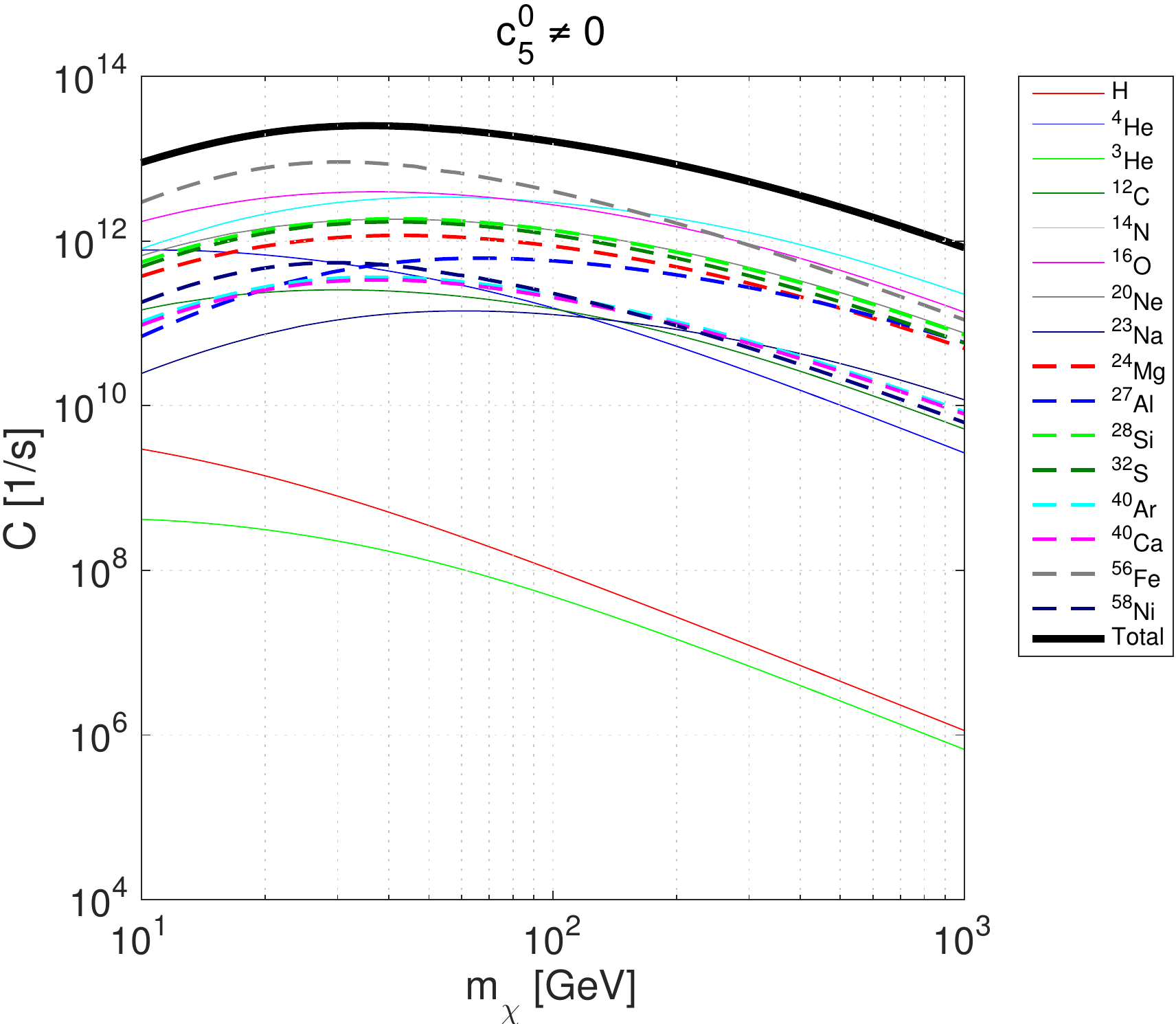}
\end{minipage}
\begin{minipage}[t]{0.49\linewidth}
\centering
\includegraphics[width=\textwidth]{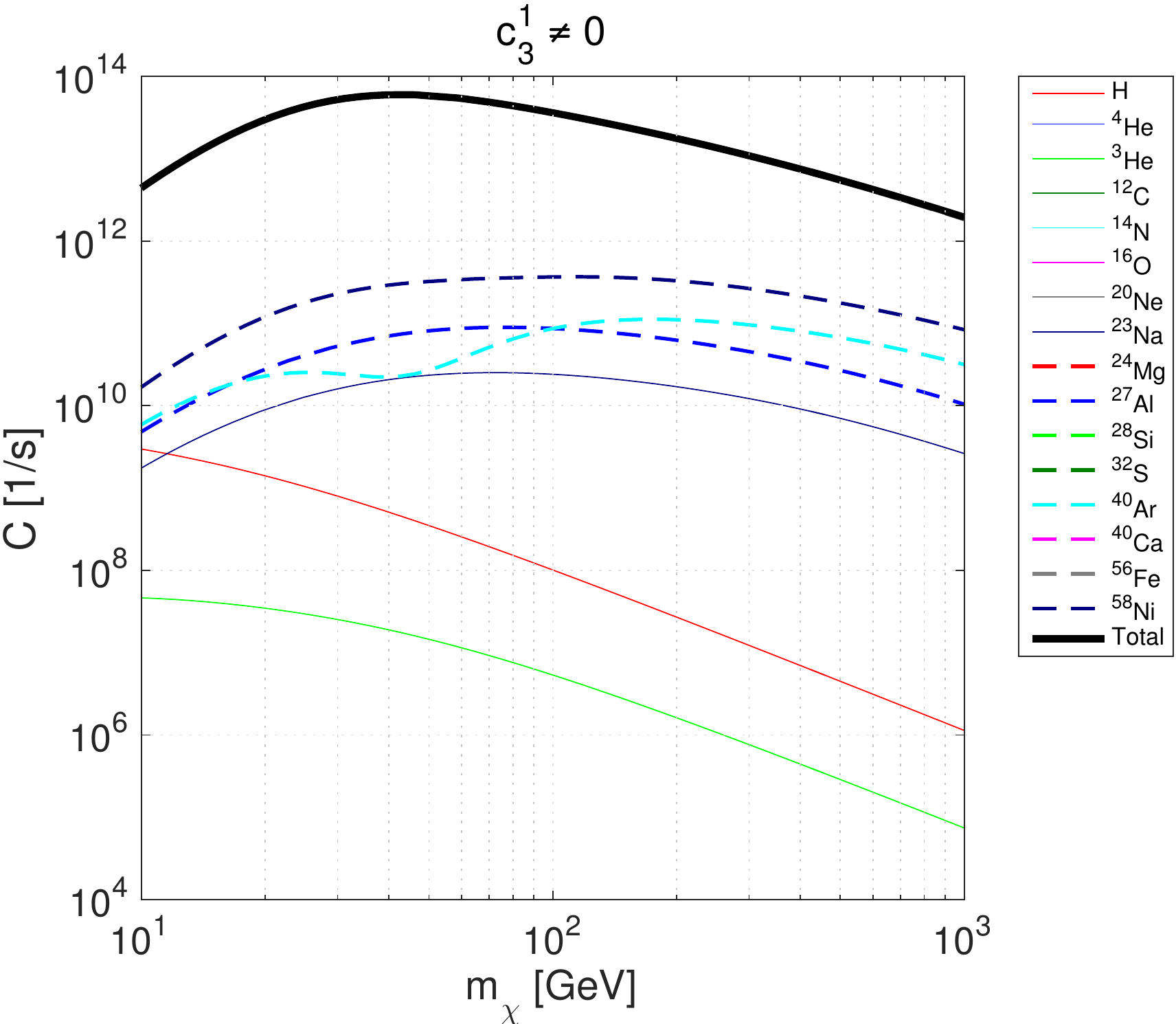}
\end{minipage}
\begin{minipage}[t]{0.49\linewidth}
\centering
\includegraphics[width=\textwidth]{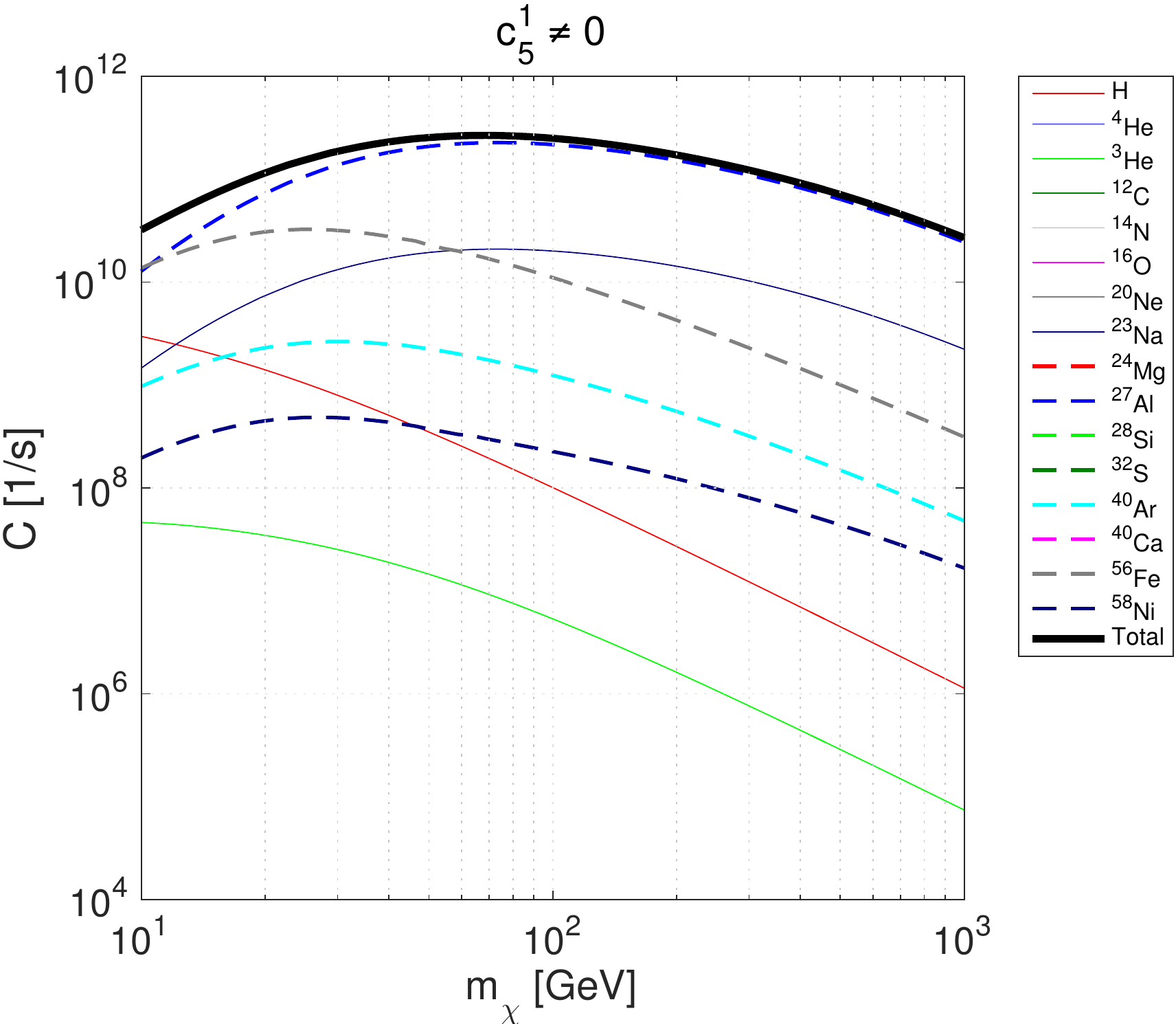}
\end{minipage}
\end{center}
\caption{Same as in Fig.~\ref{fig:c1c4}, but for the interaction operators $\hat{\mathcal{O}}_3$ and $\hat{\mathcal{O}}_5$.}
\label{fig:c3c5}
\end{figure}
\begin{figure}[t]
\begin{center}
\begin{minipage}[t]{0.49\linewidth}
\centering
\includegraphics[width=\textwidth]{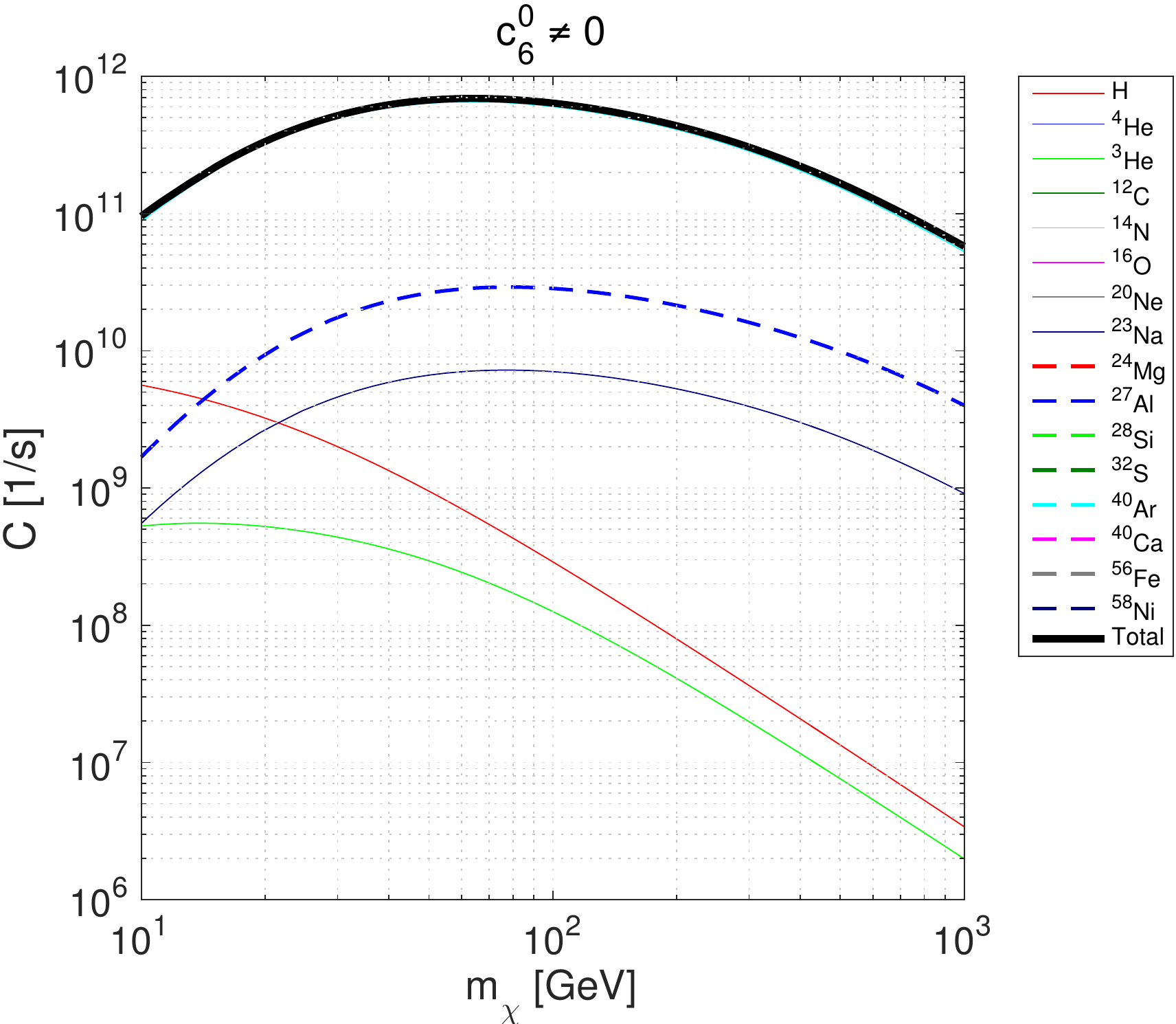}
\end{minipage}
\begin{minipage}[t]{0.49\linewidth}
\centering
\includegraphics[width=\textwidth]{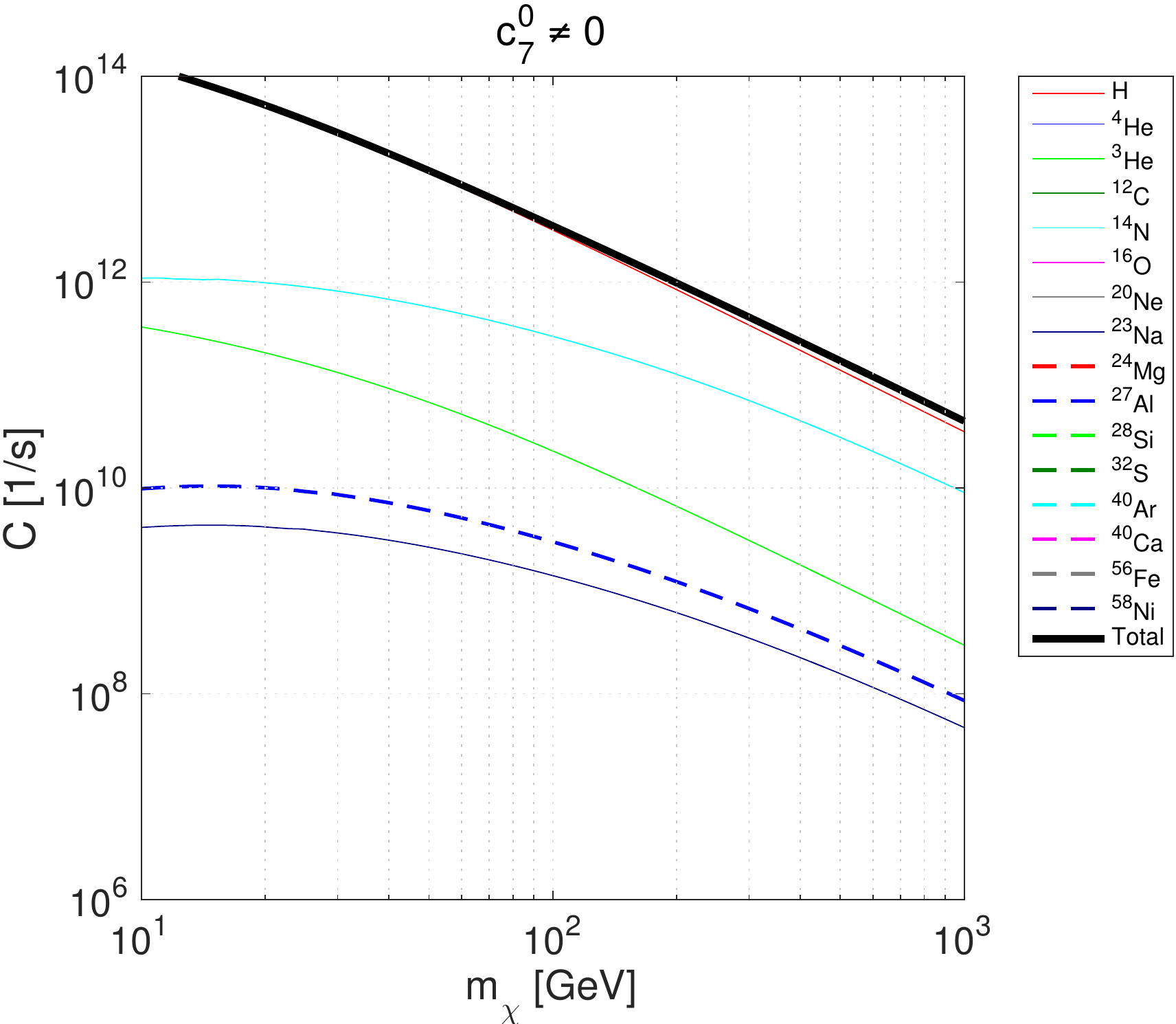}
\end{minipage}
\begin{minipage}[t]{0.49\linewidth}
\centering
\includegraphics[width=\textwidth]{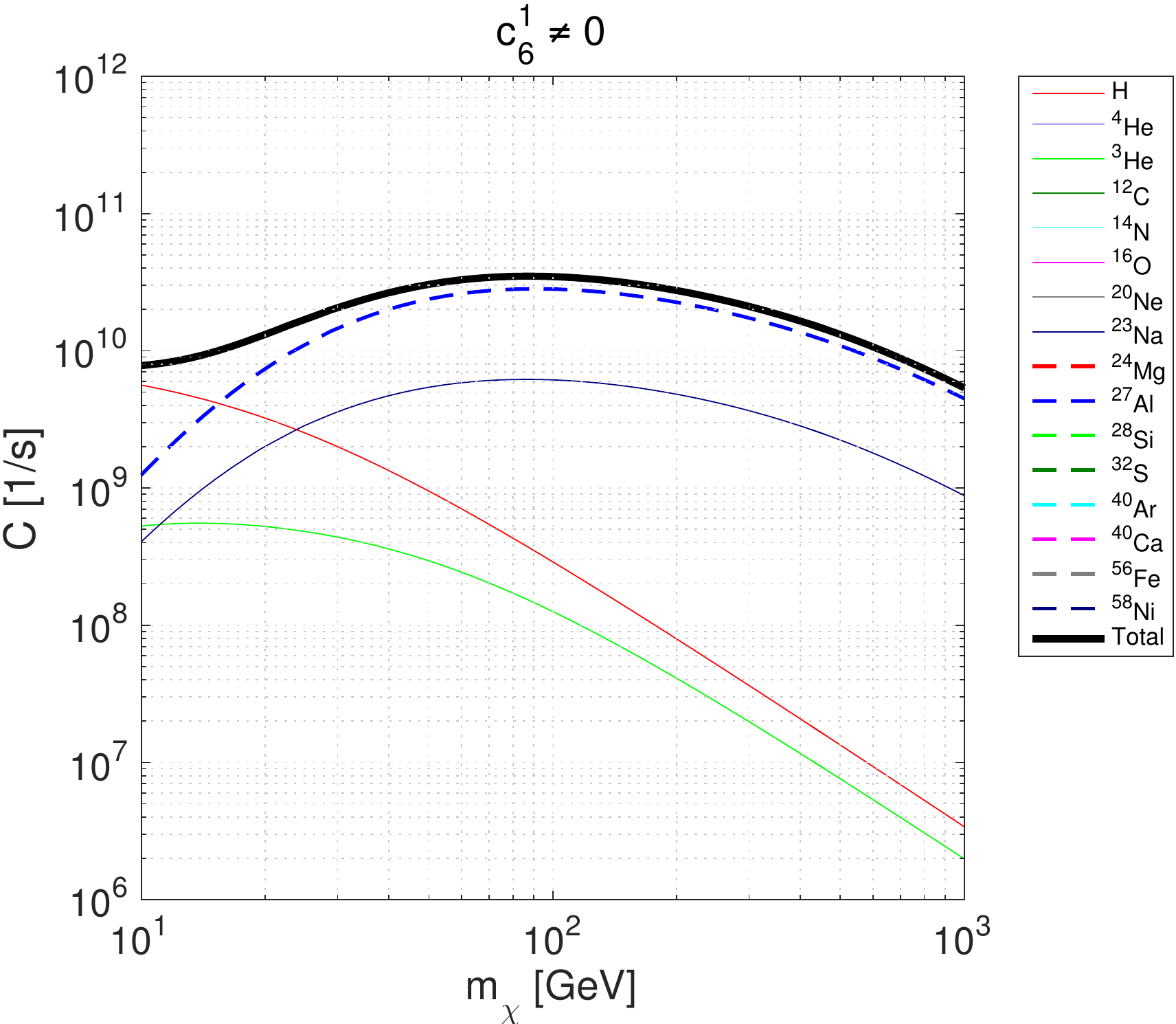}
\end{minipage}
\begin{minipage}[t]{0.49\linewidth}
\centering
\includegraphics[width=\textwidth]{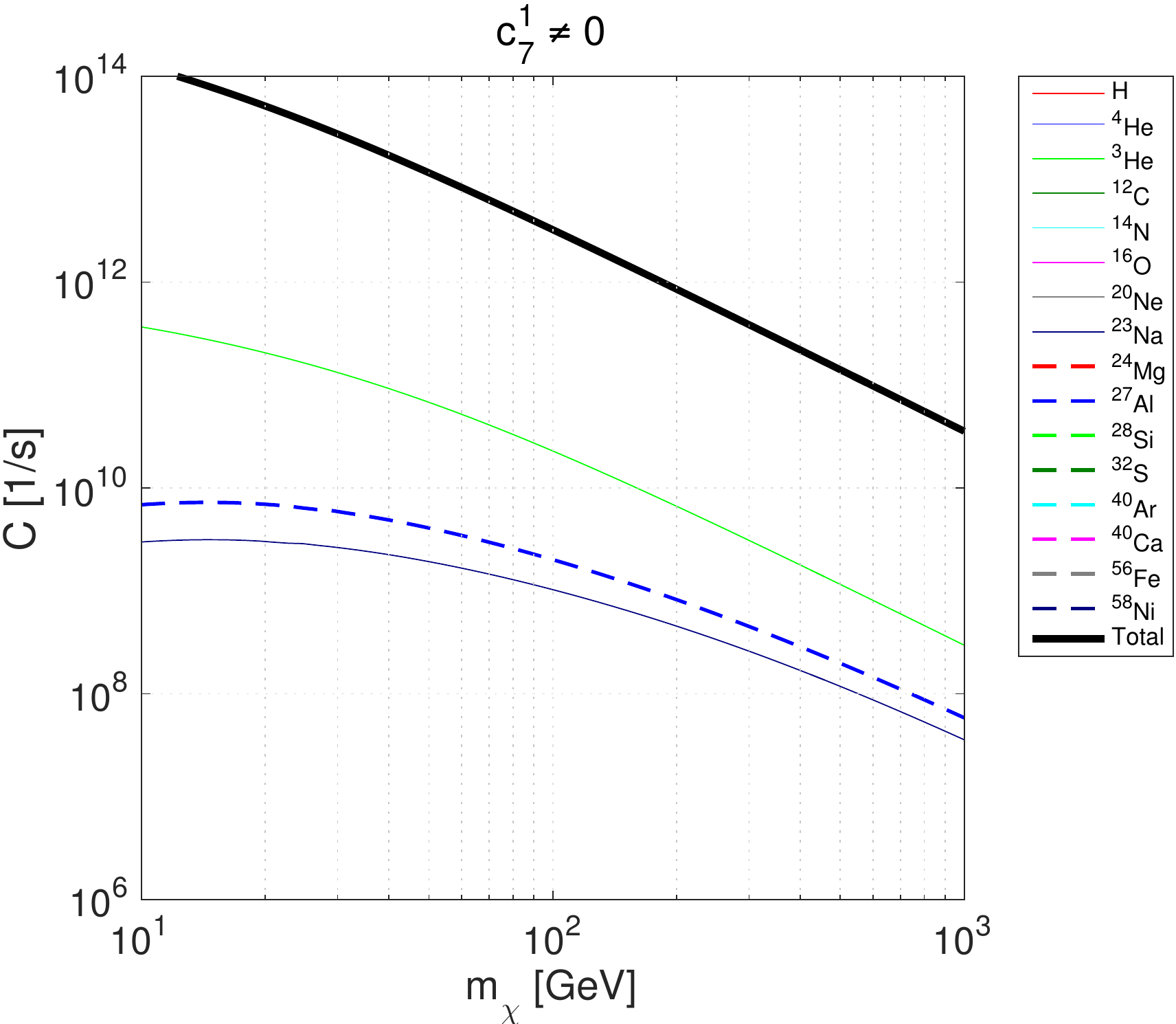}
\end{minipage}
\end{center}
\caption{Same as in Fig.~\ref{fig:c1c4}, but for the interaction operators $\hat{\mathcal{O}}_6$ and $\hat{\mathcal{O}}_7$.}
\label{fig:c6c7}
\end{figure}
\begin{figure}[t]
\begin{center}
\begin{minipage}[t]{0.49\linewidth}
\centering
\includegraphics[width=\textwidth]{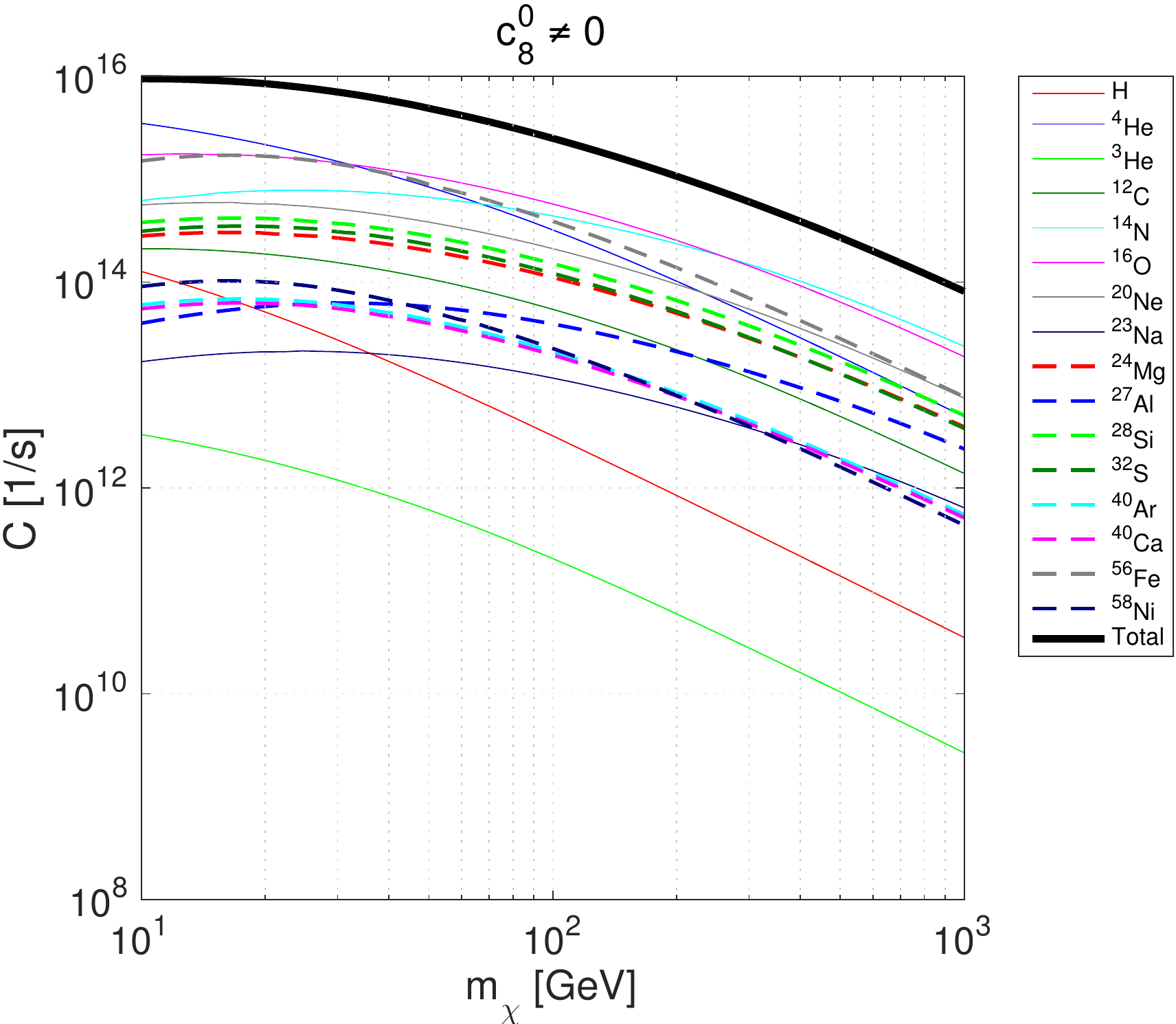}
\end{minipage}
\begin{minipage}[t]{0.49\linewidth}
\centering
\includegraphics[width=\textwidth]{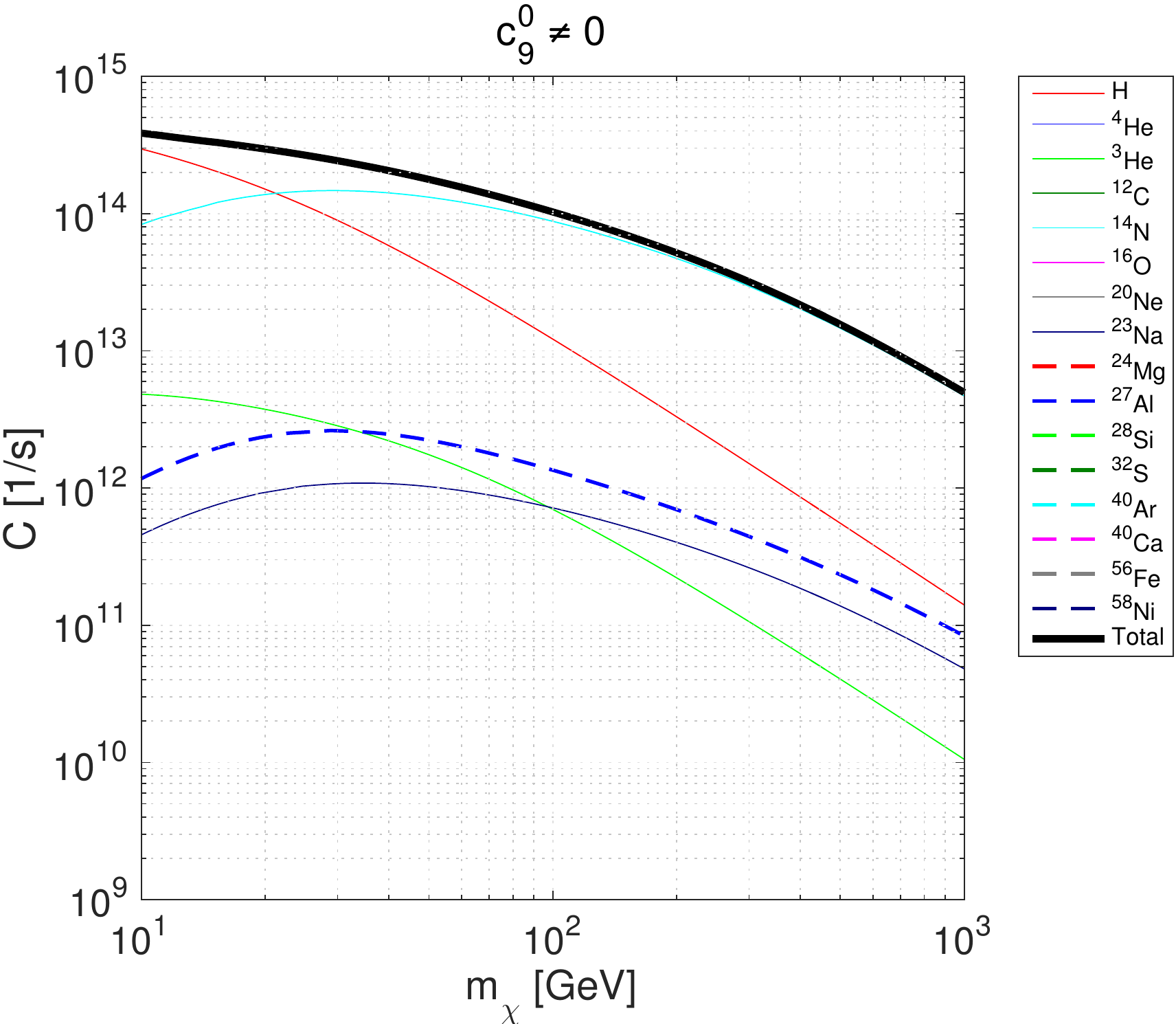}
\end{minipage}
\begin{minipage}[t]{0.49\linewidth}
\centering
\includegraphics[width=\textwidth]{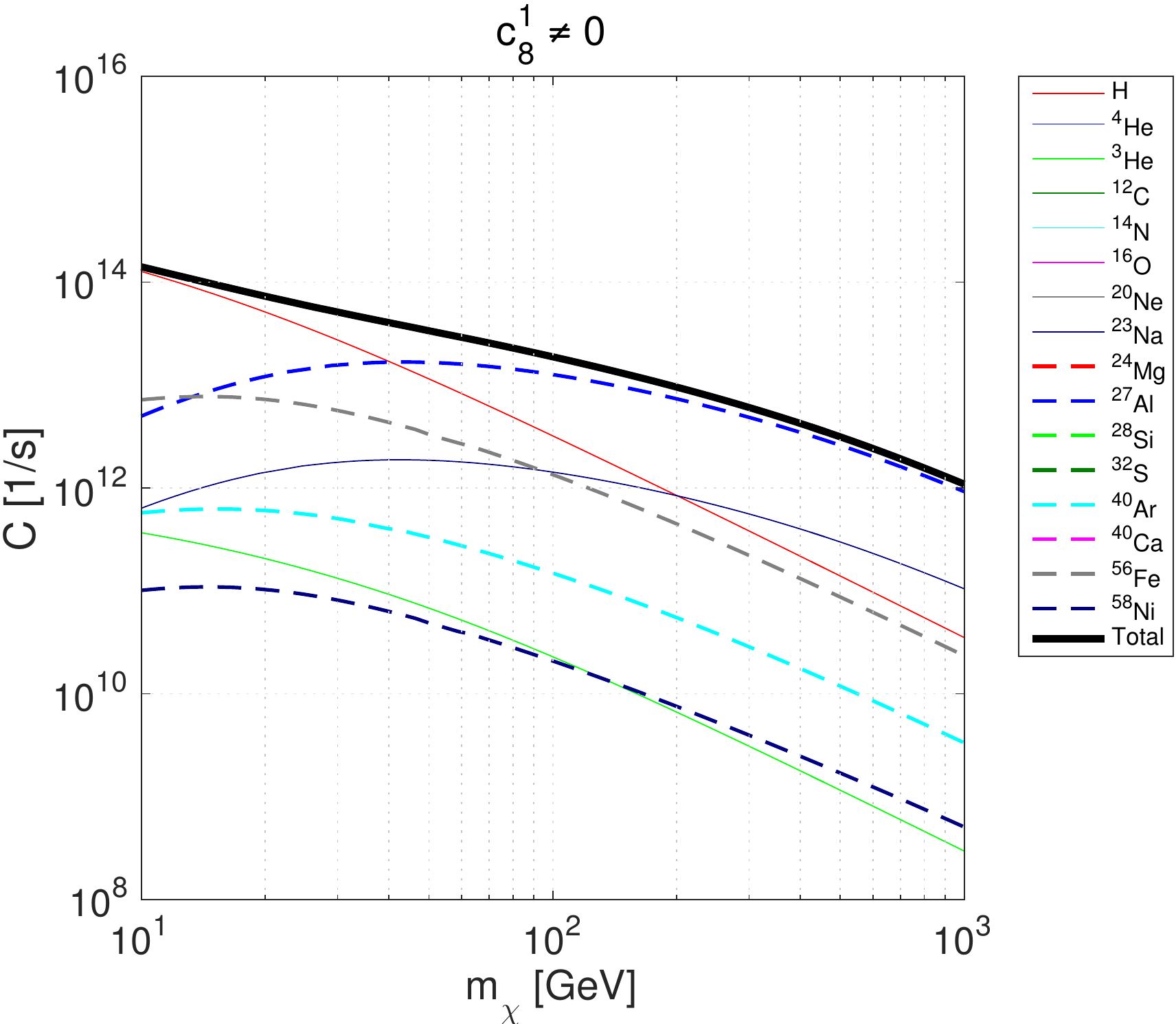}
\end{minipage}
\begin{minipage}[t]{0.49\linewidth}
\centering
\includegraphics[width=\textwidth]{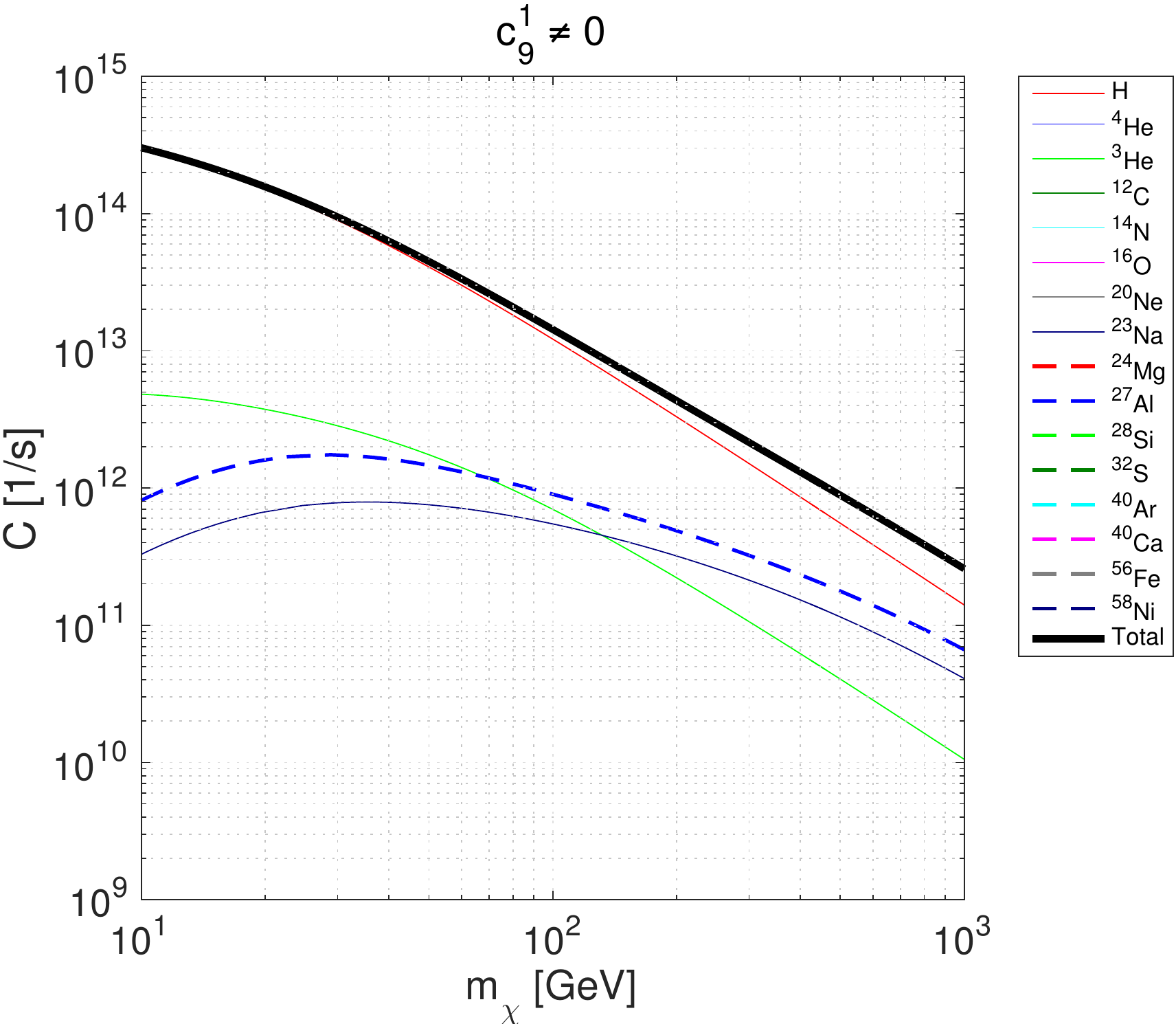}
\end{minipage}
\end{center}
\caption{Same as in Fig.~\ref{fig:c1c4}, but for the interaction operators $\hat{\mathcal{O}}_8$ and $\hat{\mathcal{O}}_9$.}
\label{fig:c8c9}
\end{figure}
\begin{figure}[t]
\begin{center}
\begin{minipage}[t]{0.49\linewidth}
\centering
\includegraphics[width=\textwidth]{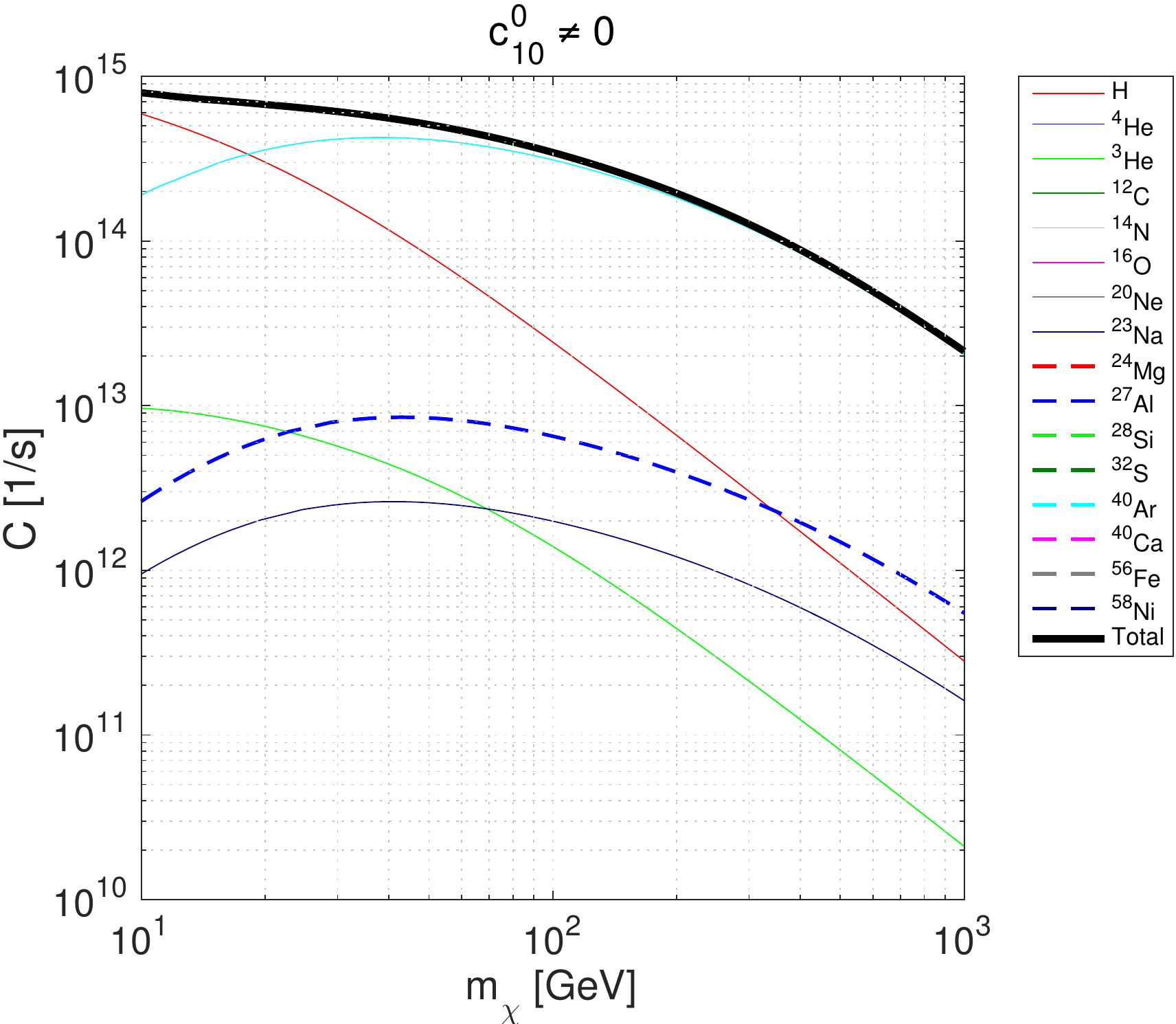}
\end{minipage}
\begin{minipage}[t]{0.49\linewidth}
\centering
\includegraphics[width=\textwidth]{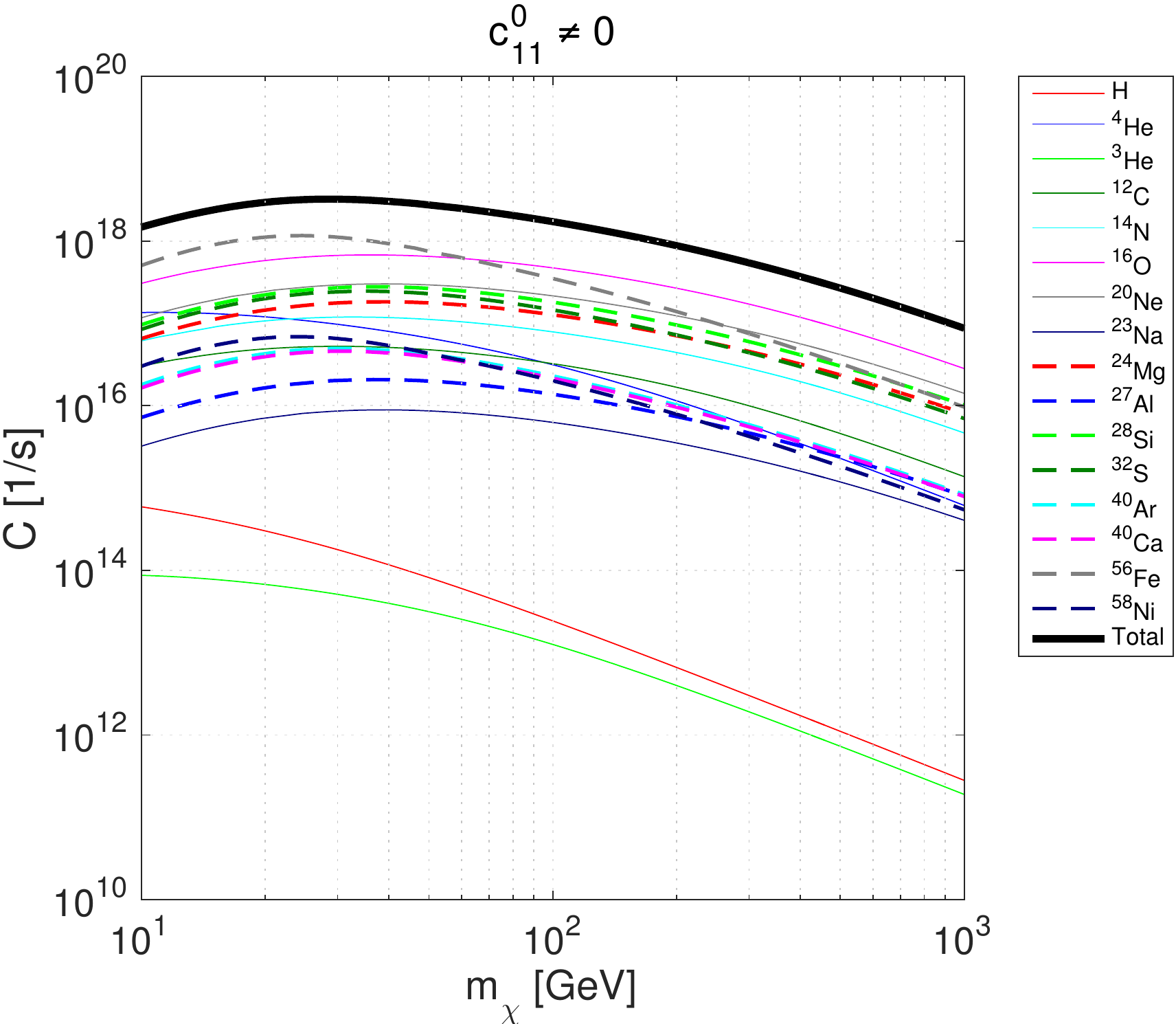}
\end{minipage}
\begin{minipage}[t]{0.49\linewidth}
\centering
\includegraphics[width=\textwidth]{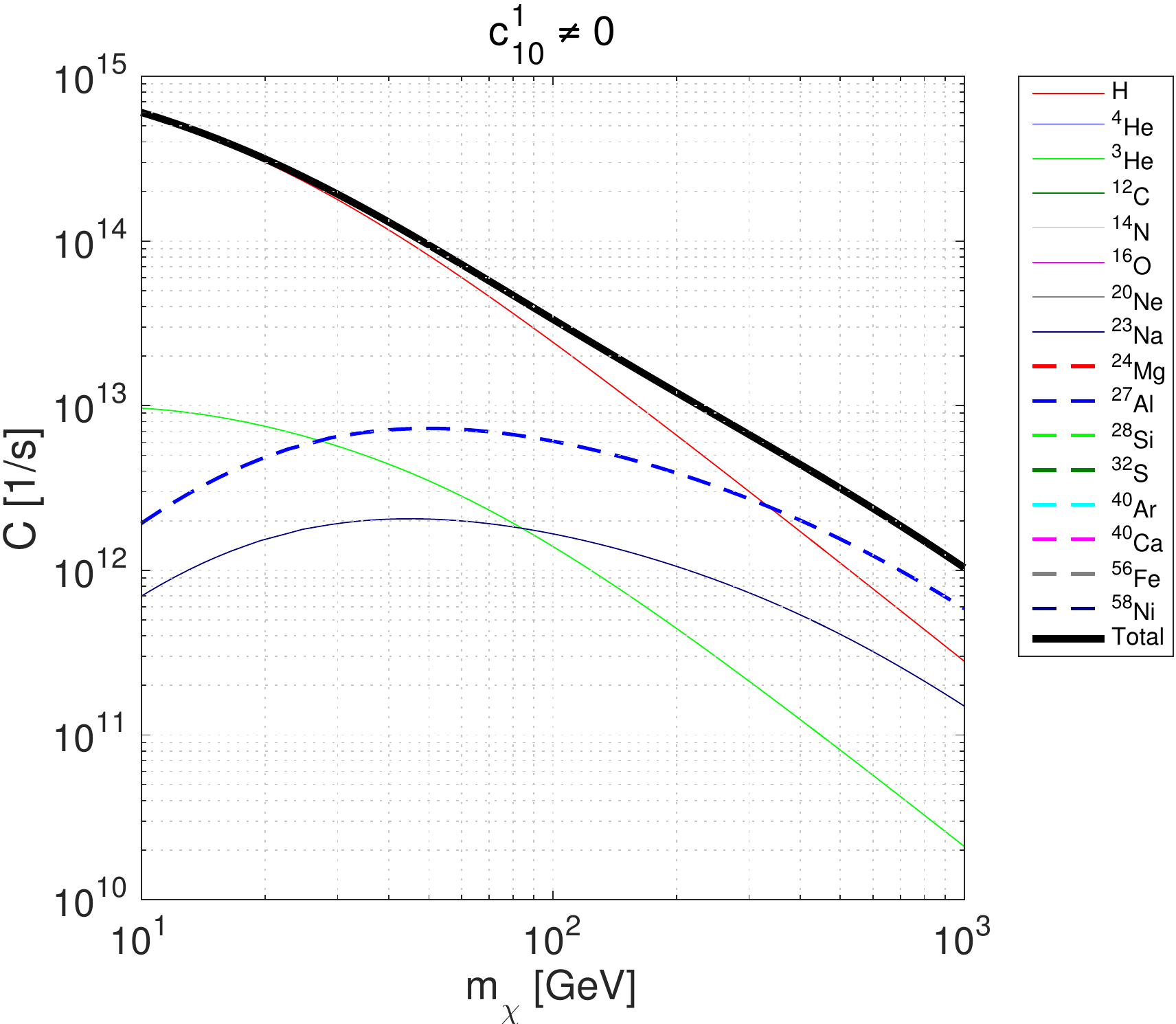}
\end{minipage}
\begin{minipage}[t]{0.49\linewidth}
\centering
\includegraphics[width=\textwidth]{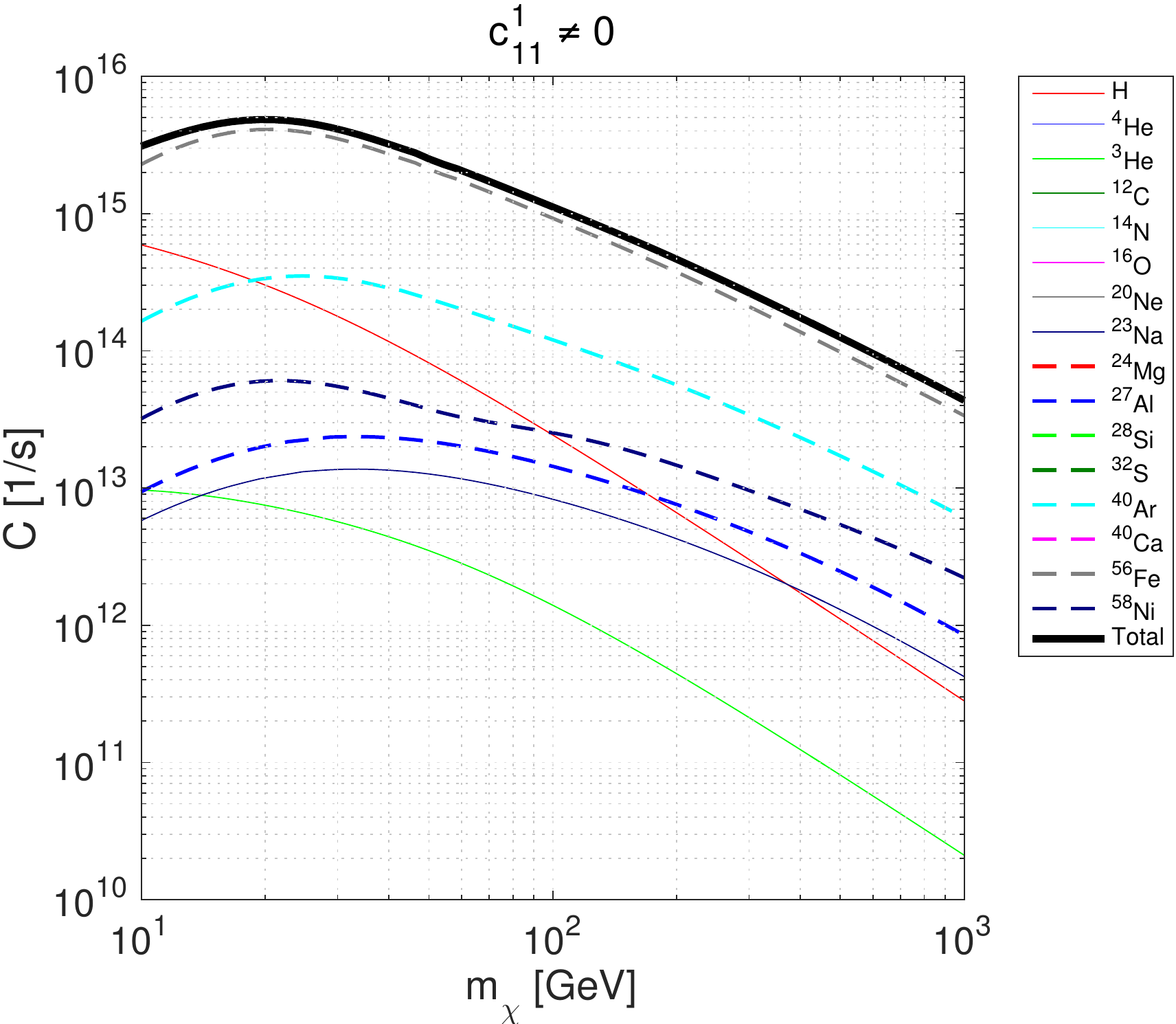}
\end{minipage}
\end{center}
\caption{Same as in Fig.~\ref{fig:c1c4}, but for the interaction operators $\hat{\mathcal{O}}_{10}$ and $\hat{\mathcal{O}}_{11}$.}
\label{fig:c10c11}
\end{figure}
\begin{figure}[t]
\begin{center}
\begin{minipage}[t]{0.49\linewidth}
\centering
\includegraphics[width=\textwidth]{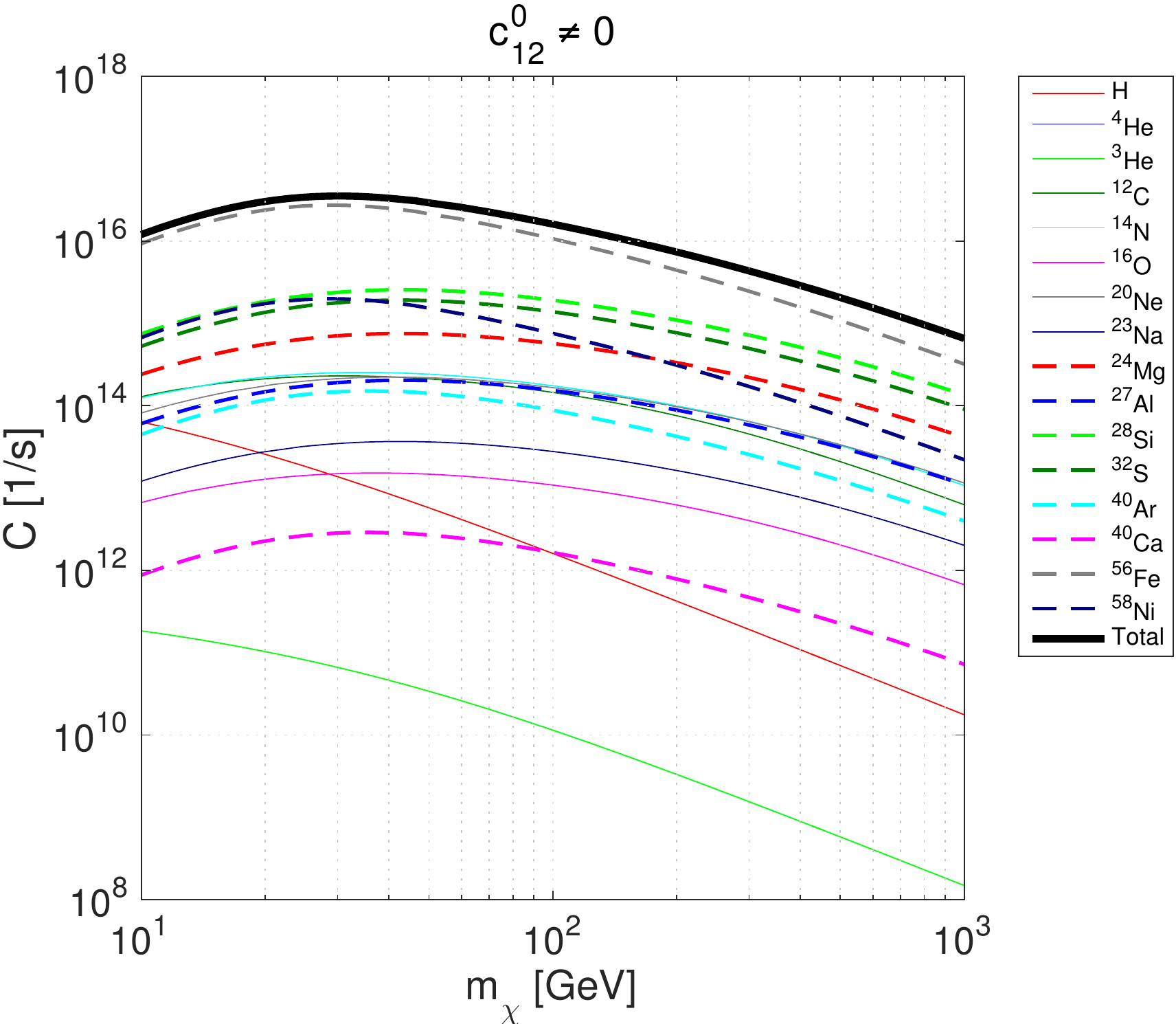}
\end{minipage}
\begin{minipage}[t]{0.49\linewidth}
\centering
\includegraphics[width=\textwidth]{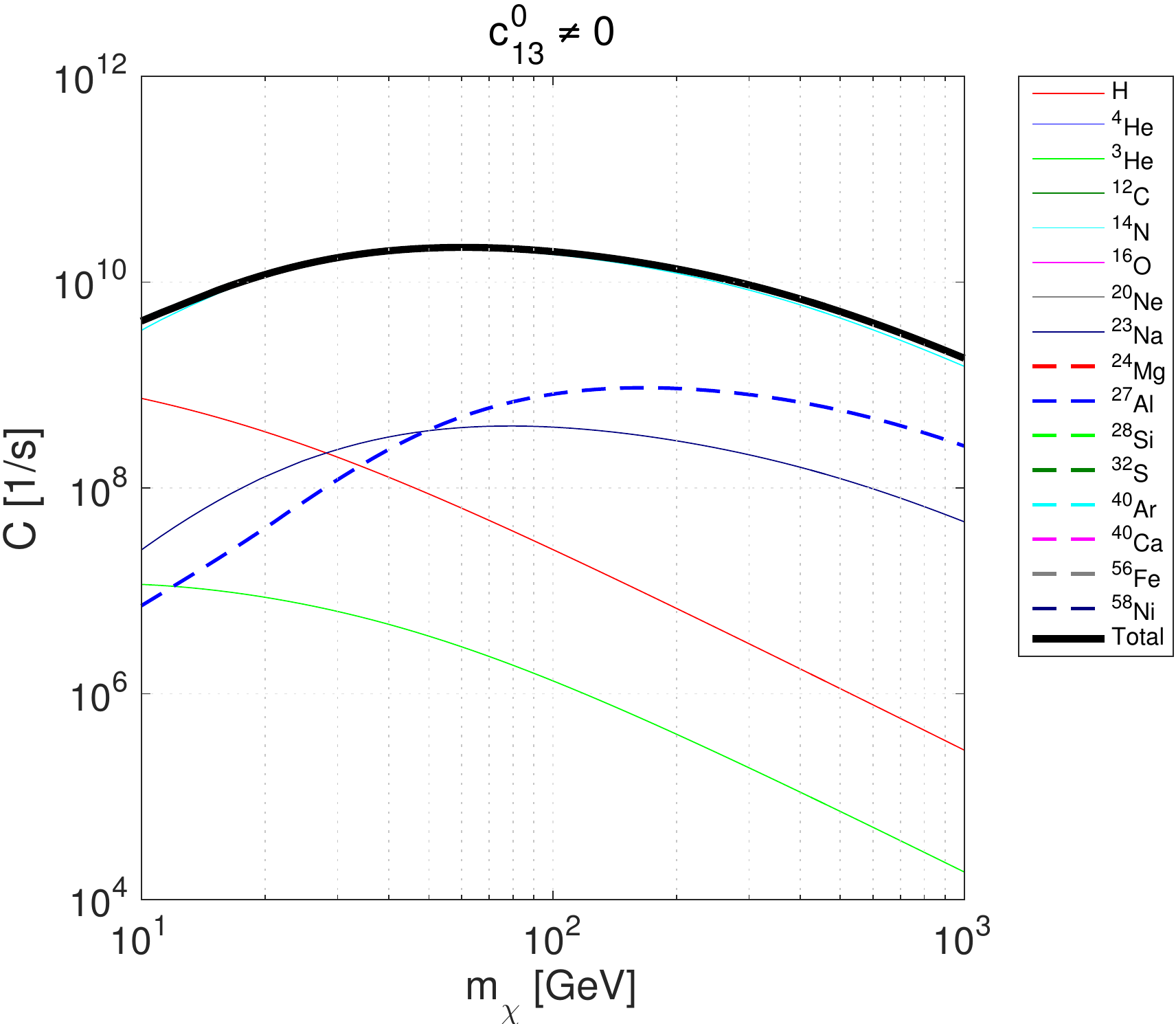}
\end{minipage}
\begin{minipage}[t]{0.49\linewidth}
\centering
\includegraphics[width=\textwidth]{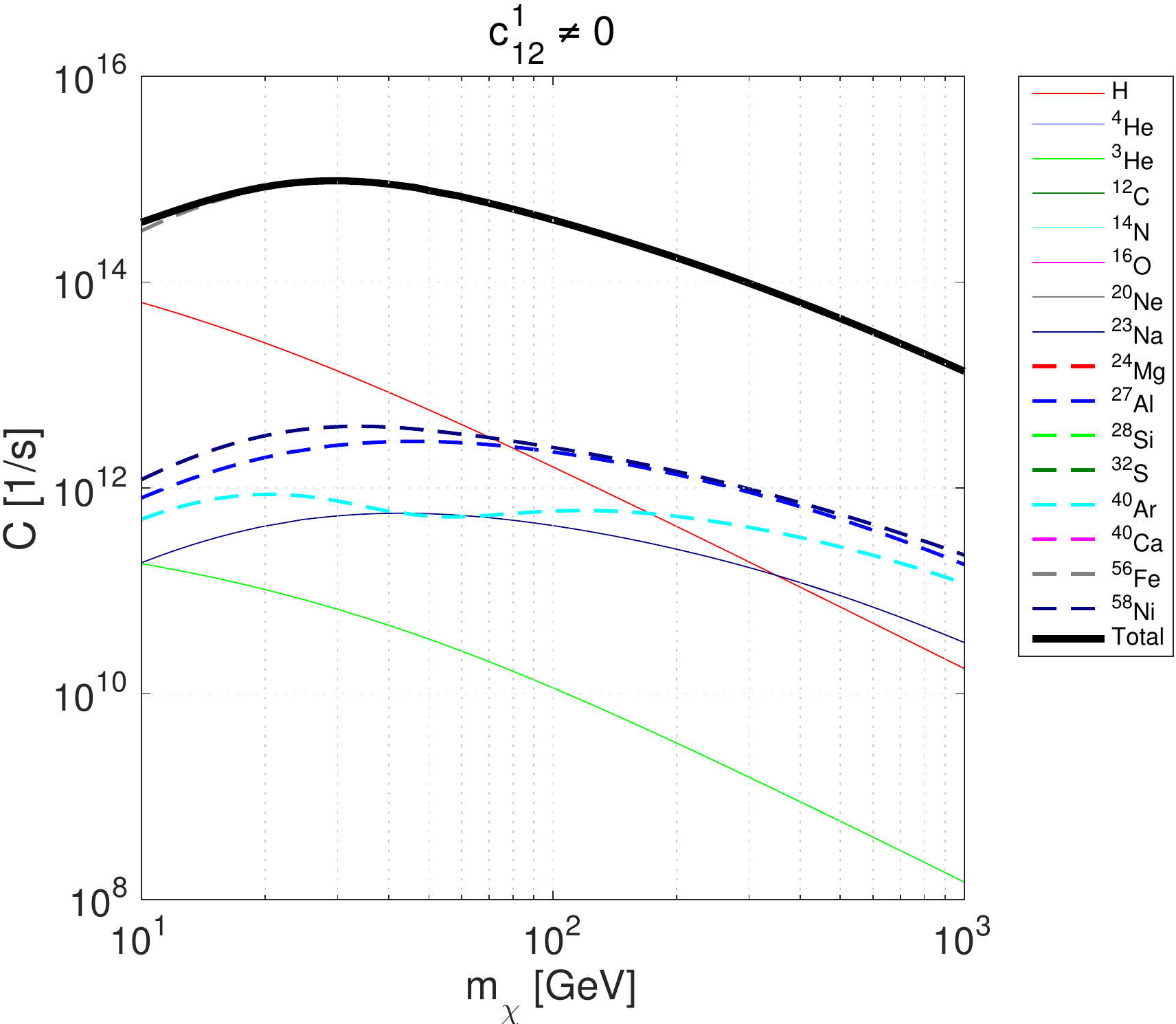}
\end{minipage}
\begin{minipage}[t]{0.49\linewidth}
\centering
\includegraphics[width=\textwidth]{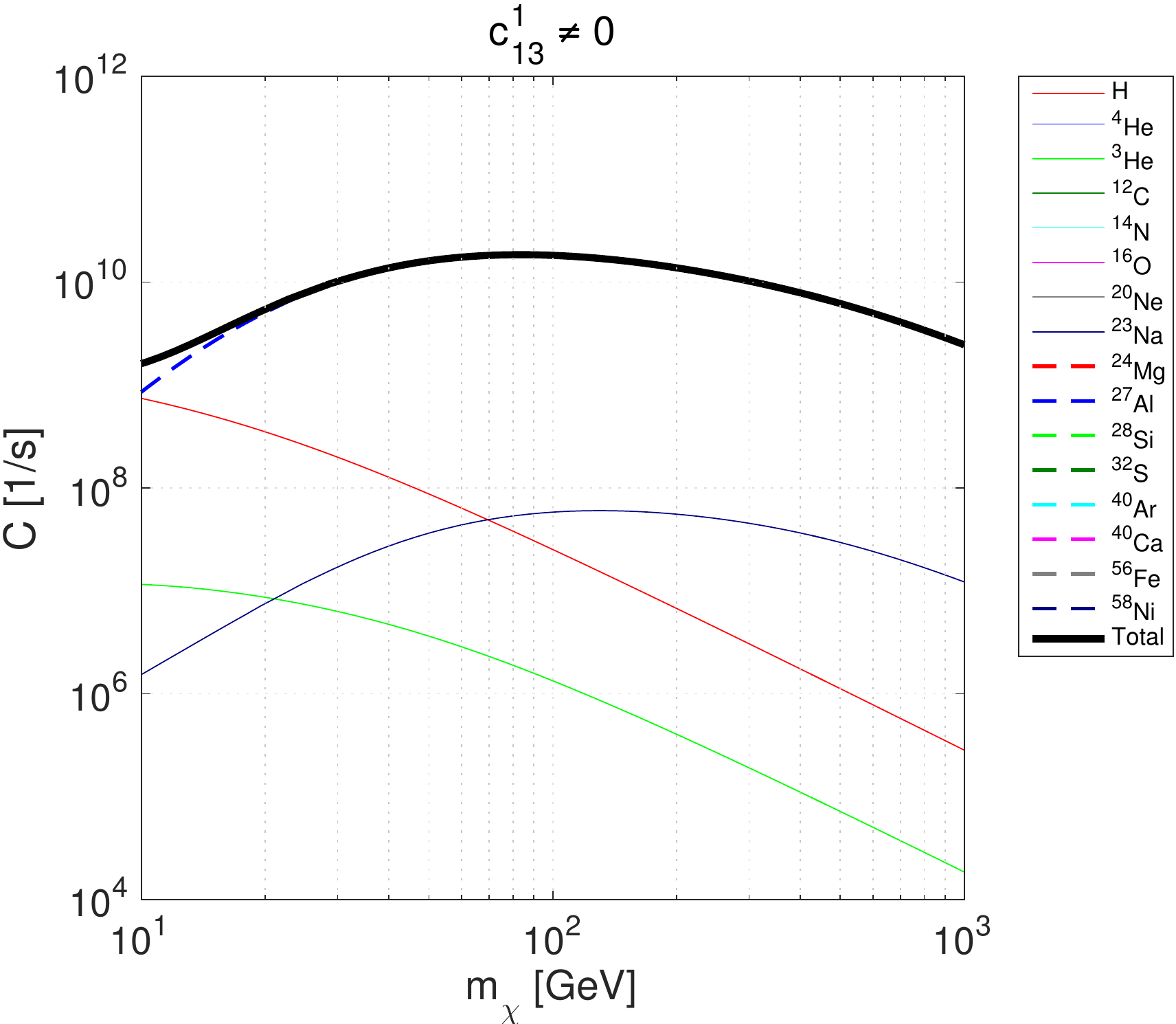}
\end{minipage}
\end{center}
\caption{Same as in Fig.~\ref{fig:c1c4}, but for the interaction operators $\hat{\mathcal{O}}_{12}$ and $\hat{\mathcal{O}}_{13}$.}
\label{fig:c12c13}
\end{figure}
\begin{figure}[t]
\begin{center}
\begin{minipage}[t]{0.49\linewidth}
\centering
\includegraphics[width=\textwidth]{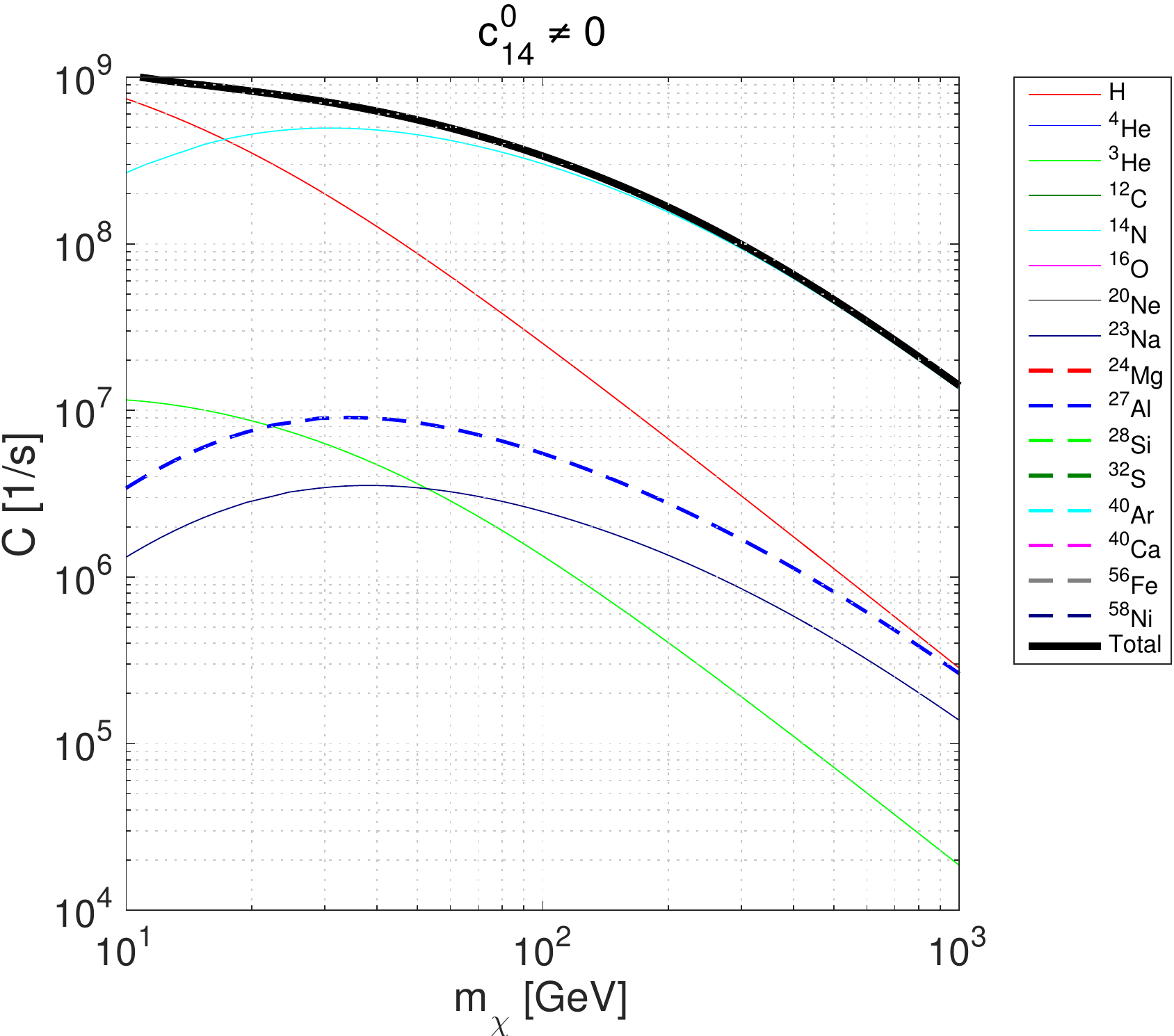}
\end{minipage}
\begin{minipage}[t]{0.49\linewidth}
\centering
\includegraphics[width=\textwidth]{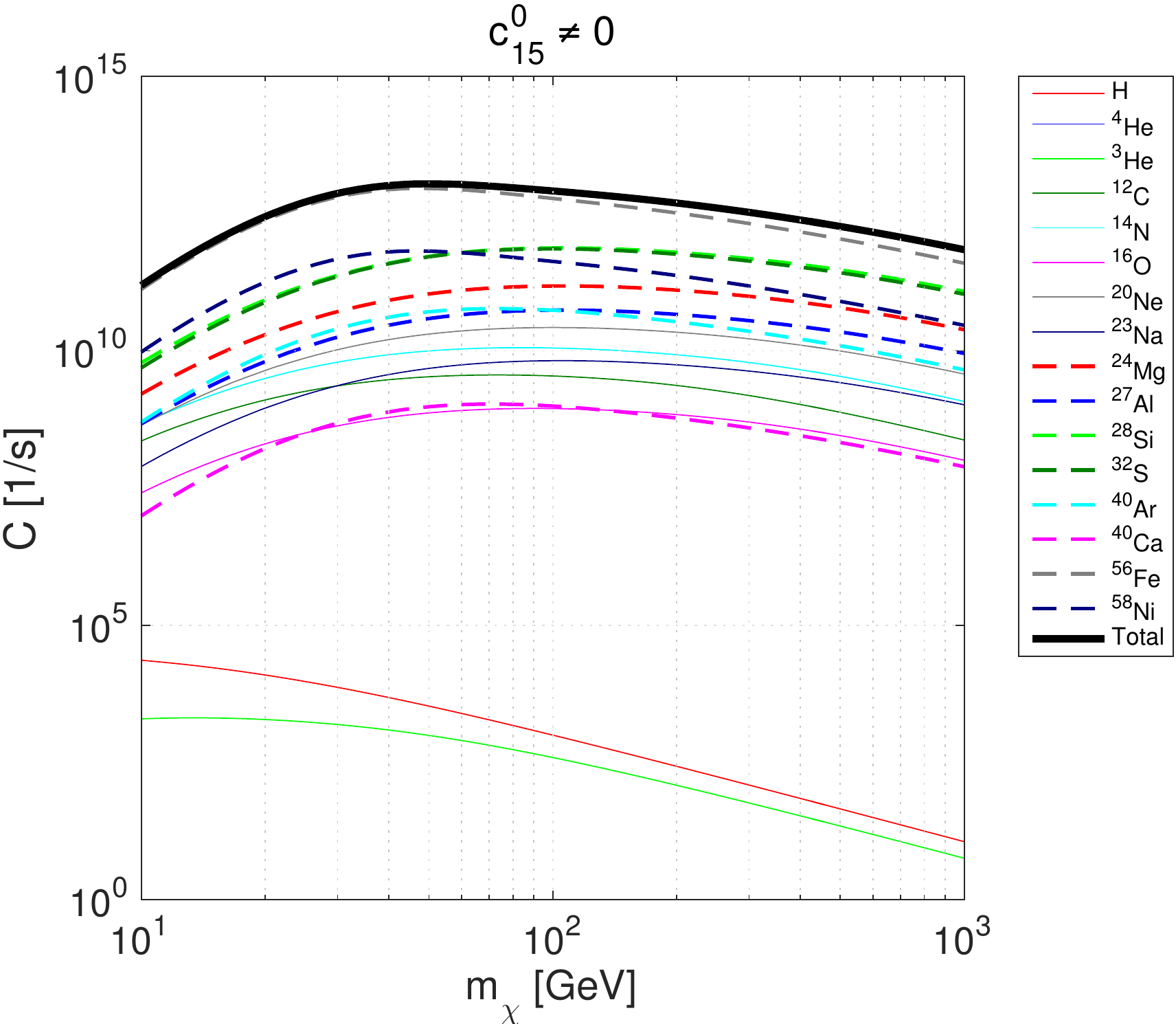}
\end{minipage}
\begin{minipage}[t]{0.49\linewidth}
\centering
\includegraphics[width=\textwidth]{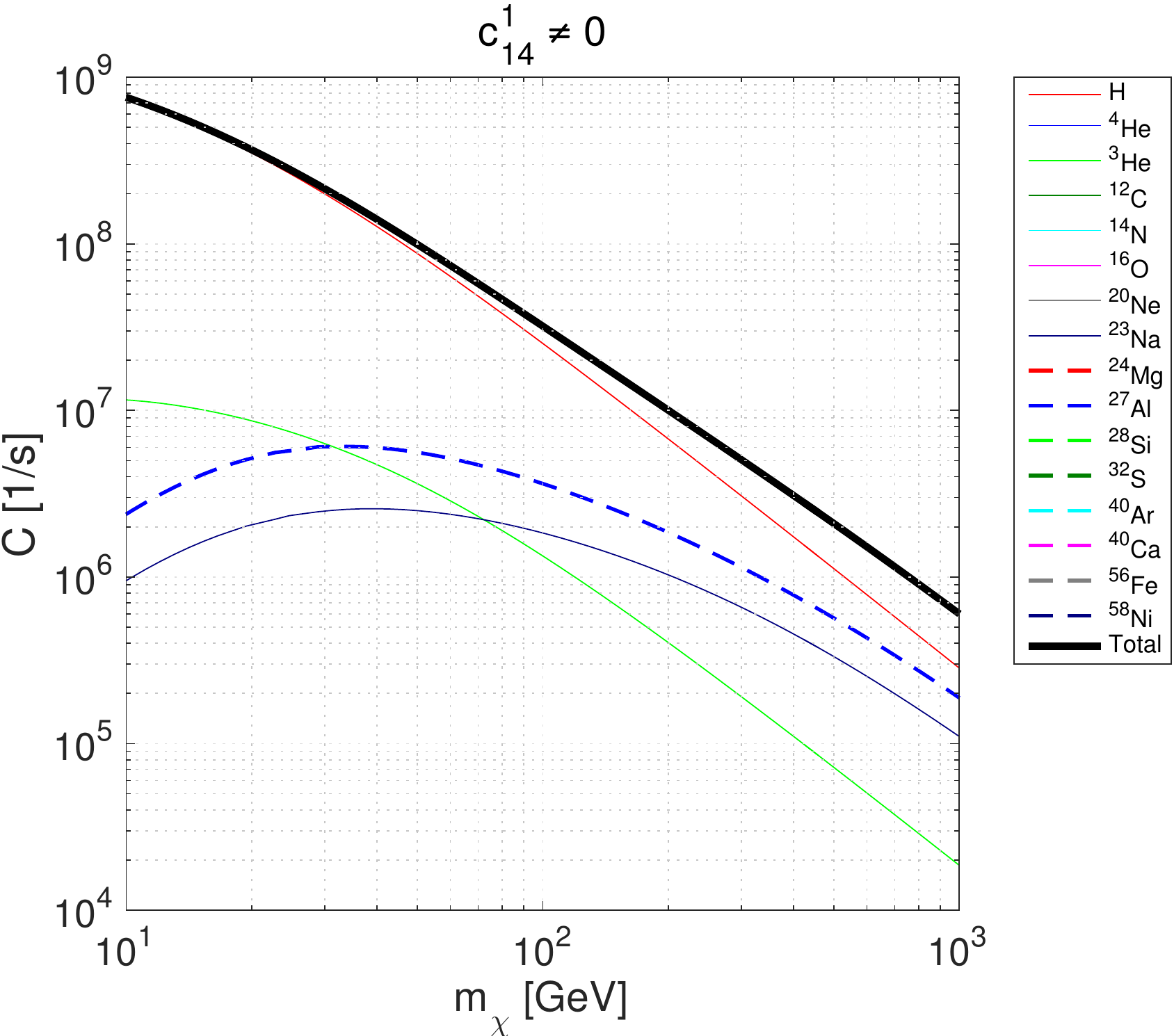}
\end{minipage}
\begin{minipage}[t]{0.49\linewidth}
\centering
\includegraphics[width=\textwidth]{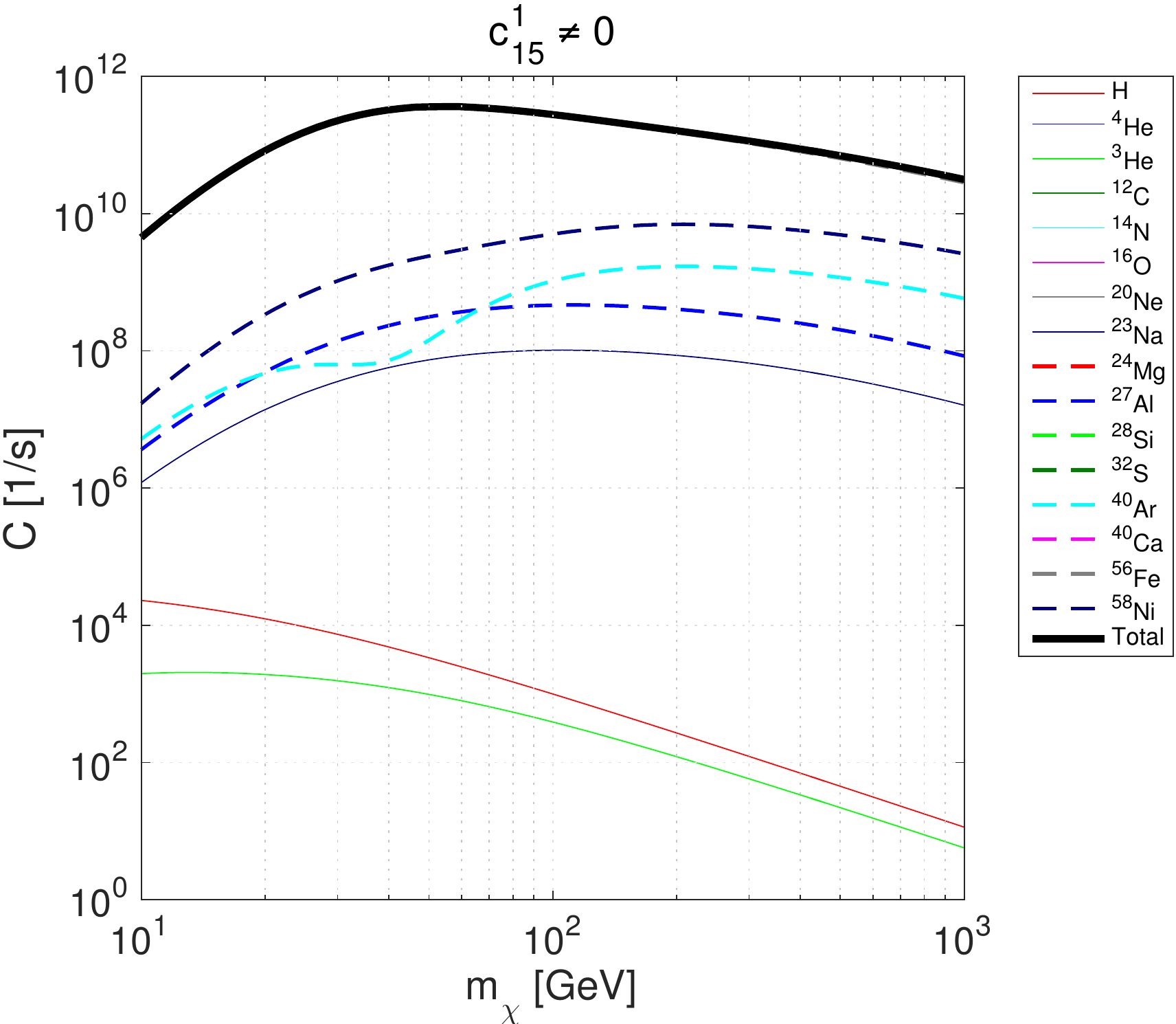}
\end{minipage}
\end{center}
\caption{Same as in Fig.~\ref{fig:c1c4}, but for the interaction operators $\hat{\mathcal{O}}_{14}$ and $\hat{\mathcal{O}}_{15}$.}
\label{fig:c14c15}
\end{figure}

\section{Numerical evaluation of the capture rate}
\label{sec:rate}
In this section we numerically evaluate the dark matter capture rate by the Sun, Eq.~(\ref{eq:rate}), using the nuclear response functions derived in the previous section, and collected in analytic form in Appendix~\ref{sec:appNuc}. We study one operator at the time, and for each interaction operator in Tab.~\ref{tab:operators}, we separately consider the corresponding isoscalar and isovector coupling constants. In the figures, we report the dark matter capture rate as a function of the dark matter particle mass, varying $m_\chi$ in the range 10 - 1000 GeV. When a coupling constant is different from zero, it takes the reference value of $10^{-3}\,m_v^{-2}$, with $m_v=246.2$~GeV. Using the same interaction strength in all panels allows for a straightforward comparison between capture rates associated with different operators. For definiteness, we assume a spin $j_\chi=1/2$ for the dark matter particle.

\subsection{Constant spin-independent and spin-dependent interactions}
We start with the capture rate for the interaction operators $\hat{\mathcal{O}}_1$ and $\hat{\mathcal{O}}_4$, corresponding to constant, i.e. velocity and momentum independent, dark matter-nucleon interactions. 
Fig.~\ref{fig:c1c4} shows the capture rate $C$ as a function of $m_\chi$ for the two operators. The top panels refer to the couplings constants $c_1^0$ and $c_4^0$, whereas the bottom panels correspond to $c_1^1$ and $c_4^1$. In the plots we report the total capture rate (thick black line), and partial capture rates specific to the 16 most abundant elements in the Sun. Conventions for colors and lines are those in the legends.

In the case $c_1^0\neq0$ many elements contribute to $C$ in a comparable manner. The leading contributions come from $^{4}$He, $^{16}$O, and $^{56}$Fe, with an additional sizable contribution due to $^{20}$Ne for $m_\chi\gtrsim 400$ GeV. For $c_4^0\neq0$ the most effective isotopes in capturing dark matter are H and $^{14}$N, though the latter significantly contributes for $m_\chi\gtrsim100$ GeV only. Similarly, in the case $c_1^1\neq0$ the most important element is H, though also $^{56}$Fe gives a sizable contribution to $C$ for large values of $m_\chi$. Finally, for $c_4^1\neq0$ only H is relevant in the dark matter capture by the Sun.

Fig.~\ref{fig:comp} compares the isoscalar rates of Fig.~\ref{fig:c1c4} with the spin-independent and spin-dependent capture rates computed by {\sffamily darksusy}. For constant spin-independent interactions, corresponding to the $\hat{\mathcal{O}}_1$ operator, {\sffamily darksusy} uses a simplified version of Eq.~(\ref{eq:omega}), namely
\begin{equation}
\Omega_{v}^{-}(w)= \sum_i n_i w \,\Theta\left( \frac{\mu_i}{\mu^2_{+,i}} - \frac{u^2}{w^2} \right)\frac{1}{E_k}\int_{E_k u^2/w^2}^{E_k \mu_i/\mu_{+,i}^2} {\rm d}E\,\sigma_i\frac{\mu_{+,i}^2}{\mu_i} \exp(-2y)\,,
\label{eq:omega2}
\end{equation}
where $\sigma_i$ is the total dark matter-nucleus scattering cross-section in the limit of zero momentum transfer, $y=(bq/2)^2$, and 
\begin{equation}
b=\sqrt{\frac{2}{3}} \left[ 0.91 \left(\frac{m_i}{\rm GeV}\right)^{1/3} + 0.3 \right]~{\rm fm} \,,
\end{equation}
which allows to analytically compute $\Omega_{v}^{-}(w)$. In the case of constant spin-dependent interactions, corresponding to the $\hat{\mathcal{O}}_4$ operator, {\sffamily darksusy} calculates the capture rate for H only, and neglects other elements. 
Other interaction operators are not included in the program, and cannot be used for comparison.

The left panel of Fig.~\ref{fig:comp} shows the ratio of the capture rate of this work for $c_1^0\neq0$ to the capture rate computed with {\sffamily darksusy} for spin-independent interactions. We report the ratio of total capture rates, and the ratio of partial rates specific to different elements in the Sun. Whereas for elements like $^{56}$Fe and $^{58}$Ni the two rates differ up to 25\% for $m_\chi\simeq 1$~TeV, the total rate computed with our nuclear response functions and the one obtained from Eq.~(\ref{eq:omega2}) differ by at most 8\%. We conclude that for constant spin-independent dark matter-nucleon interactions, the capture rate is only moderately affected by the use of refined nuclear response functions. 

The capture rate for constant spin-dependent dark matter interactions computed with {\sffamily darksusy} is systematically smaller than the capture rate of this work for $c_4^0\neq0$. This effect is however important for dark matter masses larger than $100$~GeV only. Neglecting elements heavier than H, and in particular $^{14}$N, induces an error on the total capture rate of about 25\% for $m_\chi\simeq 1$~TeV, as shown in the right panel of Fig.~\ref{fig:comp}. 

In summary, the capture rate for the operators $\hat{\mathcal{O}}_1$ and $\hat{\mathcal{O}}_4$ found with the nuclear response functions of this work does not dramatically differ from that of previous studies. However, in the future errors at the 10-20\% level on the capture rate induced by simplistic form factors might non negligibly alter the interpretation of a hypothetical signal in terms of dark matter particle mass and interaction properties.

\subsection{Velocity and momentum dependent interactions}
We now move on to our results for the capture rate of the operators $\hat{\mathcal{O}}_i$, $i=3,5\dots,15$. We report these results in Figs.~\ref{fig:c3c5}, \ref{fig:c6c7}, \ref{fig:c8c9}, \ref{fig:c10c11}, \ref{fig:c12c13}, and \ref{fig:c14c15}, which show total and partial capture rates as a function of the dark matter particle mass. In each panel the thick black line represents the total capture rate, whereas partial rates correspond to colored lines, as explained in the legends. Inspection of these figures shows that the most important element in the determination of $C$ significantly depends on the dark matter-nucleon interaction operator, on whether the coupling is of isoscalar or isovector type, and on the value of $m_\chi$. Elements that contribute the most to the capture rate for at least one interaction operator, and in a specific dark matter particle mass range are H, $^{4}$He, $^{14}$N, $^{16}$O, $^{27}$Al and $^{56}$Fe. The existence of a variegated sample of elements important in the dark matter capture process shows the significance of detailed nuclear structure calculations. This conclusion is in particular true for interaction operators that favor dark matter couplings to nuclei heavier, and more complex than H or $^{4}$He. 

The properties of the 6 nuclear response operators in Eqs.~(\ref{eq:S1S2}) and (\ref{eq:MDP}), and the solar nuclear abundances in Tab.~\ref{tab:massfrac} determine the most important isotopes for a given operator. In the small momentum transfer limit the response operator $M_{LM;\tau}$ measures the mass number $A$ of the nucleus, and is therefore larger for heavy elements, like for instance $^{56}$Fe. Operators coupling via $M_{LM;\tau}$ are $\hat{\mathcal{O}}_1$, $\hat{\mathcal{O}}_{5}$, $\hat{\mathcal{O}}_{8}$, and $\hat{\mathcal{O}}_{11}$, though with different velocity and momentum suppressions. For these operators a compromise between nuclear abundance and mass number determines the most relevant elements in the capture process. The response operators $\Sigma'_{LM;\tau}$ and $\Sigma''_{LM;\tau}$ measure the nucleon spin content of the nucleus, and favor nuclei with unpaired protons or neutrons, like H and $^{14}$N. These isotopes are important for operators like $\hat{\mathcal{O}}_4$, $\hat{\mathcal{O}}_{6}$, $\hat{\mathcal{O}}_7$, $\hat{\mathcal{O}}_{9}$, $\hat{\mathcal{O}}_{10}$, $\hat{\mathcal{O}}_{13}$, and $\hat{\mathcal{O}}_{14}$, that couple via $\Sigma'_{LM;\tau}$ and $\Sigma''_{LM;\tau}$. Similar interpretations exist for the remaining nuclear response operators. For instance, $\Delta_{LM;\tau}$ measures the nucleon angular momentum content of the nucleus, and $\Phi^{\prime \prime}_{LM;\tau}$ the content of nucleon spin-orbit coupling in the nucleus~\cite{Fitzpatrick:2012ib}. It can be shown that $\Delta_{LM;\tau}$ favors nuclei with an unpaired nucleon in a non s-shell orbit, whereas $\Phi^{\prime \prime}_{LM;\tau}$ favors heavy elements with orbits of large angular momentum not fully occupied~\cite{Fitzpatrick:2012ib}. For these reasons  the element $^{56}$Fe is particularly important for interaction operators that generate the nuclear response operator $\Phi^{\prime \prime}_{LM;\tau}$, like $\hat{\mathcal{O}}_3$, $\hat{\mathcal{O}}_{12}$, and $\hat{\mathcal{O}}_{15}$.

For elements up to $^{16}$O, we assume the solar abundances reported in Ref.~\cite{Bahcall:2004pz}. For heavier elements we consider the abundances of Ref.~\cite{Grevesse:1998bj}. These are the abundances implemented in the {\sffamily darksusy} program, which we use to calculate the average mass fractions in Tab.~\ref{tab:massfrac}. Capture rates linearly depend on the radial number densities $n_i$ (see Eq.~(\ref{eq:omega})), which are in turn proportional to the corresponding mass fractions. Assuming a different solar model, i.e. different mass fractions, would impact our results accordingly. Conservative relative uncertainties on the solar abundances are listed in Tab.~4 of~\cite{Serenelli:2012zw}, and range from 11.8\% for $^{56}$Fe to 45.3\% for $^{20}$Ne. Smaller uncertainties are quoted in~\cite{Asplund:2009fu}. Elements not included in Tab.~\ref{tab:massfrac} (heavier or lighter than $^{58}$Ni) have abundances at least a factor of a few smaller than $^{58}$Ni, and are neglected in all present calculations. Whether the corresponding nuclear response functions can compensate for the small abundances of these isotopes, is an interesting question that we leave for future work.
\begin{table}
  \centering
  \begin{tabular}[!h]{|c|c|c|c|}
\hline
    Element& Average mass fraction &  Element& Average mass fraction\\
\hline
\hline
H& 0.684&${}^{24}$Mg&7.30$\times10^{-4}$\\   
\hline
${}^{4}$He& 0.298 &${}^{27}$Al&6.38$\times10^{-5}$\\  
\hline
${}^3$He& 3.75$\times10^{-4}$ &${}^{28}$Si&7.95$\times10^{-4}$\\
\hline
${}^{12}$C& 2.53$\times10^{-3}$&${}^{32}$S&5.48$\times10^{-4}$ \\
\hline
${}^{14}$N& 1.56$\times10^{-3}$&${}^{40}$Ar&8.04$\times10^{-5}$\\ 
\hline
${}^{16}$O&8.50$\times10^{-3}$&${}^{40}$Ca&7.33$\times10^{-5}$\\
\hline
${}^{20}$Ne&1.92$\times10^{-3}$&${}^{56}$Fe&1.42$\times10^{-3}$\\
\hline
${}^{23}$Na&3.94$\times10^{-5}$&${}^{58}$Ni&8.40$\times10^{-5}$\\
\hline
  \end{tabular}
  \caption{List of average mass fractions for the 16 most abundant elements in the Sun as implemented in the {\sffamily darksusy} program~\cite{Gondolo:2004sc}. The underlying solar model is introduced in~\cite{Bahcall:2004pz}.}
\label{tab:massfrac}
\end{table}

Also the behavior of the capture rate as a function of the dark matter particle mass strongly depends on the nature of the dark matter-nucleon interaction. In the log-log planes of Figs.~\ref{fig:c3c5}, \ref{fig:c6c7}, \ref{fig:c8c9}, \ref{fig:c10c11}, \ref{fig:c12c13}, and \ref{fig:c14c15}, we observe steeply decreasing lines, e.g. for $c_1^1\neq 0$, roughly flat lines, e.g. for $c_{11}^0\neq 0$, bumps, e.g. for $c_3^0\neq 0$, and even more complex behaviors, like in the case of $c_6^1\neq 0$. Different factors intervene in determining the exact dependence of the capture rate on $m_\chi$, including what element dominates the capture process, its nuclear structure and the resulting nuclear response functions, and the intrinsic momentum/relative velocity dependence of the operator in analysis. 

\begin{figure}[t]
\begin{center}
\includegraphics[width=0.7\textwidth]{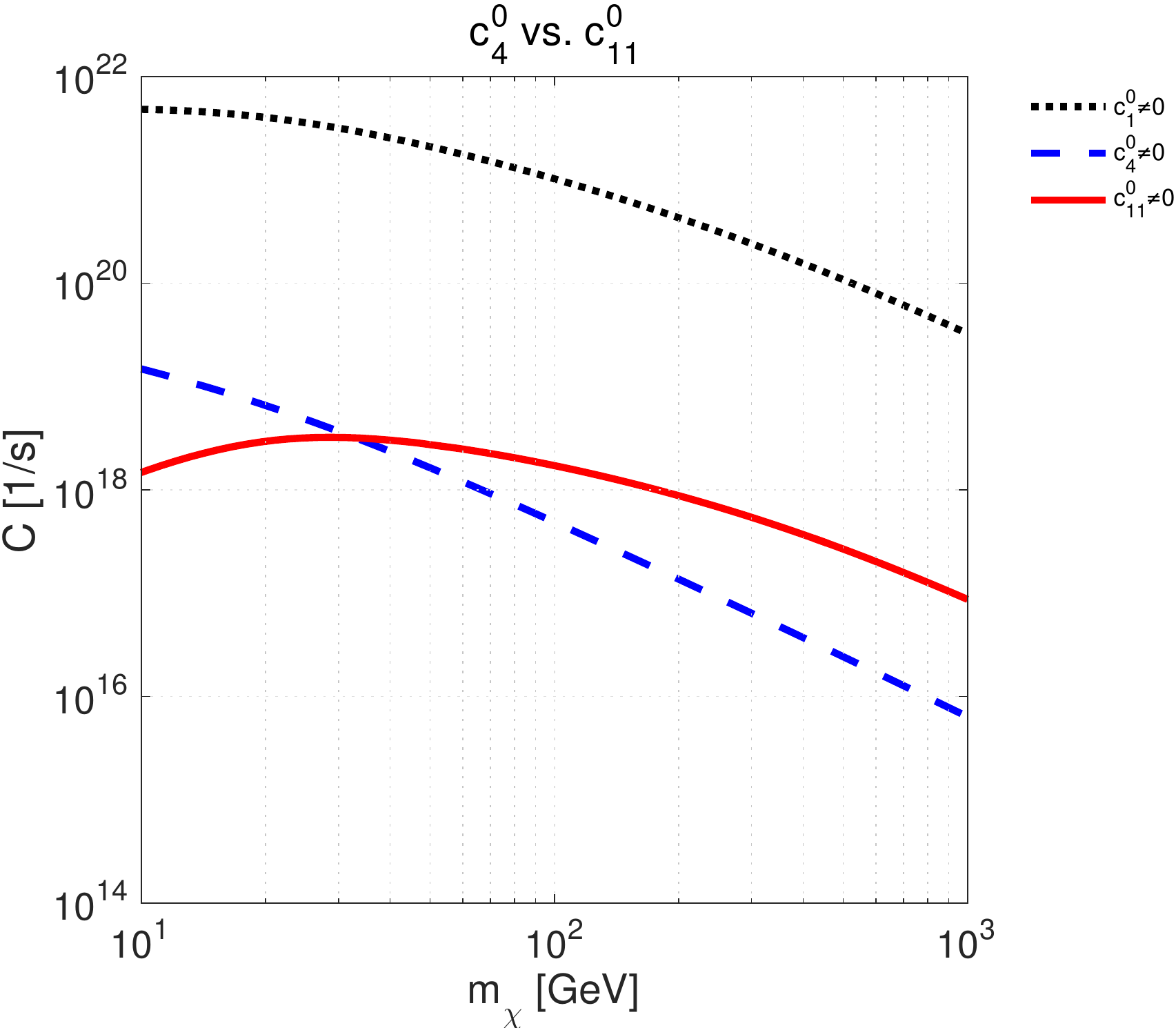}
\end{center}
\caption{Total capture rate for the interaction operators $\hat{\mathcal{O}}_1$, $\hat{\mathcal{O}}_4$, and $\hat{\mathcal{O}}_{11}$. In the three cases we assume the same value for the isoscalar coupling constant, i.e. $10^{-3}\,m_v^{-2}$, with $m_v=246.2$~GeV (we set to zero the isovector coupling constant). The operator $\hat{\mathcal{O}}_{11}$, though never included in experimental analyses, generates a capture rate larger than that associated with the operator $\hat{\mathcal{O}}_4$ for $m_\chi\gtrsim30$~GeV.}
\label{fig:c1c4c11}
\end{figure}

Another important result of this work is to observe that the operators $\hat{\mathcal{O}}_1$ and $\hat{\mathcal{O}}_4$ do not necessarily dominate the dark matter capture process. We find that the operator $\hat{\mathcal{O}}_{11} = i{\bf{\hat{S}}}_\chi\cdot{\hat{\bf{q}}}/m_N$ generates a total dark matter capture rate larger than that associated with the operator $\hat{\mathcal{O}}_4$ for values of the dark matter particle mass larger than approximately 30 GeV. This result is clearly illustrated in Fig.~\ref{fig:c1c4c11}, where we compare the total dark matter capture rate as a function of $m_\chi$ for the operators $\hat{\mathcal{O}}_1$, $\hat{\mathcal{O}}_4$ and $\hat{\mathcal{O}}_{11}$ assuming isoscalar interactions. As in the previous figures, we consider the same value of the coupling constant, i.e. $10^{-3}\,m_v^{-2}$, for the three operators. The relative strength of the three interactions in Fig.~\ref{fig:c1c4c11} is hence determined by the matrix elements of the nuclear response operators $M_{LM;\tau}(q)$, $\Sigma'_{LM;\tau}(q)$ and $\Sigma''_{LM;\tau}(q)$ when evaluated for the most abundant elements in the Sun, and by the intrinsic momentum/relative velocity dependence of the three operators. Notice that the response operator $M_{LM;\tau}(q)$ affects the cross-sections generated by $\hat{\mathcal{O}}_1$ and $\hat{\mathcal{O}}_{11}$, whereas a linear combination of $\Sigma'_{LM;\tau}(q)$ and $\Sigma''_{LM;\tau}(q)$ determines the cross-section associated with $\hat{\mathcal{O}}_{4}$.

\section{Conclusions}
\label{sec:conc}
We have calculated the 8 nuclear response functions generated in the dark matter scattering by nuclei, i.e. Eq.~(\ref{eq:W}), for the 16 most abundant elements in the Sun. We have carried out this calculation within the general effective theory of isoscalar and isovector dark matter-nucleon interactions mediated by a heavy spin-0 or spin-1 particle. This theory predicts 14 isoscalar and 14 isovector dark matter-nucleon interaction operators with a non trivial dependence on velocity and momentum transfer. In contrast, current experimental searches for dark matter focus on 2 {\it constant} spin-independent and spin-dependent interaction operators only.

We have used the nuclear response functions found in this work to calculate the rate of dark matter capture by the Sun for the 14 isoscalar and the 14 isovector dark matter-nucleon interactions separately. We find that different elements contribute to the dark matter capture rate in a significant manner. H, $^{4}$He, $^{14}$N, $^{16}$O, $^{27}$Al and $^{56}$Fe generate the leading contribution for at least one interaction operator, and in a specific dark matter particle mass range. Nuclear structure calculations, like those performed in this work, are hence crucial to accurately compute the rate of dark matter capture by the Sun, in particular for interaction operators that favor dark matter couplings to nuclei heavier, and more complex than H or $^{4}$He. 

Another important result found in this work concerns the operator $\hat{\mathcal{O}}_{11} = i{\bf{\hat{S}}}_\chi\cdot{\hat{\bf{q}}}/m_N$, which couples to the nuclear vector charge operator. For $m_\chi\gtrsim30$~GeV, this operator generates a capture rate larger than the rate induced by the operator $\hat{\mathcal{O}}_{4} = {\bf{\hat{S}}}_\chi\cdot{\bf{\hat{S}}}_N $, i.e. the constant spin-dependent operator commonly considered in dark matter searches at neutrino telescopes. This result was not known previously, and should be kept in mind in the analysis of dark matter induced neutrino signals from the Sun. It is however not unexpected, in that $\hat{\mathcal{O}}_{11}$ is independent of the nucleon spin, i.e. $C\propto A^2$, and of the transverse relative velocity operator.

Our findings significantly extends previous investigations, where the dark matter capture rate was calculated for constant dark matter-nucleon interactions only (see however~\cite{Liang:2013dsa} for an interesting exception), and using simplistic nuclear form factors. The nuclear response functions obtained in this work are listed in analytic form in Appendix~\ref{sec:appNuc}, and can be used in model independent analyses of dark matter induced neutrino signals from the Sun. 

\appendix

\acknowledgments This work has partially been funded through a start-up grant of the University of G\"ottingen. R.C. acknowledges partial support from the European Union FP7 ITN INVISIBLES (Marie Curie Actions, PITN-GA-2011-289442).

\section{Dark matter response functions}
\label{sec:appDM}
Below, we list the dark matter response functions that appear in Eq.~(\ref{eq:sigma}). The notation is the same used in the body of the paper.
\begin{eqnarray}
 R_{M}^{\tau \tau^\prime}\left(v_T^{\perp 2}, {q^2 \over m_N^2}\right) &=& c_1^\tau c_1^{\tau^\prime } + {j_\chi (j_\chi+1) \over 3} \left[ {q^2 \over m_N^2} v_T^{\perp 2} c_5^\tau c_5^{\tau^\prime }+v_T^{\perp 2}c_8^\tau c_8^{\tau^\prime }
+ {q^2 \over m_N^2} c_{11}^\tau c_{11}^{\tau^\prime } \right] \nonumber \\
 R_{\Phi^{\prime \prime}}^{\tau \tau^\prime}\left(v_T^{\perp 2}, {q^2 \over m_N^2}\right) &=& {q^2 \over 4 m_N^2} c_3^\tau c_3^{\tau^\prime } + {j_\chi (j_\chi+1) \over 12} \left( c_{12}^\tau-{q^2 \over m_N^2} c_{15}^\tau\right) \left( c_{12}^{\tau^\prime }-{q^2 \over m_N^2}c_{15}^{\tau^\prime} \right)  \nonumber \\
 R_{\Phi^{\prime \prime} M}^{\tau \tau^\prime}\left(v_T^{\perp 2}, {q^2 \over m_N^2}\right) &=&  c_3^\tau c_1^{\tau^\prime } + {j_\chi (j_\chi+1) \over 3} \left( c_{12}^\tau -{q^2 \over m_N^2} c_{15}^\tau \right) c_{11}^{\tau^\prime } \nonumber \\
  R_{\tilde{\Phi}^\prime}^{\tau \tau^\prime}\left(v_T^{\perp 2}, {q^2 \over m_N^2}\right) &=&{j_\chi (j_\chi+1) \over 12} \left[ c_{12}^\tau c_{12}^{\tau^\prime }+{q^2 \over m_N^2}  c_{13}^\tau c_{13}^{\tau^\prime}  \right] \nonumber \\
   R_{\Sigma^{\prime \prime}}^{\tau \tau^\prime}\left(v_T^{\perp 2}, {q^2 \over m_N^2}\right)  &=&{q^2 \over 4 m_N^2} c_{10}^\tau  c_{10}^{\tau^\prime } +
  {j_\chi (j_\chi+1) \over 12} \left[ c_4^\tau c_4^{\tau^\prime} + \right.  \nonumber \\
 && \left. {q^2 \over m_N^2} ( c_4^\tau c_6^{\tau^\prime }+c_6^\tau c_4^{\tau^\prime })+
 {q^4 \over m_N^4} c_{6}^\tau c_{6}^{\tau^\prime } +v_T^{\perp 2} c_{12}^\tau c_{12}^{\tau^\prime }+{q^2 \over m_N^2} v_T^{\perp 2} c_{13}^\tau c_{13}^{\tau^\prime } \right] \nonumber \\
    R_{\Sigma^\prime}^{\tau \tau^\prime}\left(v_T^{\perp 2}, {q^2 \over m_N^2}\right)  &=&{1 \over 8} \left[ {q^2 \over  m_N^2}  v_T^{\perp 2} c_{3}^\tau  c_{3}^{\tau^\prime } + v_T^{\perp 2}  c_{7}^\tau  c_{7}^{\tau^\prime }  \right]
       + {j_\chi (j_\chi+1) \over 12} \left[ c_4^\tau c_4^{\tau^\prime} +  \right.\nonumber \\
       &&\left. {q^2 \over m_N^2} c_9^\tau c_9^{\tau^\prime }+{v_T^{\perp 2} \over 2} \left(c_{12}^\tau-{q^2 \over m_N^2}c_{15}^\tau \right) \left( c_{12}^{\tau^\prime }-{q^2 \over m_N^2}c_{15}^{\tau \prime} \right) +{q^2 \over 2 m_N^2} v_T^{\perp 2}  c_{14}^\tau c_{14}^{\tau^\prime } \right] \nonumber \\
     R_{\Delta}^{\tau \tau^\prime}\left(v_T^{\perp 2}, {q^2 \over m_N^2}\right)&=&  {j_\chi (j_\chi+1) \over 3} \left[ {q^2 \over m_N^2} c_{5}^\tau c_{5}^{\tau^\prime }+ c_{8}^\tau c_{8}^{\tau^\prime } \right] \nonumber \\
 R_{\Delta \Sigma^\prime}^{\tau \tau^\prime}\left(v_T^{\perp 2}, {q^2 \over m_N^2}\right)&=& {j_\chi (j_\chi+1) \over 3} \left[c_{5}^\tau c_{4}^{\tau^\prime }-c_8^\tau c_9^{\tau^\prime} \right].
\end{eqnarray}

\section{Single-particle matrix elements of nuclear response operators}
\label{sec:appME}
Here we list the single-particle matrix elements of the nuclear response operators in Eqs.~(\ref{eq:S1S2}) and (\ref{eq:MDP}). Eqs.~(\ref{eq:m1}), (\ref{eq:m2}), (\ref{eq:m3}), and (\ref{eq:m4}) are implemented in the {\sffamily Mathematica} package of Ref.~\cite{Anand:2013yka}. 

Only 4 independent nuclear response operators are actually generated in the dark matter-nucleus scattering, and need to be considered in order to evaluate the dark matter-nucleus scattering cross-section. These are $M_{JM}(q{\bf{r}}_i)$, ${\bf{M}}_{JL}^M(q{\bf{r}}_i)\cdot{\vec{\sigma}(i)}$, ${\bf{M}}^M_{JL}(q{\bf{r}}_i)\cdot\frac{1}{q} \overrightarrow{\nabla}$, and ${\bf{M}}^M_{JJ+1}(q{\bf{r}}_i)\cdot\left({\vec{\sigma}(i)}\times\frac{1}{q} \overrightarrow{\nabla}\right)$. This result follows from the identities
\begin{eqnarray}
\Sigma'_{JM;\tau}(q) &=& \sum_{i=1}^{A} \left[-\sqrt{\frac{J}{2J+1}}{\bf{M}}_{JJ+1}^{M}(q {\bf{r}}_i)  + \sqrt{\frac{J+1}{2J+1}}{\bf{M}}_{JJ-1}^{M}(q {\bf{r}}_i) \right] \cdot \vec{\sigma}(i) t^{\tau}_{(i)}\nonumber\\
\Sigma''_{JM;\tau}(q) &=&\sum_{i=1}^{A} \left[\sqrt{\frac{J+1}{2J+1}} {\bf{M}}_{JJ+1}^{M}(q {\bf{r}}_i) + \sqrt{\frac{J}{2J+1}}{\bf{M}}_{JJ-1}^{M}(q {\bf{r}}_i) \right] \cdot \vec{\sigma}(i) t^{\tau}_{(i)} \,.
\label{eq:S1S2bis}
\end{eqnarray}
The reduced single-particle matrix elements of the four independent nuclear response operators are given in the following, where to simplify the equations we use the notation $\langle \boldsymbol{\alpha}||\cdot|| \boldsymbol{\beta} \rangle \equiv \langle n_\alpha(l_\alpha1/2)j_\alpha||\cdot|| n_\beta(l_\beta1/2)j_\beta\rangle$, and $[\lambda]=\sqrt{2\lambda+1}$, for any index $\lambda$. They read as follows
\begin{eqnarray}
\label{eq:m1}
  \langle \boldsymbol{\alpha}||M_{J}(q{\bf{r}}_i)||\boldsymbol{\beta} \rangle &=&
\frac{1}{\sqrt{4\pi}}(-1)^{J+j_\beta+1/2}[l_\alpha][l_\beta][j_\alpha][j_\beta][J] \nonumber\\
&\times& \begin{Bmatrix}
    l_\alpha&j_\alpha&\frac{1}{2}\\
    j_\beta&l_\beta&J
\end{Bmatrix}
\begin{pmatrix}
    l_\alpha&J&l_\beta\\
    0&0&0
\end{pmatrix}
\Braket{n_\alpha l_\alpha j_\alpha |j_{J}(\rho)|n_\beta l_\beta j_\beta}\,\nonumber\\
%&&\\
\end{eqnarray}
\begin{eqnarray}
\label{eq:m2}
  \langle \boldsymbol{\alpha}||{\bf{M}}_{JL}(q{\bf{r}}_i)\cdot{\vec{\sigma}(i)}||\boldsymbol{\beta}\rangle &=& 
  \sqrt{\frac{3}{2\pi}}(-1)^{l_\alpha}[l_\alpha][l_\beta][j_\alpha][j_\beta][L][J] \nonumber\\
  &\times&
  \begin{Bmatrix}
    l_\alpha&l_\beta&L\\
    \frac{1}{2}&\frac{1}{2}&1\\
    j_\alpha&j_\beta&J
  \end{Bmatrix}
  \begin{pmatrix}
    l_\alpha&L&l_\beta\\
    0&0&0
  \end{pmatrix}\Braket{n_\alpha l_\alpha j_\alpha|j_{L}(\rho)|n_\beta l_\beta j_\beta} \nonumber \\
  %&&\\
\end{eqnarray}
\begin{eqnarray}
\label{eq:m3}
  \langle \boldsymbol{\alpha}||{\bf{M}}_{JL}(q{\bf{r}}_i)\cdot\frac{1}{q} \overrightarrow{\nabla}||\boldsymbol{\beta}\rangle &=& 
  \frac{1}{\sqrt{4\pi}} (-1)^{L+j_\beta+1/2}[l_\alpha][j_\alpha][j_\beta][L][J] 
  \begin{Bmatrix}
    l_\alpha&j_\alpha&\frac{1}{2}\\
    j_\beta&l_\beta&J
  \end{Bmatrix}\nonumber\\
&\times&\Bigg[-\sqrt{l_\beta+1}[l_\beta+1]
\begin{Bmatrix}
    L&1&J\\
    l_\beta&l_\alpha&l_\beta+1
  \end{Bmatrix}
  \begin{pmatrix}
    l_\alpha&L&l_\beta+1\\
    0&0&0
  \end{pmatrix} \nonumber\\
 &\times& \langle n_\alpha l_\alpha j_\alpha |j_{L}(\rho)\left(\frac{\text{d}}{\text{d}\rho}-\frac{l_\beta}{\rho}\right)| n_\beta l_\beta j_\beta\rangle + \nonumber\\
 &+&\sqrt{l_\beta}[l_\beta-1]
\begin{Bmatrix}
    L&1&J\\
    l_\beta&l_\alpha&l_\beta-1
  \end{Bmatrix}
  \begin{pmatrix}
    l_\alpha&L&l_\beta-1\\
    0&0&0
  \end{pmatrix}\nonumber\\
  &\times&\langle n_\alpha l_\alpha j_\alpha |j_{L}(\rho)\left(\frac{\text{d}}{\text{d}\rho}+\frac{l_\beta+1}{\rho}\right)|n_\beta l_\beta j_\beta \rangle \Bigg] \nonumber\\
\end{eqnarray}
\begin{eqnarray}
\label{eq:m4}
\langle \boldsymbol{\alpha}||{\bf{M}}_{JL}(q{\bf{r}}_i)&\cdot&\left({\vec{\sigma}(i)}\times\frac{1}{q} \overrightarrow{\nabla}\right)||\boldsymbol{\beta}\rangle = \nonumber\\ 
&=& (-1)^{l_\alpha}\frac{6i}{\sqrt{4\pi}}[l_\alpha][j_\alpha][j_\beta][J][L] \nonumber\\
&\times& \Bigg\{-[l_\beta+1]\sqrt{l_\beta+1}
\begin{pmatrix}
    l_\alpha&L&l_\beta+1\\
    0&0&0
  \end{pmatrix}
 \langle n_\alpha l_\alpha j_\alpha |j_{L}(\rho)\left(\frac{\text{d}}{\text{d}\rho}-\frac{l_\beta}{\rho}\right)|n_\beta l_\beta j_\beta \rangle  \nonumber\\
&\times&\Bigg[ \sum_{L'} [L']^{2} (-1)^{J-L'}
  \begin{Bmatrix}
    L&1&L'\\
    1&J&1
  \end{Bmatrix}\begin{Bmatrix}
    L&1&L'\\
    l_\beta&l_\alpha&l_\beta+1
  \end{Bmatrix}\begin{Bmatrix}
    l_\alpha&l_\beta&L'\\
    \frac{1}{2}&\frac{1}{2}&1\\
    j_\alpha&j_\beta&J
  \end{Bmatrix} \Bigg] \nonumber\\
&+& [l_\beta-1]\sqrt{l_\beta}
\begin{pmatrix}
    l_\alpha&L&l_\beta-1\\
    0&0&0
  \end{pmatrix}
  \langle n_\alpha l_\alpha j_\alpha|j_{L}(\rho)\left(\frac{\text{d}}{\text{d}\rho}+\frac{l_\beta+1}{\rho}\right)|n_\beta l_\beta j_\beta\rangle \nonumber \\
&\times& \Bigg[  \sum_{L'} [L']^{2} (-1)^{J-L'}
  \begin{Bmatrix}
    L&1&L'\\
    1&J&1
  \end{Bmatrix}\begin{Bmatrix}
    L&1&L'\\
    l_\beta&l_\alpha&l_\beta-1
  \end{Bmatrix}\begin{Bmatrix}
    l_\alpha&l_\beta&L'\\
    \frac{1}{2}&\frac{1}{2}&1\\
    j_\alpha&j_\beta&J
  \end{Bmatrix} 
\Bigg]\Bigg\} \,. \nonumber\\
\end{eqnarray}
Eqs.~(\ref{eq:m1}), (\ref{eq:m2}), and (\ref{eq:m3}) also appear in the calculation of nuclear matrix elements for electroweak lepton-nucleus interactions.
The latter expression is instead needed to evaluate the matrix elements of the nuclear response operators $\tilde{\Phi}'$ and $\Phi''$, specific to the dark matter-nucleus scattering. 
Different combinations of Wigner $3j$, $6j$ and $9j$ symbols appear in the equations above, which also depend on residual radial matrix elements of spherical Bessel functions and of their derivatives at $\rho=qr_i$. 
In the case of the harmonic oscillator single-particle basis, these radial matrix elements can be analytically evaluated as follows 
\begin{align*}
  &\langle n_\alpha l_\alpha j_\alpha | j_{L}(\rho)| n_\beta l_\beta j_\beta \rangle =\\ &\qquad=\frac{2^{L}}{(2L+1)!!}y^{L/2}e^{-y}\sqrt{(n_\beta-1)!(n_\alpha-1)!\Gamma(n_\alpha+l_\alpha+\frac{1}{2})\Gamma(n_\beta+l_\beta+\frac{1}{2})}\\
&\qquad\times\sum_{k=0}^{n_\beta-1}\sum_{k'=0}^{n_\alpha-1}\frac{(-1)^{k+k'}}{k!k'!}\frac{1}{(n_\beta-1-k)!(n_\alpha-1-k')!}\nonumber\\
&\qquad\times\frac{\Gamma[\frac{1}{2}(l_\beta+l_\alpha+L+2k+2k'+3)]}{\Gamma[l_\beta+k+\frac{3}{2}]\Gamma[l_\alpha+k'+\frac{3}{2}]}
\leftidx{_{1}}F_{1}[\frac{1}{2}(L-l_\beta-l_\alpha-2k-2k');L+\frac{3}{2};y],
\end{align*}
\begin{align*}
&\langle n_\alpha l_\alpha j_\alpha |j_{L}(\rho)\left(\frac{\text{d}}{\text{d}r}-\frac{l_\beta}{r_i}\right)|n_\beta l_\beta j_\beta \rangle =\\ &\qquad=\frac{2^{L-1}}{(2L+1)!!}y^{(L-1)/2}e^{-y}\sqrt{(n_\beta-1)!(n_\alpha-1)!\Gamma(n_\alpha+l_\alpha+\frac{1}{2})\Gamma(n_\beta+l_\beta+\frac{1}{2})}\\
&\qquad\times\sum_{k=0}^{n_\beta-1}\sum_{k'=0}^{n_\alpha-1}\frac{(-1)^{k+k'}}{k!k'!}\frac{1}{(n_\beta-1-k)!(n_\alpha-1-k')!} \frac{\Gamma[\frac{1}{2}(l_\beta+l_\alpha+L+2k+2k'+2)]}{\Gamma[l_\beta+k+\frac{3}{2}]\Gamma[l_\alpha+k'+\frac{3}{2}]}\\
&\qquad\times\left\{-\frac{1}{2}(l_\beta+l_\alpha+L+2k+2k'+2)\leftidx{_{1}}F_{1}[\frac{1}{2}(L-l_\beta-l_\alpha-2k-2k'-1);L+\frac{3}{2};y]\right.\\
&\left.\qquad +(2k)\leftidx{_{1}}F_{1}[\frac{1}{2}(L-l_\beta-l_\alpha-2k-2k'+1);L+\frac{3}{2};y]\right\},
\end{align*}
\begin{align*}
&\langle n_\alpha l_\alpha j_\alpha |j_{L}(\rho)\left(\frac{\text{d}}{\text{d}r_i}+\frac{l_\beta+1}{r_i}\right)| n_\beta l_\beta j_\beta \rangle =\\
&\qquad = \frac{2^{L-1}}{(2L+1)!!}y^{(L-1)/2}e^{-y}\sqrt{(n_\beta-1)!(n_\alpha-1)!\Gamma(n_\alpha+l_\alpha+\frac{1}{2})\Gamma(n_\beta+l_\beta+\frac{1}{2})}\\
&\qquad\times\sum_{k=0}^{n_\beta-1}\sum_{k'=0}^{n_\alpha-1}\frac{(-1)^{k+k'}}{k!k'!}\frac{1}{(n_\beta-1-k)!(n_\alpha-1-k')!}\frac{\Gamma[\frac{1}{2}(l_\beta+l_\alpha+L+2k+2k'+2)]}{\Gamma[l_\beta+k+\frac{3}{2}]\Gamma[l_\alpha+k'+\frac{3}{2}]}\\
&\qquad\times\left\{-\frac{1}{2}(l_\beta+l_\alpha+L+2k+2k'+2)\leftidx{_{1}}F_{1}[\frac{1}{2}(L-l_\beta-l_\alpha-2k-2k'-1);L+\frac{3}{2};y]\right.\\
&\left.\qquad +(2l_\beta+2k+1)\leftidx{_{1}}F_{1}[\frac{1}{2}(L-l_\beta-l_\alpha-2k-2k'+1);L+\frac{3}{2};y]\right\},
\end{align*}
where $y=(qb/2)^2$, and $\leftidx{_{1}}F_{1}$ is the confluent hypergeometric function.

\section{Nuclear response functions}
\label{sec:appNuc}
Below, we only list nuclear response functions different from zero. 

\subsection*{Hydrogen (H)}
\begin{flalign}
W^{00}_{M}(y)&= 0.0397887 & W^{00}_{\Sigma^{\prime\prime}}(y)&= 0.0397887 & W^{00}_{\Sigma^\prime}(y)&= 0.0795775 &\nonumber\\
W^{11}_{M}(y)&= 0.0397887 & W^{11}_{\Sigma^{\prime\prime}}(y)&= 0.0397887 & W^{11}_{\Sigma^\prime}(y)&= 0.0795775 &\nonumber\\ 
W^{10}_{M}(y)&= 0.0397887 & W^{10}_{\Sigma^{\prime\prime}}(y)&= 0.0397887 & W^{10}_{\Sigma^\prime}(y)&= 0.0795775 &\nonumber\\ 
W^{01}_{M}(y)&= 0.0397887 & W^{01}_{\Sigma^{\prime\prime}}(y)&= 0.0397887 & W^{01}_{\Sigma^\prime}(y)&= 0.0795775 &\nonumber\\
\end{flalign}

\subsection*{Helium ($^3$He)}
\begin{flalign}
W^{00}_{M}(y)&= 0.358099 e^{-2y}& W^{00}_{\Sigma^{\prime\prime}}(y)&= 0.0397887  e^{-2y} &W^{00}_{\Sigma^\prime}(y)&= 0.0795775  e^{-2y} &\nonumber\\
W^{11}_{M}(y)&= 0.0397887 e^{-2y}& W^{11}_{\Sigma^{\prime\prime}}(y)&= 0.0397887 e^{-2y} &W^{11}_{\Sigma^\prime}(y)&= 0.0795775 e^{-2y} &  \nonumber\\ 
W^{10}_{M}(y)&= 0.119366 e^{-2y} &W^{10}_{\Sigma^{\prime\prime}}(y)&= -0.0397887 e^{-2y} &W^{10}_{\Sigma^\prime}(y)&= -0.0795775 e^{-2y} &  \nonumber\\ 
W^{01}_{M}(y)&= 0.119366 e^{-2y}& W^{01}_{\Sigma^{\prime\prime}}(y)&= -0.0397887e^{-2y}& W^{01}_{\Sigma^\prime}(y)&= -0.0795775 e^{-2y} &\nonumber\\
\end{flalign}

\subsection*{Helium ($^4$He)}
\begin{flalign}
W^{00}_{M}(y)&= 0.31831 e^{-2y}& \nonumber\\
\end{flalign}

\subsection*{Carbon ($^{12}$C)}
\begin{flalign}
W^{00}_{M}(y)&= 0.565882 e^{-2y} (2.25 - y)^2& \nonumber\\ 
W^{00}_{\Phi^{\prime\prime}}(y)&= 0.0480805 e^{-2y} & \nonumber\\
W^{00}_{M\Phi^{\prime\prime}}(y)&= e^{-2y} (-0.371134 + 0.164948 y) & \nonumber\\
\end{flalign}

\subsection*{Nitrogen ($^{14}$N)}
\begin{flalign}
W^{00}_{M}(y) &= e^{-2y} (11.6979 - 11.1409 y + 2.67574 y^2) &\nonumber \\
W^{00}_{\Sigma^{\prime\prime}}(y) &= 0.0230079 e^{-2y} (1.20986 + y)^2 &\nonumber \\
W^{00}_{\Sigma^\prime}(y) &= 0.134532 e^{-2y} (0.707578 - y)^2 \nonumber \\
W^{00}_{\Phi^{\prime\prime}}(y)&= 0.0905048 e^{-2y} &\nonumber \\
W^{00}_{\tilde{\Phi}^\prime}(y)&= 0.00126432 e^{-2y} &\nonumber \\
W^{00}_{\Delta}(y)&= 0.0424075 e^{-2y} &\nonumber \\
W^{00}_{M\Phi^{\prime\prime}}(y)&= e^{-2y} (-1.02414 + 0.483267 y)&\nonumber \\
W^{00}_{\Sigma^{\prime}\Delta}(y)&= e^{-2y} (0.0534451 - 0.0755325 y) &\nonumber\\
\end{flalign}

\subsection*{Oxygen ($^{16}$O)}
\begin{flalign}
W^{00}_{M}(y)&=  0.000032628 e^{-2y} (395.084 - 200.042 y + y^2)^2 &\nonumber\\ 
W^{00}_{\Phi^{\prime\prime}}(y)&= 0.000032628 e^{-2y} (3.66055 - y)^2 &\nonumber\\
W^{00}_{M\Phi^{\prime\prime}}(y)&= e^{-2y} (-0.0471874 + 0.0367831 y - 0.00664641 y^2 + 0.000032628 y^3) &\nonumber\\
\end{flalign}

\subsection*{Neon ($^{20}$Ne)}
\begin{flalign}
W^{00}_{M}(y)&= 0.0431723 e^{-2y} (13.5766 - 9.05108 y + y^2)^2  &\nonumber\\ 
W^{00}_{\Phi^{\prime\prime}}(y)&= 0.00348077 e^{-2y} (2.50001 - y)^2 &\nonumber\\
W^{00}_{M\Phi^{\prime\prime}}(y)&= e^{-2y} (-0.416077 + 0.443815 y - 0.1416 y^2 + 0.0122586 y^3)&\nonumber\\
\end{flalign}

\subsection*{Magnesium ($^{24}$Mg)}
\begin{flalign}
W^{00}_{M}(y)&= 0.123467 e^{-2y} (9.63385 - 7.49299 y + y^2)^2 &\nonumber\\ 
W^{00}_{\Phi^{\prime\prime}}(y)&= 0.0260816 e^{-2y} (2.5 - y)^2 &\nonumber\\
W^{00}_{M\Phi^{\prime\prime}}(y)&= e^{-2y} (-1.36673 + 1.6097 y - 0.567072 y^2 + 0.056747 y^3)&\nonumber\\
\end{flalign}

\subsection*{Sodium ($^{23}$Na)}
\begin{flalign}
W^{00}_{M}(y)&= e^{-2y} (42.0965 - 63.4498 y + 32.5913 y^2 - 6.57878 y^3 + 
   0.483166 y^4) &\nonumber\\ 
W^{11}_{M}(y)&=e^{-2y} (0.0795776 - 0.212207 y + 0.182941 y^2 - 0.0543892 y^3 + 
   0.00523012 y^4)  &\nonumber\\
W^{10}_{M}(y)&= e^{-2y} (-1.83028 + 3.81972 y - 2.50445 y^2 + 0.597822 y^3 - 
   0.04545 y^4) &\nonumber\\
W^{01}_{M}(y)&= e^{-2y} (-1.83028 + 3.81972 y - 2.50445 y^2 + 0.597822 y^3 - 
   0.04545 y^4)  &\nonumber\\
W^{00}_{\Sigma^{\prime\prime}}(y)&= e^{-2y} (0.0126672 - 0.0262533 y + 0.0401886 y^2 - 0.010514 y^3 + 
   0.00078605 y^4)&\nonumber\\
W^{11}_{\Sigma^{\prime\prime}}(y)&= e^{-2y} (0.00917577 - 0.0167053 y + 0.0332751 y^2 - 0.00765719 y^3 + 
   0.000597676 y^4)&\nonumber\\
W^{10}_{\Sigma^{\prime\prime}}(y)&=e^{-2y} (0.0107811 - 0.020986 y + 0.0360971 y^2 - 0.00876213 y^3 + 
   0.000626718 y^4) &\nonumber\\
W^{01}_{\Sigma^{\prime\prime}}(y)&=e^{-2y} (0.0107811 - 0.020986 y + 0.0360971 y^2 - 0.00876213 y^3 + 
   0.000626718 y^4) &\nonumber\\
W^{00}_{\Sigma^\prime}(y)&= e^{-2y} (0.0253345 - 0.0750847 y + 0.100235 y^2 - 0.0384261 y^3 + 
   0.00466396 y^4)&\nonumber\\
W^{11}_{\Sigma^\prime}(y)&=e^{-2y} (0.0183515 - 0.0567009 y + 0.0887794 y^2 - 0.0374699 y^3 + 
   0.00477955 y^4) &\nonumber\\
W^{10}_{\Sigma^\prime}(y)&= e^{-2y} (0.0215622 - 0.0652627 y + 0.0941439 y^2 - 0.0379511 y^3 + 
   0.00472138 y^4)&\nonumber\\
W^{01}_{\Sigma^\prime}(y)&=e^{-2y} (0.0215622 - 0.0652627 y + 0.0941439 y^2 - 0.0379511 y^3 + 
   0.00472138 y^4) &\nonumber\\
W^{00}_{\Phi^{\prime\prime}}(y)&= e^{-2y} (0.612149 - 0.49308 y + 0.107832 y^2)&\nonumber\\
W^{11}_{\Phi^{\prime\prime}}(y)&= e^{-2y} (0.00940911 - 0.00747826 y + 0.00163204 y^2)&\nonumber\\
W^{10}_{\Phi^{\prime\prime}}(y)&= e^{-2y} (-0.075893 + 0.060682 y - 0.0110124 y^2)&\nonumber\\
W^{01}_{\Phi^{\prime\prime}}(y)&= e^{-2y} (-0.075893 + 0.060682 y - 0.0110124 y^2)&\nonumber\\
W^{00}_{\tilde{\Phi}^\prime}(y)&= e^{-2y} (0.000495589 - 0.00010394 y + 0.00000544981y^2 )&\nonumber\\%+ 
   %1.93866*10^-20 y^3 + 1.7241*10^-35 y^4)&\nonumber\\
W^{11}_{\tilde{\Phi}^\prime}(y)&= e^{-2y} (0.00000616583 + 0.00008381 y + 0.0002848 y^2 )&\nonumber\\%+ 
  % 1.65458*10^-20 y^3 + 2.40313*10^-37 y^4)&\nonumber\\
W^{10}_{\tilde{\Phi}^\prime}(y)&= e^{-2y} (-0.0000552785 - 0.000369894 y + 0.0000393968 y^2 )&\nonumber\\%+ 
   %7.12174*10^-20 y^3 + 2.03549*10^-36 y^4)&\nonumber\\
W^{01}_{\tilde{\Phi}^\prime}(y)&= e^{-2y} (-0.0000552785 - 0.000369894 y + 0.0000393968 y^2 )&\nonumber\\%+ 
   %7.12174*10^-20 y^3 + 2.03549*10^-36 y^4)&\nonumber\\
W^{00}_{\Delta}(y)&=e^{-2y} (0.0335711 - 0.0268568 y + 0.00656896 y^2) &\nonumber\\
W^{11}_{\Delta}(y)&=e^{-2y} (0.00772326 - 0.00617861 y + 0.0021619 y^2) &\nonumber\\
W^{10}_{\Delta}(y)&=e^{-2y} (0.0161021 - 0.0128817 y + 0.00362952 y^2) &\nonumber\\
W^{01}_{\Delta}(y)&=e^{-2y} (0.0161021 - 0.0128817 y + 0.00362952 y^2) &\nonumber\\
W^{00}_{M\Phi^{\prime\prime}}(y)&= e^{-2y} (-5.07498 + 5.86765 y - 2.09908 y^2 + 0.226345 y^3)&\nonumber\\ 
W^{11}_{M\Phi^{\prime\prime}}(y)&= e^{-2y} (-0.0273574 + 0.0474719 y - 0.0213121 y^2 + 0.00280825 y^3)&\nonumber\\
W^{10}_{M\Phi^{\prime\prime}}(y)&=e^{-2y} (0.220651 - 0.382932 y + 0.17682 y^2 - 0.0226015 y^3) &\nonumber\\
W^{01}_{M\Phi^{\prime\prime}}(y)&=e^{-2y} (0.62922 - 0.727336 y + 0.243236 y^2 - 0.0210943 y^3) &\nonumber\\
W^{00}_{\Sigma^{\prime}\Delta}(y)&= e^{-2y} (-0.0291634 + 0.0548817 y - 0.0305345 y^2 + 0.00476387 y^3)&\nonumber\\
W^{11}_{\Sigma^{\prime}\Delta}(y)&= e^{-2y} (-0.0119052 + 0.0231539 y - 0.0164035 y^2 + 0.00310235 y^3) &\nonumber\\
W^{10}_{\Sigma^{\prime}\Delta}(y)&= e^{-2y} (-0.024821 + 0.0482732 y - 0.02884 y^2 + 0.00481368 y^3)&\nonumber\\
W^{01}_{\Sigma^{\prime}\Delta}(y)&= e^{-2y} (-0.013988 + 0.0263236 y - 0.0171362 y^2 + 0.00306717 y^3)&\nonumber\\
\end{flalign}

\subsection*{Aluminium ($^{27}$Al)}
\begin{flalign}
W^{00}_{M}(y)&= e^{-2y} (87.0146 - 146.097 y + 83.5367 y^2 - 18.5981 y^3 + 
   1.43446 y^4) &\nonumber\\ 
W^{11}_{M}(y)&= e^{-2y} (0.119366 - 0.31831 y + 0.337291 y^2 - 0.132526 y^3 + 
   0.018155 y^4) &\nonumber\\
W^{10}_{M}(y)&=  e^{-2y} (-3.22283 + 7.00266 y - 4.92756 y^2 + 1.33587 y^3 - 
   0.11524 y^4) &\nonumber\\
W^{01}_{M}(y)&= e^{-2y} (-3.22283 + 7.00266 y - 4.92756 y^2 + 1.33587 y^3 - 
   0.11524 y^4) &\nonumber\\
W^{00}_{\Sigma^{\prime\prime}}(y)&= e^{-2y} (0.0309465 - 0.0367242 y + 0.0265347 y^2 - 0.00241606 y^3 + 
   0.0110011 y^4)&\nonumber\\
W^{11}_{\Sigma^{\prime\prime}}(y)&= e^{-2y} (0.0218834 - 0.00944476 y + 0.011506 y^2 + 0.000953537 y^3 + 
   0.0104813 y^4)&\nonumber\\
W^{10}_{\Sigma^{\prime\prime}}(y)&= e^{-2y} (0.0260233 - 0.0210567 y + 0.0158643 y^2 + 0.000606077 y^3 + 
   0.0105713 y^4)&\nonumber\\
W^{01}_{\Sigma^{\prime\prime}}(y)&= e^{-2y} (0.0260233 - 0.0210567 y + 0.0158643 y^2 + 0.000606077 y^3 + 
   0.0105713 y^4)&\nonumber\\
W^{00}_{\Sigma^\prime}(y)&=e^{-2y} (0.0618929 - 0.210848 y + 0.244466 y^2 - 0.0942682 y^3 + 
   0.0243737 y^4) &\nonumber\\
W^{11}_{\Sigma^\prime}(y)&=e^{-2y} (0.0437667 - 0.165622 y + 0.221193 y^2 - 0.101991 y^3 + 
   0.0277477 y^4) &\nonumber\\
W^{10}_{\Sigma^\prime}(y)&= e^{-2y} (0.0520466 - 0.18713 y + 0.233007 y^2 - 0.0985082 y^3 + 
   0.0259327 y^4)&\nonumber\\
W^{01}_{\Sigma^\prime}(y)&= e^{-2y} (0.0520466 - 0.18713 y + 0.233007 y^2 - 0.0985082 y^3 + 
   0.0259327 y^4)&\nonumber\\
W^{00}_{\Phi^{\prime\prime}}(y)&= e^{-2y} (2.80498 - 2.24306 y + 0.455491 y^2)&\nonumber\\
W^{11}_{\Phi^{\prime\prime}}(y)&= e^{-2y} (0.021493 - 0.0156159 y + 0.00596886 y^2)&\nonumber\\
W^{10}_{\Phi^{\prime\prime}}(y)&= e^{-2y} (-0.180417 + 0.137389 y - 0.0239615 y^2)&\nonumber\\
W^{01}_{\Phi^{\prime\prime}}(y)&=e^{-2y} (-0.180417 + 0.137389 y - 0.0239615 y^2) &\nonumber\\
W^{00}_{\tilde{\Phi}^\prime}(y)&= e^{-2y} (0.0000680703 - 0.000376682 y + 0.00340251 y^2 )&\nonumber\\%- 
   %4.74722\times10^{-20} y^3 + 2.44131\times10^{-36} y^4)&\nonumber\\
W^{11}_{\tilde{\Phi}^\prime}(y)&= e^{-2y} (0.0149622 - 0.00563307 y + 0.00440385 y^2) &\nonumber\\%- 
  %7.12645\times10^{-20} y^3 + 2.3947\times10^{-36} y^4)&\nonumber\\
W^{10}_{\tilde{\Phi}^\prime}(y)&= e^{-2y} (-0.0010092 + 0.00298228 y + 0.00281525 y^2) &\nonumber\\%+ 
  %8.43958\times10^{-20} y^3 - 2.3947\times10^{-36} y^4)&\nonumber\\
W^{01}_{\tilde{\Phi}^\prime}(y)&= e^{-2y} (-0.0010092 + 0.00298228 y + 0.00281525 y^2) &\nonumber\\%+ 
   %8.43958\times10^{-20} y^3 - 2.3947\times10^{-36} y^4)&\nonumber\\
W^{00}_{\Delta}(y)&= e^{-2y} (0.126043 - 0.100835 y + 0.0237577 y^2)&\nonumber\\
W^{11}_{\Delta}(y)&= e^{-2y} (0.05736 - 0.045888 y + 0.012102 y^2)&\nonumber\\
W^{10}_{\Delta}(y)&= e^{-2y} (0.0850285 - 0.0680228 y + 0.016845 y^2)&\nonumber\\
W^{01}_{\Delta}(y)&= e^{-2y} (0.0850285 - 0.0680228 y + 0.016845 y^2)&\nonumber\\
W^{00}_{M\Phi^{\prime\prime}}(y)&= e^{-2y} (-15.6228 + 19.3589 y - 7.23234 y^2 + 0.79705 y^3) &\nonumber\\ 
W^{11}_{M\Phi^{\prime\prime}}(y)&= e^{-2y} (-0.0370794 + 0.0852545 y - 0.0449284 y^2 + 0.00866992 y^3)&\nonumber\\
W^{10}_{M\Phi^{\prime\prime}}(y)&= e^{-2y} (0.578632 - 1.00438 y + 0.491252 y^2 - 0.0730693 y^3)&\nonumber\\
W^{01}_{M\Phi^{\prime\prime}}(y)&= e^{-2y} (1.00112 - 1.15934 y + 0.40275 y^2 - 0.0364952 y^3)&\nonumber\\
W^{00}_{\Sigma^{\prime}\Delta}(y)&= e^{-2y} (-0.0883243 + 0.185775 y - 0.104001 y^2 + 0.0163635 y^3)&\nonumber\\
W^{11}_{\Sigma^{\prime}\Delta}(y)&= e^{-2y} (-0.0501045 + 0.114845 y - 0.0729898 y^2 + 0.0131315 y^3)&\nonumber\\
W^{10}_{\Sigma^{\prime}\Delta}(y)&= e^{-2y} (-0.0742731 + 0.170242 y - 0.105744 y^2 + 0.0188197 y^3)&\nonumber\\
W^{01}_{\Sigma^{\prime}\Delta}(y)&= e^{-2y} (-0.0595834 + 0.125323 y - 0.0717204 y^2 + 0.011398 y^3)&\nonumber\\
\end{flalign}

\subsection*{Silicon ($^{28}$Si)}
\begin{flalign}
W^{00}_{M}(y)&= 0.281695 e^{-2y} (7.44089 - 6.37784 y + y^2)^2 &\nonumber\\ 
W^{00}_{\Phi^{\prime\prime}}(y)&=0.0739103 e^{-2y} (2.5 - y)^2 &\nonumber\\
W^{00}_{M\Phi^{\prime\prime}}(y)&= e^{-2y} (-2.68415 + 3.37434 y - 1.281 y^2 + 0.144292 y^3)&\nonumber\\
\end{flalign}

\subsection*{Sulfur ($^{32}$S)} 

\begin{flalign}
W^{00}_{M}(y)&=  0.580305 e^{-2y} (5.92494 - 5.43118 y + y^2)^2&\nonumber\\ 
W^{00}_{\Phi^{\prime\prime}}(y)&= 0.0765941 e^{-2y} (2.5 - y)^2&\nonumber\\
W^{00}_{M\Phi^{\prime\prime}}(y)&= e^{-2y} (-3.12284 + 4.11173 y - 1.6721 y^2 + 0.210827 y^3)&\nonumber\\
\end{flalign}

\subsection*{Argon ($^{40}$Ar)}
\begin{flalign}
W^{00}_{M}(y)&= e^{-2y} (31.8294- 65.9618 y + 48.5834 y^2 - 15.194 y^3 + 
   1.9036 y^4 - 0.0595886 y^5 &\nonumber \\ &+ 0.000544329 y^6) &\nonumber\\ 
W^{11}_{M}(y)&= e^{-2y} (0.318304 - 1.06524 y + 1.24846 y^2 - 0.62249 y^3 + 
   0.141618 y^4 - 0.0138797 y^5 &\nonumber\\ &+ 0.000480513 y^6) &\nonumber\\
W^{10}_{M}(y)&= e^{-2y} (-3.18299 + 8.62425 y - 8.02539 y^2 + 3.19316 y^3 - 
   0.554467 y^4 + 0.0353797 y^5 &\nonumber\\ &- 0.000511426 y^6) &\nonumber\\
W^{01}_{M}(y)&=  e^{-2y} (-3.18299 + 8.62425 y - 8.02539 y^2 + 3.19316 y^3 - 
   0.554467 y^4 + 0.0353797 y^5 &\nonumber\\ &- 0.000511426 y^6) &\nonumber\\
W^{00}_{\Phi^{\prime\prime}}(y)&= e^{-2y} (0.299629 - 0.373798 y + 0.154895 y^2 - 0.0238983 y^3 + 
   0.00122474 y^4)&\nonumber\\
W^{11}_{\Phi^{\prime\prime}}(y)&= e^{-2y} (0.00414999 - 0.0181474 y + 0.0240755 y^2 - 0.00926264 y^3 + 
   0.00108115 y^4)&\nonumber\\
W^{10}_{\Phi^{\prime\prime}}(y)&= e^{-2y} (-0.0352627 + 0.0990955 y - 0.0683453 y^2 + 0.0161561 y^3 - 
   0.00115071 y^4)&\nonumber\\
W^{01}_{\Phi^{\prime\prime}}(y)&= e^{-2y} (-0.0352627 + 0.0990955 y - 0.0683453 y^2 + 0.0161561 y^3 - 
   0.00115071 y^4) &\nonumber\\
W^{00}_{M\Phi^{\prime\prime}}(y)&= e^{-2y} (-3.08821 + 5.12625 y - 2.89248 y^2 + 0.653386 y^3 - 
   0.0526576 y^4 &\nonumber\\ &+ 0.000816493 y^5)&\nonumber\\ 
W^{11}_{M\Phi^{\prime\prime}}(y)&= e^{-2y} (-0.036345 + 0.140282 y - 0.171917 y^2 + 0.0770456 y^3 - 
   0.0134973 y^4 &\nonumber\\ &+ 0.000720769 y^5)&\nonumber\\
W^{10}_{M\Phi^{\prime\prime}}(y)&= e^{-2y} (0.308826 - 0.709394 y + 0.515378 y^2 - 0.153134 y^3 + 
   0.0185641 y^4 &\nonumber\\& - 0.000767139 y^5)&\nonumber\\
W^{01}_{M\Phi^{\prime\prime}}(y)&= e^{-2y} (0.363444 -1.17124 y + 1.09117 y^2 - 0.373592 y^3 + 
   0.0452762 y^4 &\nonumber\\& - 0.000767139 y^5)&\nonumber\\
   \end{flalign}

\subsection*{Calcium ($^{40}$Ca)}
\begin{flalign}
W^{00}_{M}(y)&=  0.000016743 e^{-2y} (1378.8 - 1387.54 y + 281.953 y^2 - y^3)^2&\nonumber\\
W^{00}_{\Phi^{\prime\prime}}(y)&= 0.0000376718 e^{-2y} (13.117 - 8.74678 y + y^2)^2&\nonumber\\
W^{00}_{M\Phi^{\prime\prime}}(y)&= e^{-2y} (-0.454214 + 0.759976 y - 0.432314 y^2 + 0.0971138 y^3 - 
   0.00730079 y^4 &\nonumber\\ &+ 0.0000251146 y^5)&\nonumber\\
\end{flalign}

\subsection*{Iron ($^{56}$Fe)}

\begin{flalign}
W^{00}_{M}(y)&=  e^{-2y} (62.3888 - 160.428 y + 152.644 y^2 - 67.2779 y^3 + 
   14.478 y^4 - 1.43665 y^5 &\nonumber\\&+ 0.0525291 y^6) &\nonumber\\ 
W^{11}_{M}(y)&= e^{-2y} (0.318309 - 1.27323 y + 1.99188 y^2 - 1.54562 y^3 + 
   0.622264 y^4 - 0.122277 y^5 &\nonumber\\ &+ 0.00921525 y^6) &\nonumber\\
W^{10}_{M}(y)&= e^{-2y} (-4.45633 + 14.6422 y - 18.2579 y^2 + 10.8919 y^3 - 
   3.2296 y^4 + 0.446836 y^5 &\nonumber\\& - 0.0220016 y^6) &\nonumber\\
W^{01}_{M}(y)&= e^{-2y} (-4.45633 + 14.6422 y - 18.2579 y^2 + 10.8919 y^3 - 
   3.2296 y^4 + 0.446836 y^5 &\nonumber\\ &- 0.0220016 y^6)  &\nonumber\\
W^{00}_{\Phi^{\prime\prime}}(y)&= e^{-2y} (4.22872 - 6.76595 y + 3.79067 y^2 - 0.867433 y^3 + 
   0.069506 y^4)&\nonumber\\
W^{11}_{\Phi^{\prime\prime}}(y)&= e^{-2y} (0.143378 - 0.229404 y + 0.144606 y^2 - 0.0422756 y^3 + 
   0.00486921 y^4)&\nonumber\\
W^{10}_{\Phi^{\prime\prime}}(y)&=e^{-2y} (-0.778655 + 1.24585 y - 0.741661 y^2 + 0.194658 y^3 - 
   0.0183967 y^4) &\nonumber\\
W^{01}_{\Phi^{\prime\prime}}(y)&=e^{-2y} (-0.778655 + 1.24585 y - 0.741661 y^2 + 0.194658 y^3 - 
   0.0183967 y^4) &\nonumber\\
W^{00}_{M\Phi^{\prime\prime}}(y)&= e^{-2y} (-16.2427 + 33.8776 y - 25.2342 y^2 + 8.30471 y^3 - 
   1.20334 y^4 + 0.0604243 y^5)&\nonumber\\ 
W^{11}_{M\Phi^{\prime\prime}}(y)&= e^{-2y} (-0.213631 + 0.598168 y - 0.622338 y^2 + 0.308014 y^3 - 
   0.0735211 y^4 &\nonumber\\ &+ 0.00669858 y^5)&\nonumber\\
W^{10}_{M\Phi^{\prime\prime}}(y)&=e^{-2y} (1.16019 - 3.24853 y + 3.31473 y^2 - 1.54264 y^3 + 
   0.325833 y^4- 0.0253084 y^5) &\nonumber\\
W^{01}_{M\Phi^{\prime\prime}}(y)&= e^{-2y} (2.99085 - 6.23805 y + 4.81422 y^2 - 1.74483 y^3 + 
   0.288128 y^4 - 0.015993 y^5) &\nonumber\\
\end{flalign}

\subsection*{Nickel ($^{58}$Ni)}
\begin{flalign}
W^{00}_{M}(y)&=  e^{-2y} (66.9246 - 175.389 y + 169.877 y^2 - 76.127 y^3 + 16.6597 y^4 - 1.6839 y^5 \nonumber\\
&+ 0.0628067 y^6)\nonumber\\
W^{11}_{M}(y)&= e^{-2y} (0.0795762 - 0.318305 y + 0.548985 y^2 - 0.503018 y^3 + 0.250492 y^4\nonumber\\
& - 0.0603789 y^5 + 0.00545169 y^6)\nonumber\\
W^{10}_{M}(y)&= e^{-2y} (-2.30773 + 7.63937 y - 10.3404 y^2 + 6.95311 y^3 - 2.30652 y^4 + 0.350525 y^5 \nonumber\\
&- 0.0185041 y^6)\nonumber\\
W^{01}_{M}(y)&= e^{-2y} (-2.30773 + 7.63937 y - 10.3404 y^2 + 6.95311 y^3 - 2.30652 y^4 + 0.350525 y^5 \nonumber\\
&- 0.0185041 y^6)\nonumber\\
W^{00}_{\Phi^{\prime\prime}}(y)&= e^{-2y} (5.4697 - 8.75152 y + 4.88454 y^2 - 1.10715 y^3 + 0.0875404 y^4)\nonumber\\
W^{11}_{\Phi^{\prime\prime}}(y)&= e^{-2y} (0.00977975 - 0.0156476 y + 0.0136707 y^2 - 0.00592935 y^3 + 0.00140426 y^4)\nonumber\\
W^{10}_{\Phi^{\prime\prime}}(y)&= e^{-2y} (-0.231284 + 0.370054 y - 0.264922 y^2 + 0.0935201 y^3 - 0.0110873 y^4)\nonumber\\
W^{01}_{\Phi^{\prime\prime}}(y)&= e^{-2y} (-0.231284 + 0.370054 y - 0.264922 y^2 + 0.0935201 y^3 - 0.0110873 y^4)\nonumber\\
W^{00}_{M\Phi^{\prime\prime}}(y)&=e^{-2y} (-19.1326 + 40.3764 y - 30.3339 y^2 + 10.0435 y^3 - 1.4629 y^4 + 
 0.0741493 y^5)\nonumber\\
W^{11}_{M\Phi^{\prime\prime}}(y)&=e^{-2y} (-0.0278969 + 0.0781112 y - 0.0956406 y^2 + 0.0607914 y^3 - 0.0211633 y^4 \nonumber\\
&+ 0.00276687 y^5)\nonumber\\
W^{10}_{M\Phi^{\prime\prime}}(y)&=e^{-2y} (0.659741 - 1.84727 y + 2.0953 y^2 - 1.10461 y^3 + 0.25912 y^4 - 0.0218459 y^5)\nonumber\\
W^{01}_{M\Phi^{\prime\prime}}(y)&= e^{-2y} (0.809015 - 1.7073 y + 1.48687 y^2 - 0.692274 y^3 + 0.145722 y^4 \nonumber\\
&- 0.00939131 y^5)
\end{flalign}

\providecommand{\newblock}{}

%\bibliography{ref}{}
%\bibliographystyle{iopart-num}

\end{document}